\documentclass[showpacs,superscriptaddress,preprintnumbers,amsmath,amssymb,nofootinbib]{revtex4}

\usepackage{amsmath,amssymb,latexsym}
\usepackage{graphicx,bm}
\usepackage{dcolumn}    
\usepackage{epsf,psfig}
\begin{document}
%
%
%
%
\newcommand{\sinc}{ {\mathrm{sinc}} } 
\newcommand{\hf}{ {\hat{f}} }
\newcommand{\Rdots}{\rotatebox[origin=c]{-45}{\vdots}}
%
%
%
%
%
\title{ Probing anisotropies of gravitational-wave backgrounds
  \\
  with a space-based interferometer II:
  \\
  \it{ 
  perturbative reconstruction of a low-frequency skymap
  } }

\author{Atsushi Taruya} \email{ataruya_at_utap.phys.s.u-tokyo.ac.jp}
\affiliation{ Research Center for the Early Universe~(RESCEU), School
  of Science, The University of Tokyo, Tokyo 113-0033, Japan }

\author{Hideaki Kudoh} \email{kudoh_at_utap.phys.s.u-tokyo.ac.jp}
\affiliation{ Department of Physics, The University of Tokyo, Tokyo
  113-0033, Japan }

\preprint{UTAP-531, RESCEU-34/05}
\begin{abstract}
We present a perturbative reconstruction method to make a skymap of 
gravitational-wave backgrounds (GWBs) observed via space-based 
interferometer. In the presence of anisotropies in GWBs, the 
cross-correlated signals of observed GWBs are inherently time-dependent 
due to the non-stationarity of the gravitational-wave detector. Since 
the cross-correlated signal is obtained through an all-sky integral 
of primary signals convolving with the antenna pattern function of 
gravitational-wave detectors, the non-stationarity of cross-correlated 
signals, together with full knowledge of antenna pattern functions, 
can be used to reconstruct an intensity map of the GWBs. 
Here, we give two simple methods to reconstruct a skymap of GWBs 
based on the perturbative expansion in low-frequency regime. 
The first one is based on harmonic-Fourier representation of data 
streams and the second is based on ``direct'' time-series data. 
The latter method enables us to create a skymap in a direct manner.   
The reconstruction technique is demonstrated for the case of the 
Galactic gravitational wave background observed via planned space 
interferometer, LISA. Although the angular resolution of 
low-frequency skymap is rather restricted, the methodology presented 
here would be helpful in discriminating the GWBs of galactic origins 
by those of the extragalactic and/or cosmological origins.
\end{abstract}

\maketitle

\section{Introduction}
\label{sec:intro}

The gravitational-wave background, incoherent superposition of
gravitational waves coming from many unresolved point-sources and/or 
diffuse-sources would be a cosmological gold mine to probe the dark side 
of the Universe. While these signals are generally random and act as 
{\it confusion noises} with respect to a periodic gravitational-wave
signal, the statistical properties of gravitational-wave backgrounds
(GWBs) carry valuable information such as physical processes of 
gravitational-wave emission, source distribution and
populations. Moreover, the extremely early universe beyond the
last-scattering surface of cosmic microwave background (CMB) can be
directly explored using GWB. Therefore, GWBs may be regarded as an
ultimate cosmological tool alternative to the CMB.

Currently, several grand-based gravitational-wave detectors are now
under scientific operation, and the search for gravitational waves
enters a new era. There are several kinds of target GW sources in these 
detectors, such as, coalescence of neutron star binaries and
core-collapsed supernova. On the other hand, planned space-based
detector, LISA (Laser Interferometer Space Antenna) and the
next-generation detectors, e.g., DECIGO \cite{Seto:2001qf} and 
BBO \cite{BBO:2003} aim at detecting gravitational waves in
low-frequency band $0.1$mHz --$0.1$Hz, in which many detectable
candidates for GWBs exist. Among them, the GWB originated from the 
inflationary epoch may be detected directly in this band 
(e.g., \cite{Ungarelli:2000jp,Smith:2005mm}). 
Therefore, for future application to cosmology, various implications for 
these GWBs should deserve consideration both from the theoretical and
the observational viewpoint.

One fundamental issue to access a new subject of cosmology is to make a 
skymap of GWB as a first step. Similar to the case of the CMB, the 
intensity map of the GWBs is of particular interest and it provides
valuable information, which plays a key role to clarify the origin of
GWBs. Importantly, the low-frequency GWBs observed by LISA are expected
to be anisotropic due to the contribution of Galactic binaries to the 
confusion noise (\cite{Hils:1990hg,Bender:1997bc,Nelemans:2001hp}, see
also recent numerical simulations  
\cite{Benacquista:2003th,Edlund:2005ye,Timpano:2005gm}).
Thus, the GWB skymap is potentially useful to discriminate the
individual backgrounds from many superposed GWBs, as well as to identify 
the underlying physical processes. The basic idea to create a skymap of
GWB is to use the directional information obtained through the
time-modulation of the correlation signals, which is caused by the
motion of gravitational-wave detector. As detector's antenna pattern
sweeps out the sky, the amplitude of the gravitational-wave signal would 
gradually change in time if the intensity distribution of GWB is
anisotropic. Using this, a method to explore anisotropies of GWB has
been proposed \cite{Giampieri:1997,Giampieri:1997ie,Allen:1997gp}. Later, 
the methodology was applied to study the anisotropic GWB observed by
space interferometer, LISA 
\cite{Ungarelli:2001xu,Cornish:2001hg,Cornish:2002bh,Seto:2004ji,Seto:2004np}.

In the previous paper \cite{Kudoh:2004he}, which is referred to as 
{\it Paper I} in the present paper, we have investigated the directional 
sensitivity of space interferometer to the anisotropy of GWB. 
Particularly focusing on the geometric properties of antenna pattern 
functions and their angular power, we found that the angular sensitivity 
to the anisotropic GWBs is severely restricted by the data combination
and the symmetry of detector configuration. As a result, in the case of
the single LISA detector, detectable multipole moments are limited to 
$\ell\lesssim 8$--$10$ with the effective strain sensitivity 
$h\sim 10^{-20}$ Hz$^{-1/2}$. 
This is marked contrast to the angular sensitivity to the chirp signals 
emitted from point sources, in which the angular resolution can reach at 
a level of a square degree or even better than that 
\cite{Cutler:1997ta,Moore:1999zw,Peterseim:1997ic,Takahashi:2003wm}.

Despite the poor resolution of LISA detector with respect to GWBs, 
making a skymap of GWBs is still an important general issue and thus 
needs to be investigated. In the present paper, we continue to
investigate the map-making problem and consider how the intensity map of the 
GWBs is reconstructed from the time-modulation signals observed via
space interferometer. In particular, we are interested in the 
low-frequency GWB, the wavelength of which is longer than the arm-length 
of the detector. In such case, the frequency dependence of the detector 
response becomes simpler and a perturbative scheme based on the
low-frequency expansion of the antenna pattern functions can be
applied. Owing to the least-squares approximation, we present a robust 
reconstruction method. The methodology is quite general and is also
applicable to the map-making problem in the case of the ground detectors.
We demonstrate how the present reconstruction method works well in a
specific example of GWB source, i.e., Galactic confusion-noise background. 
With a sufficient high signal-to-noise ratio for each anisotropic
components of GWB, we show that the space interferometer 
LISA can create the low-frequency skymap of Galactic GWB with angular 
resolution $\ell\leq5$. 
Since the resultant skymap is obtained in a non-parametric way without 
any assumption of source distributions, we hope that despite the poor 
angular resolution, the present methodology will be helpful to give a
tight constraint on the luminosity distribution of GWBs if we combine 
it with the other observational techniques.

The organization of the paper is as follows. In Sec.
\ref{sec:detection of anisotropy}, a brief discussion on the detection
and the signal processing of anisotropic GWBs is presented, together
with the detector response of space interferometer. Sec.\ref{sec:method} 
describes the details of the reconstruction of a GWB skymap in 
low-frequency regime. Owing to the least-squares approximation and the 
low-frequency expansion, a perturbative reconstruction scheme is developed.
Related to this (in the case of LISA), and we give an important remark
on the degeneracy between some multipole moments (Appendix 
\ref{appendix:on_the_degeneracy}). In Sec.\ref{sec:demonstration}, the 
reconstruction method is demonstrated in the case of the Galactic GWB. 
The signal-to-noise ratios for anisotropy of GWB are evaluated and the 
feasibility to make a skymap of GWB is discussed.  Finally, 
Section \ref{sec:conclusion} is devoted to a summary and conclusion.

\section{Basic formalism}
\label{sec:detection of anisotropy}

\subsection{Correlation analysis}
\label{subsec:correlation}

Let us begin with briefly reviewing the signal processing of 
gravitational-wave backgrounds based on the correlation analysis 
\cite{Kudoh:2004he}. Stochastic gravitational-wave backgrounds are
described by incoherent superposition of plane gravitational waves 
${\mathbf h}=h_{ij}$ given by  
\begin{eqnarray}
{\mathbf h}(t,\mathbf{x})= \sum_{A=+,\times} \int_{-\infty}^{+\infty} df
\int d\mathbf{\Omega}\,\,e^{i\,\,2\pi\,f(t-\mathbf{\Omega}\cdot \mathbf{x})}
\,\tilde{h}_A(f,\mathbf{\Omega})\,\mathbf{e}^A(\mathbf{\Omega}).
\end{eqnarray}
The Fourier amplitude $\tilde{h}_A (f,\mathbf{\Omega})$ of the
gravitational waves for the two polarization modes 
${\mathbf e}^A$ ($A=+,\times$) is assumed to be characterized by a
stationary 
random process with zero mean $\big\langle  \widetilde{h}_A \big\rangle =0 $. 
The power spectral density $S_h$ is then defined by 
\begin{eqnarray}
    \left\langle 
        \widetilde{h}_A^*  (f, {\bf{\Omega}})  
        {{\widetilde{h}}_{A'} }  (f', {\bf{\Omega}}')  
    \right\rangle  
    &=& 
    \frac{1}{2} \delta(f-f') 
    \frac{ \delta^2 (  {\bf{\Omega}} - {\bf{\Omega}}')}{4\pi}
    \delta_{AA'} S_h(|f|, \,\mathbf{\Omega}), 
    \label{eq:h*h gaussian process}
\end{eqnarray}
where ${\mathbf{\Omega}}$ is the direction of a propagating plane wave. 
Note that the statistical independence between two different directions
in the sky is implicitly assumed in equation (\ref{eq:h*h gaussian
process}), which might not be generally correct. Actually, CMB skymap
exhibits a large-angle correlation between the different skies, which
reflects the primordial density fluctuations. For the GWB of our
interest, the wavelength of the tensor fluctuations detected via space 
interferometers is much shorter than the cosmological scales and thereby 
the assumption put in equation (\ref{eq:h*h gaussian process}) 
is practically valid.

The detection of a gravitational-wave background is achieved through the 
correlation analysis of two data streams. The planned space
interferometer, LISA and also the next generation detectors DECIGO/BBO 
constitute several spacecrafts, each of which exchanges laser beams with 
the others. Combining these laser pulses, it is possible to synthesize
the various output streams which are sensitive (or insensitive) to the 
gravitational-wave signal. The output stream for a specific combination 
$I$ denoted by $s_I(t)$ is generally described by a sum of the 
gravitational-wave signal $h_I(t)$ and the instrumental noise $n_I(t)$ by  
\begin{equation*}
    s_I(t) = h_I(t)+ n_I(t). 
\end{equation*}
We assume that the noise $n_I(t)$ is treated as a Gaussian random
process with spectral density $S_{\rm n}(f)$ and zero mean 
$\big\langle n_I\big\rangle=0 $. The gravitational-wave signal $h_I$ is 
obtained by contracting $\mathbf{h}$ with detector's response function 
and/or detector tensor ${\mathbf D}:=D_{ij}$:  
\begin{equation}
    h_I(t) = \sum_{A=+,\times} \int_{-\infty}^{+\infty} df 
    \int d \mathbf{\Omega}  ~
    e^{i2\pi f (t-\mathbf{\Omega} \cdot \mathbf{x}_I) }
    {\textbf{D}}_I    ( {\mathbf{\Omega}} ,f;t) \,
     {\textbf : }\, \mathbf{e}^{A}(\mathbf{\Omega})\,\,
     \widetilde{h}_A(f,\mathbf{\Omega}).  
\label{eq:h_I(t)}
\end{equation}
Note that the response function explicitly depends on time. 
The time variation of response function is caused by the orbital motion
of the detector and it plays a key role in reconstructing a skymap of the GWBs. 
Typically, the orbital frequency of the detector motion is much lower
than the observed frequency and within the case, the expression 
(\ref{eq:h_I(t)}) is validated.

Provided the two output data sets, the correlation analysis is examined 
depending on the strategy of data analysis, i.e., self-correlation
analysis using the single data stream or cross-correlation analysis
using the two independent data stream. 
Defining $S_{IJ}(t) \equiv \left\langle s_I(t)s_J(t) \right\rangle$, the Fourier 
counterpart of it $\widetilde{S}_{IJ}(t,f)$, which is related with 
$S_{IJ}(t)$ by $S_{IJ}(t)=\int df \widetilde{S}_{IJ}(t,f)$, becomes
\footnote{
In Paper I, we had used $C_{IJ}$ to denote the output data 
$\left\langle s_I(t)s_J(t) \right\rangle$ itself. This coincides with 
the present definition if we neglect the instrumental noises. 
}
\begin{eqnarray} 
    \widetilde{S}_{IJ}(t,f) & = &  
      \widetilde{C}_{IJ}(t,f) + \delta_{IJ} \,\,S_n(|f|),
\label{eq:detector output S(t)}
\end{eqnarray}
where we define
\begin{eqnarray}
\widetilde{C}_{IJ}(t,f)=\int \frac{d
      \mathbf{\Omega}}{4\pi}  S_h(|f|, \mathbf{\Omega})~ 
    \mathcal{ F}_{IJ}^E(f,  \mathbf{\Omega};\,t). 
\label{eq:def_of_C(f)}
\end{eqnarray}
Here, the function $\mathcal{F}^E_{IJ}$ is the so-called antenna pattern 
function defined in an ecliptic coordinate, which is expressed in terms 
of detector's response function: 
\begin{eqnarray}
    && \mathcal{F}^E_{IJ}(f, \mathbf{\Omega};\,t)=  
    e^{ i \, 2\pi f \,{\bf \Omega \cdot}(\textbf{x}_I -\textbf{x}_J) }
    \sum_{A=+,\times} 
    F_I^{A*}(  \mathbf{\Omega},f;\,t) F_J^A( \mathbf{\Omega},f;\,t)
    \cr
    && F_I^A(  \mathbf{\Omega},f;\,t)  = 
    {\bf D}_I( \mathbf{\Omega},f;\,t)\, \textbf{:} \,
    \textbf{e} ^A (\mathbf{\Omega})
    \label{eq:def_of_antenna}
\end{eqnarray}

Apart from the second term, equation (\ref{eq:detector output S(t)})
implies that the luminosity distribution of GWBs
$S_h(f,\mathbf{\Omega})$ can be obtained by deconvolving the all-sky 
integral of antenna pattern function from the time-series data $S_{IJ}(t)$. 
To see this more explicitly, we focus on equation (\ref{eq:def_of_C(f)}) 
and decompose the antenna pattern function and the luminosity
distribution into spherical harmonics in an ecliptic coordinate, i.e., 
sky-fixed frame:
\begin{eqnarray}
    S_h(|f|,\, \mathbf{\Omega}) 
    =  \sum_{\ell, m} \,\,[p_{\ell m} (f)]^* \,\,
    Y_{\ell m}^* ( \mathbf{\Omega}), 
\cr
    \mathcal{F}^E_{IJ}(f,\, \mathbf{\Omega};\,t)
    = 
    \sum_{\ell, m} \,\, a_{\ell m}^E (f,t) \,\,
    Y_{\ell m} ( \mathbf{\Omega}). 
    \label{eq:Y_lm expansion of S and F}
\end{eqnarray}
Note that the properties of spherical harmonics yield 
$p_{\ell m}^*=(-1)^m p_{\ell,-m}$ and 
$a_{\ell m}^*=(-1)^{\ell-m}a_{\ell,-m}$, where the latter property comes from 
${\mathcal{F}}_{IJ}^* (f,  \Omega;t) = {\mathcal{F}}_{IJ} (f, - \Omega;t)$ 
\cite{Kudoh:2004he}. Substituting (\ref{eq:Y_lm expansion of S and F}) into
(\ref{eq:def_of_C(f)}) becomes
\begin{eqnarray}
    \widetilde{C}_{IJ}(t,f) = \frac{1}{4\pi}\, \sum_{\ell m}\,\, 
    \left[ p_{\ell m}(f) \right]^*  a^{E }_{\ell m}(f,\,t),  
\label{eq:C_IJ in E-frame} 
\end{eqnarray}
where we have dropped the contribution from the detector noise. 
The expression (\ref{eq:C_IJ in E-frame}) still involves the
time-dependence of antenna pattern functions. 
To eliminate this, one may further rewrite equation 
(\ref{eq:C_IJ in E-frame}) by employing the harmonic expansion in
detector's rest frame. 
We denote the multipole coefficients of the antenna pattern in
detector's rest frame by $a_{\ell m}$. 
The transformation between the detector rest frame and the sky-fixed
frame is described by a rotation matrix by the Euler angles 
$(\psi, \vartheta ,\varphi)$, whose explicit relation is expressed in 
terms of the Wigner $D$ matrices \footnote{
\label{footnote:orbital motion}
We are specifically concerned with the time dependence of directional 
sensitivity of antenna pattern functions, not the real orbital motion. 
Hence, only the time evolution of directional dependence is considered 
in the expression (\ref{eq:Wigner_formula}).}
\cite{Allen:1997gp,Cornish:2002bh, Edmonds:1957}:
\begin{eqnarray}
    a^E_{\ell m }(f,t)  = \sum_{n=-\ell}^\ell
    e^{-i \,n \,\psi}\, d^\ell_{nm}( \vartheta)\, e^{-i \,m \,\varphi}
    \,\,a_{\ell n}(f). 
    \label{eq:Wigner_formula}
\end{eqnarray}
Here the Euler rotation is defined to perform a sequence of rotation, 
starting with a rotation by $\psi$ about the original $z$ axis, followed 
by rotation by $\vartheta$ about the original $y$ axis, and ending with
a rotation by $\varphi$ about the original $z$ axis. 
The Wigner $D$ matrices for $n\ge m$ is 
\begin{eqnarray}
 d_{n m}^\ell ({\mathcal{\theta}}) =
 (-1)^{\ell-n}
 \sqrt{\frac{ (\ell+n)! (\ell-n)! }{ (\ell+m)!(\ell-m)!  } }
\left( \cos \frac{\theta}{2}\right)^{n+m}
\left(-\sin \frac{\theta}{2}\right)^{n-m} 
P_{\ell-n}^{(n+m, n-m)}(-\cos\theta)
\end{eqnarray}
with $P_n^{(a,b)}$ being Jacobi polynomial. For 
$n<m$, we have $d^{\ell}_{nm}=(-1)^{n-m} d^\ell_{mn}$.

\begin{figure}[t]
\begin{center}
\includegraphics[width=8cm,angle=0,clip]{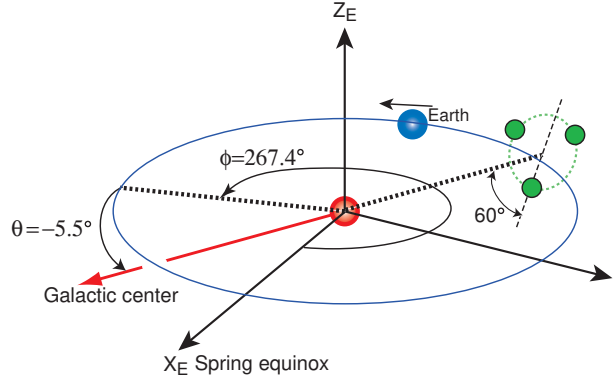}
\end{center}
\caption{
LISA configuration in the ecliptic coordinates. The Galactic center is at
RA $267.4^\circ$ and dec. $-5.5^{\circ}$ in this coordinate system.  
}
\label{fig:LISA}
\end{figure}

Let us now focus on the orbital motion of the LISA constellation (see 
Fig.\ref{fig:LISA}). 
The LISA orbital motion can be expressed by 
$\psi=-\omega t$, $\vartheta=-\pi/3$, $\varphi= \omega t$, where 
$\omega = 2\pi/T_0$ is the orbital frequency of LISA ($T_0=1$ sidereal year)
\footnote{
The relation $\psi=-\varphi$ does not necessarily hold for orbital motion of 
LISA and there may be some possibilities to impose a constant phase 
difference, i.e., $\psi=-\varphi + c$. However, for the sake of
simplicity, we put $c=0$.
}
Since the antenna pattern function is periodic in time due to the
orbital motion, the expected signals also vary in time with the same
period $T_0$. It is therefore convenient to express the output signals by 
\begin{equation}
\widetilde{C}_{IJ}(t,f)=\sum_{k=-\infty}^{+\infty}\,\,
\widetilde{C}_{IJ,k}(f)\,\,e^{i\,k\,\omega\,t}. 
\end{equation}
Using the relation (\ref{eq:Wigner_formula}) with specific parameter
set, the Fourier component $\widetilde{C}_{IJ,k}(f)$ is then given by 
\begin{eqnarray} 
    \tilde{C}_k(f) &\equiv&   \frac{1}{T_0} 
   \int^{T_0}_{0} dt~ e^{-ik \omega t}\,\, \tilde{C}_{IJ}(t,f)
\nonumber\\
&=&    \frac{1}{4\pi}
    \sum_{\ell=0}^{\infty} \sum_{m=- \ell}^{\ell-k}~ [p_{\ell m}(f)]^*\,\, 
    d^{\ell}_{(m+k),m}  \left( -\frac{\pi}{3}  \right) ~a_{\ell,(m+k)}(f)   
    \label{eq:deconvolution}
\end{eqnarray}
for $k\ge 0$. As for $k<0$, the lower and the upper limit of the sum 
over $m$ are changed to $m=-\ell-k$ and $m=\ell$, respectively.

Equation (\ref{eq:deconvolution}) as well as (\ref{eq:def_of_C(f)}) is 
the theoretical basis to reconstruct the skymap of GWBs. 
Given the output data $\tilde{C}_k(f)$ (or $\tilde{C}_{IJ}(t,f)$) 
experimentally, the task is to solve the linear system 
(\ref{eq:deconvolution}) with respect to $p_{\ell m}(f)$ for given
antenna pattern functions. One important remark deduced from equation 
(\ref{eq:deconvolution}) is that the accessible multipole coefficients 
$p_{\ell m}$ are severely restricted by the angular sensitivity of
antenna pattern functions. The important properties of the antenna
pattern functions are summarized in next subsection. 
Another important message is that the above linear systems are generally 
either over-constrained or under-determined. For the expression 
(\ref{eq:deconvolution}), if one truncates the multipole expansion with 
$\ell=\ell_\mathrm{max}$, the system consists of 
$\frac{1}{2}(\ell_{max}+1)(2\ell_{max}+1)$ unknowns and
$N(2\ell_{max}+1)$ equations (for $k \ge0$), where $N$ is the number of 
available modes of the antenna pattern functions. Thus this
deconvolution problem is, in principle, over-determined for a relatively 
small truncation multipole $\ell_{\text{max}}$, while it becomes 
under-determined for a larger value of  $\ell_{\text{max}}$. We will 
later discuss the over-determined case in
Sec.\ref{subsubsec:over-determined}, where 
$\ell_{\text{max}} = 5$ and $N \simeq 5$. 
%
%
%
%
%
%
%
\subsection{Detector response and antenna pattern functions}
\label{subsec:antenna_pattern}
%
%
%
%
%
\begin{table}[b]
\caption{\label{tab:summay_antenna}
Important properties of antenna pattern function. 
Note that the cross-correlated data are blind to isotropic GWBs. 
}
\begin{ruledtabular}
\begin{tabular}{cll}
Combination of variables & Visible multipole moments in low-frequency 
regime $(\hat{f}\lesssim1)$ & General properties  \\
\hline
(A,A), (E,E) & $\mathcal{O}(\hat{f}^2):$ $\ell=0,\,\,2,\,\,4$ 
& visible only to $\ell=$even \\
(T,T)     & $\mathcal{O}(\hat{f}^4):$ $\ell=0,\,\,4,\,\,6$ 
& visible only to $\ell=$even \\
(A,E)     & $\mathcal{O}(\hat{f}^2):$ $\ell=4$,\quad \quad 
$\mathcal{O}(\hat{f}^3):$ $\ell=3,\,\,5$
& blind to $\ell=0,\,\,1$ \\
(A,T), (E,T)  & $\mathcal{O}(\hat{f}^3):$ $\ell=1,\,\,3,\,\,5$
& blind to $\ell=0$ \\
\end{tabular}
\end{ruledtabular}
\end{table}

The output signals of space interferometer sensitive to the
gravitational waves are constructed by time-delayed combination of laser 
pulses. In the case of LISA, the technique to synthesize data streams 
canceling the laser frequency noise is known as time-delay
interferometry (TDI), which is crucial for our subsequent analysis. 
In the present paper, we use the optimal set of TDI variables
independently found by Prince et al. \cite{Prince:2002hp} and 
Nayak et al.\cite{Nayak:2002ir}, which are free from the noise correlation
\footnote{
The optimal TDI variables adopted here may not be the best TDI
combinations for the present purpose. There might be a better choice of
the signal combinations, although a dramatic improvement of the angular 
sensitivity would not be expected. }. 
A simple realization of such data set is obtained from a combination of 
Sagnac signals, which are the six-link observables using all six LISA 
oriented arms. For example, the Sagnac signal S$_1$ measures the phase 
difference accumulated by two laser beams received at space craft $1$, 
each of which travels around the LISA array in clockwise or 
counter-clockwise direction. 
The explicit expression of detector tensor for such signal, which we denote by 
${\bf D}_{\scriptscriptstyle\rm S_1}$, is given in equation (21) of 
Paper I (see also \cite{Cornish:2001bb,Armstrong:1999}).  
The analytic expression for other detector tensors ${\bf D}_{\rm S_2}$ 
and ${\bf D}_{\rm S_3}$ are also obtained by the cyclic permutation of 
the unit vectors $\mathbf{a},\,\mathbf{b}$ and $\mathbf{c}$.

Combining the three Sagnac signals, a set of optimal data combinations 
can be constructed 
\cite{Prince:2002hp,Nayak:2002ir} (see also \cite{Krolak:2004xp}): 
\begin{eqnarray}
&& {\bf D}_{\rm A} =  
\frac{1}{\sqrt{2} }( {\bf D}_{\rm S_3}- {\bf D}_{\rm S_1} ),
\nonumber
\\
&& {\bf D}_{\rm E} = 
\frac{1}{\sqrt{6}}(  
  {\bf D}_{\rm S_1} - 2{\bf D}_{ \rm S_2 } + {\bf D}_{ \rm S_3} ),
\nonumber
\\
&& {\bf D}_{\rm T} = 
\frac{1}{\sqrt{3}}( {\bf D}_{\rm S_1} 
+  {\bf D}_{\rm S_2} + {\bf D}_{\rm S_3} ).  
\label{eq:def AET mode}
\end{eqnarray}
These three detector tensors are referred to as $A,E,T$-variables, 
which generate six kinds of antenna pattern functions. 
Notice that in the equal arm-length limit, frequency dependence of these 
functions is simply expressed in terms of the normalized frequency: 
\begin{equation}
\hat{f}\equiv\frac{f}{f_*},  
\label{eq:normalized_f}
\end{equation}
where the characteristic frequency is given by 
$f_*=c/(2\pi L)$ with $L$ being the arm-length of detector.  
With the arm-length $L=5\times10^6$ km, the characteristic frequency of 
LISA becomes $f_*\simeq 9.54$ mHz. 
In what follows, we use the analytic expressions for equal arm-length 
limit to demonstrate the reconstruction of GWB skymap. 
This is sufficient for the present purpose, because we are concerned 
with a fundamental theoretical basis to map-making capability of GWBs. 
The idea provided in the present paper would allow us to examine a more 
realistic situation and the same strategy can be applied to an extended 
analysis in the same manner.

Finally, we note that the antenna pattern functions for the optimal 
combinations of TDI have several distinctive features in angular sensitivity.
In the low frequency limit ($\hat{f} \ll 1$), the antenna pattern
functions for the self-correlations and the cross-correlations are expanded as
\begin{equation}
\begin{aligned}
{\mathcal F}_{AA} (f,\,\mathbf{\Omega}) =~ 
 & {\mathcal F}_{AA}^{(2)}(\mathbf{\Omega})\, \hat{f}^2 &
 &&
 +~ &~{\mathcal F}_{AA}^{(4)}(\mathbf{\Omega})\, \hat{f}^4 &
                    + ~O(\hat{f}^6)
\\
{\mathcal F}_{TT} (f,\,\mathbf{\Omega}) =~ 
 & &
 & &
 &~{\mathcal F}_{TT}^{(4)}(\mathbf{\Omega})\, \hat{f}^4 &
   +~ O(\hat{f}^6)
\\
{\mathcal F}_{AE} (f,\,\mathbf{\Omega}) = ~ 
    &{\mathcal F}_{AE}^{(2)}(\mathbf{\Omega})\, \hat{f}^2&
  +~ &~{\mathcal F}_{AE}^{(3)}(\mathbf{\Omega})\, \hat{f}^3&
  +~ &~{\mathcal F}_{AE}^{(4)}(\mathbf{\Omega})\, \hat{f}^4&
  +~ O(f^5)
\\
{\mathcal F}_{AT} (f,\,\mathbf{\Omega}) =~ 
   & &
   &~{\mathcal F}_{AT}^{(3)}(\mathbf{\Omega})\, \hat{f}^3 &
 +~ &~{\mathcal F}_{AT}^{(4)}(\mathbf{\Omega})\, \hat{f}^4&
 + ~O(\hat{f}^5)
\end{aligned}
\label{eq:expand_antenna}
\end{equation}
The frequency dependence of ${\mathcal F}_{EE}$ and ${\mathcal F}_{ET}$
are the same as for  ${\mathcal F}_{AA}$ and ${\mathcal F}_{AT}$, respectively. 
Since the leading term of the TT-correlation is $O(\hat{f}^4)$, the 
$TT$-correlation becomes insensitive to the gravitational waves in the 
low-frequency regime. We will use all correlated data set except for the 
$TT$-correlation. In Table \ref{tab:summay_antenna}, we summarize the 
important properties for the multipole moments of antenna pattern functions.  
Also, in Appendix \ref{appendix:Mulipole coefficients}, employing the 
perturbative approach based on the low-frequency approximation 
$\hat{f}\ll1$, the spherical harmonic expansion for antenna pattern 
functions are analytically computed, which will be useful in subsequent analysis.  
%
%
%
%
%
%
%
%
%
%
\section{Perturbative reconstruction method 
for GWB skymap}
\label{sec:method}
%
%
%
%
%
%
%
%
%
We are in position to discuss the methodology to reconstruct a skymap of 
GWB based on the expression (\ref{eq:deconvolution}) (or 
(\ref{eq:def_of_C(f)})). 
Since we are specifically concerned with low-frequency sources observed
via LISA, it would be helpful to employ a perturbative approach using
the low-frequency expansion of antenna pattern function. 
In Sec.\ref{subsec:general_scheme}, owing to the expression 
(\ref{eq:deconvolution}), a perturbative reconstruction method is presented. 
Sec.\ref{subsec:time-domain} discusses alternative reconstruction method 
based on the time-series representation (\ref{eq:def_of_C(f)}).

\subsection{General scheme}
\label{subsec:general_scheme}

Hereafter, we focus on the GWBs in the low-frequency band of the
detector, the wavelength of which is typically longer than the
arm-length of the gravitational detector, i.e., $\hat{f}\lesssim 1$. 
Without loss of generality, we restrict our attention to the
reconstruction of a GWB skymap in a certain narrow frequency range, 
$f\sim f+ \Delta f$, within which a separable form of the GWB spectrum 
is a good assumption:  
\begin{equation}
S_h(f,\mathbf{\Omega}) = H(f)\,\,P(\mathbf{\Omega}). 
\label{eq:S_h_separable}
\end{equation}
Further, for our interest of the narrow bandwidth, it is reasonable to 
assume that the spectral density $H(f)$ is described by a power-law 
form as $H(f)= {\mathcal N} \, f^{\alpha}$. 
In the reconstruction analysis discussed below, the spectral index 
$\alpha$ is assumed to be determined beforehand from theoretical 
prediction and/or experimental constraint 
\footnote{In our general scheme, we do not assume a priori the 
normalization factor ${\mathcal N}$ in the function $H(f)$, which 
should be simultaneously determined with the reconstruction of an 
intensity distribution $P(\mathbf{\Omega})$. 
In the low-frequency approximation, $\hat{f}\ll1$, however, there 
exists a degeneracy between the monopole and the quadrupole 
components and one cannot correctly determine the normalization 
${\mathcal N}$. (See Appendix.\ref{appendix:on_the_degeneracy}.) }.

In the low frequency approximation up to the order 
$\mathcal{O}(\hat{f}^3)$, we have five output signals which 
respond to the GWBs, i.e., $AA$-, $EE$-, $AE$-, $AT$- and
$ET$-correlations, each output of which is represented by equation 
(\ref{eq:deconvolution}). 
Collecting these linear equations and arraying them appropriately, 
the linear algebraic equations can be symbolically written in a matrix form as  
\begin{eqnarray}
{\mathbf {c}}(f)  = {\mathbf {A}}(f)  \cdot {\mathbf {p}}.    
\label{eq: c=Ap}
\end{eqnarray}
In the above expression, while the vector ${\mathbf {p}}$ represents 
the unknowns consisting of the multipole coefficients of GWBs 
$p_{\ell m}$, the vector ${\mathbf {c}}(f)$ contains the correlation signals 
$\tilde{C}_{k}(f)$. 
Here, the frequency dependence of the function $H(f)$ has been already 
factorized and thereby the vector ${\mathbf {p}}$ only contains the 
information about the angular distribution. 
The matrix ${\mathbf {A}}(f)$ is the known quantity consisting of the 
multipole coefficients of each antenna pattern and the Wigner $D$ matrices
(see Appendix \ref{appendix:SVD} for an explicit example).

As we have already mentioned, the linear systems (\ref{eq: c=Ap}) 
become either over-determined or under-determined system. 
In most of our treatment in the low-frequency regime, the linear systems 
(\ref{eq: c=Ap}) tend to become over-determined, but this is not always 
correct depending on the amplitude of GWB spectrum  
$H(f)$ (see Sec.\ref{subsec:on_snr}). 
In any case, the matrix $\mathbf{A}$ would not be a square matrix and
the number of components of the vector $\mathbf{c}$ does not coincide
with the one in the vector $\mathbf{p}$. 
In the over-determined case, while there is a hope to get a unique solution 
$\mathbf{p}$ to produce the gravitational-wave signal 
$\mathbf{c}$, it seems practically difficult due to the errors
associated with the instrumental noise and/or numerical analysis. 
Hence, instead of pursuit of a rigorous solution, it would be better to 
focus on the issue how to get an approximate solution by a simple and 
systematic method.

In the case of our linear system, the approximate solution for the 
multipole coefficient $\mathbf{p}_{\rm approx}$ can be obtained from 
the least-squares method in the following form:  
\begin{equation}
\mathbf{p}_{\rm\scriptscriptstyle approx} 
= \mathbf{A}^+ \cdot \mathbf{c}.
\label{eq:p=AC}
\end{equation}
The matrix $\mathbf{A}^+$ is called the pseudo-inverse matrix of 
Moore-Penrose type, whose explicit expression can be uniquely determined 
from the singular value decomposition (SVD)
\cite{Press:NRC++}. According to the theorem of linear algebra, the
matrix $\mathbf{A}$, whose number of rows is greater than or equal to
its number of columns, can be generally written as 
$\mathbf{A}= U^T\cdot \mbox{diag}[w_i]\cdot V $. Here, the matrices $U$ and 
$V$ are orthonormal matrices which satisfy 
$U^{\dagger}\,U=V^{\dagger}\,V=\mathbf{1}$, 
where the quantity with subscript $^{\dagger}$ represents the Hermite 
conjugate variable. The quantity $\mbox{diag}[w_i]$ represents an
diagonal matrix with singular values $w_i$ with respect to the matrix 
$\mathbf{A}$. Then the pseudo-inverse matrix becomes 
\begin{equation}
    \mathbf{A}^+ = V^T\cdot \mbox{diag}[w_i^{-1}]\cdot U. 
\label{eq:pseudo-inverse_A}
\end{equation}
It should be stressed that the explicit form of the pseudo-inverse matrix 
$\mathbf{A}^+$ is characterized only by the angular dependence of
antenna pattern functions.

In principle, the least-squares method by SVD can work well and all the 
accessible multipole moments of GWB would be obtained as long as the
antenna pattern functions have the corresponding sensitivity to each 
detectable multipole moment. As we mentioned in 
Sec.\ref{subsec:antenna_pattern}, however, the angular power of antenna 
pattern function depends sensitively on the frequency. In the
low-frequency regime, the frequency dependence of the non-vanishing 
multipole moments appears at 
$\mathcal{O}(\hat{f}^2)$ for 
$\ell=0,\,\,2$ and $4$, and $\mathcal{O}(\hat{f}^3)$ for 
$\ell=1,\,\,3$ and $5$ (see Table \ref{tab:summay_antenna}). 
In this respect, by a naive application of the least-squares method, 
it is difficult to extract the information about $\ell=$odd modes of 
GWBs because the singular values of the matrix $\mathbf{A}$ are
dominated by the lowest-order contribution of the antenna pattern functions.

For a practical and a reliable estimate of the odd multipoles in 
low-frequency regime, the least-squares method should be applied
combining with the perturbative scheme described below.  Let us recall
from the expression (\ref{eq:expand_antenna}) that the matrix
$\mathbf{A}$ can be expanded as 
\begin{eqnarray}
\mathbf{A} = \hat{f}^2\,\,\mathbf{A}^{(2)} + 
\hat{f}^4\,\,\mathbf{A}^{(4)} + \cdots
\label{eq:expand_mat_A1}
\end{eqnarray}
for a matrix consisting of the self-correlation signals 
$\mathcal{F}_{AA}$ and $\mathcal{F}_{EE}$,  
\begin{eqnarray}
\mathbf{A} = \hat{f}^2\,\,\mathbf{A}^{(2)} + \hat{f}^3\,\,\mathbf{A}^{(3)} +
\hat{f}^4\,\,\mathbf{A}^{(4)} +  \cdots
\label{eq:expand_mat_A2}
\end{eqnarray}
for a matrix consisting of the cross-correlation signal 
$\mathcal{F}_{AE}$, and 
\begin{eqnarray}
\mathbf{A} = \hat{f}^3\,\,\mathbf{A}^{(3)} +
\hat{f}^4\,\,\mathbf{A}^{(4)} +  \cdots
\label{eq:expand_mat_A3}
\end{eqnarray}
for a matrix consisting of the cross-correlation signals  
$\mathcal{F}_{AT}$ and $\mathcal{F}_{ET}$.  
Then the resultant matrices $\mathbf{A}^{(i)}$ become independent of the 
frequency. The above perturbative expansion implies that the output
signal $\mathbf{c}$ is also expanded in powers of $\hat{f}$. Since the 
frequency dependence of the function 
$H(f)$ has been already factorized in equation (\ref{eq: c=Ap}), we have 
\begin{eqnarray}
\mathbf{c}(f) =
\left\{
\begin{array}{lll}
   & \hat{f}^2 ~ \mathbf{c}^{(2)} + 
    \hat{f}^4 \mathbf{c}^{(4)} + \cdots, 
    \quad\, & \text{for AA-, EE-correlations} 
\\ 
\\
    &  \hat{f}^2 ~ \mathbf{c}^{(2)} + 
    \hat{f}^3 ~ \mathbf{c}^{(3)} + 
    \hat{f}^4 \mathbf{c}^{(4)} + \cdots,
    \quad\, & \text{for AE-correlation} 
\\
\\
   &  \hat{f}^3 ~ \mathbf{c}^{(3)} + 
    \hat{f}^4 \mathbf{c}^{(4)} + \cdots,
    \quad\, &
    \text{for AT-,ET-correlations} 
\end{array}
\right.
\label{eq:expand_vec_c}
\end{eqnarray}
Substituting the terms (\ref{eq:expand_mat_A1})-(\ref{eq:expand_mat_A3}) and 
(\ref{eq:expand_vec_c}) into the expression (\ref{eq: c=Ap}) and 
collecting the terms of each order of $\hat{f}$, we have 
\begin{equation}
\mathbf{c}^{(i)} = \mathbf{A}^{(i)}\,\cdot\,\mathbf{p}^{(i)}, 
\quad \quad (i=2,\,\,3,\,\,4,\,\,\cdots)
\label{eq:perturbative_c=Ap}
\end{equation}
where the subscript $^{(i)}$ means the quantity consisting of the order 
$\mathcal{O}(\hat{f}^i)$ terms. 
The vector $\mathbf{p}^{(i)}$ represents the accessible multipole moments 
$p_{\ell m}$ to which the antenna pattern functions become sensitive 
in this order. For example, the vector $\mathbf{p}^{(2)}$ contains the 
$\ell=0,\,2$ and $4$ modes, while the vector $\mathbf{p}^{(3)}$  have 
the multipole moments with $\ell=1,\,3$ and $5$. Since the expression 
(\ref{eq:perturbative_c=Ap}) has no explicit frequency dependence, we 
can reliably apply the least-squares solution by the SVD to reconstruct 
the odd modes of GWBs, as well as the even modes:  
\begin{equation}
\mathbf{p}^{(i)}_{\rm\scriptscriptstyle approx } 
= [\mathbf{A}^{(i)}]^+\,\cdot\,\mathbf{c}^{(i)}.
\label{eq:perturbative_p=AC}
\end{equation}
Each step of the perturbative reconstruction scheme is summarized in Fig.
\ref{fig:flowchart} (see Appendix \ref{appendix:SVD} in more details).

\begin{figure}[htbp]
\begin{center}
\includegraphics[width=13cm,angle=0,clip]{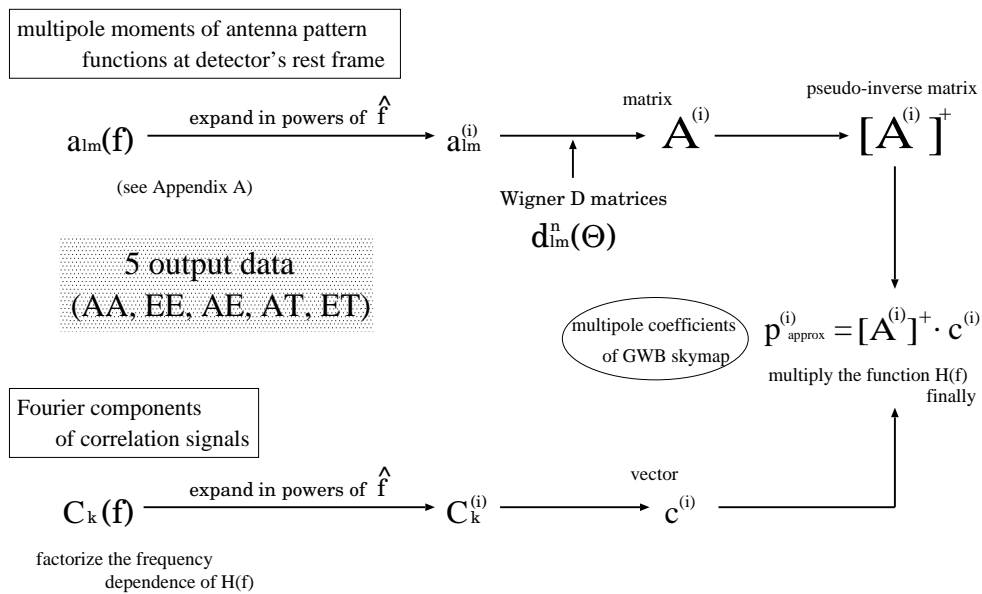}
\end{center}
    \caption{Flowchart of perturbative reconstruction scheme based on
      the harmonic-Fourier representation. 
    }
     \label{fig:flowchart}
\end{figure}

Finally, we note that the low-frequency reconstruction method presented
here assumes the perturbative expansion form of the output signal $\mathbf{c}$. 
To determine the coefficients $\mathbf{c}^{(i)}$ in equation 
(\ref{eq:expand_vec_c}), one must know the frequency dependence of the 
vector $\mathbf{c}$  in the narrow bandwidth $f\sim f+\Delta f$, which 
can be achieved by analyzing the multi-frequency data.  
One important remark is that the signal-to-noise ratio in each
reconstructed multipoles might be influenced by the sampling frequencies 
and/or the data analysis strategy. This point will be discussed later in Sec.
\ref{subsec:reconstruction}. 
%
%
%
%
%
%
%
%
%
%
\subsection{Reconstruction based on the time-series representation}
\label{subsec:time-domain}
%
%
%
%
%
So far, we have discussed the reconstruction method based on the 
harmonic-Fourier representation (\ref{eq:deconvolution}).  
The main advantage of the harmonic-Fourier representation is that 
it gives a simple algebraic equation suitable for applying the 
least-squares solution by SVD. 
However, the harmonic-Fourier representation implicitly assumes that 
the space interferometer orbits around the sun under keeping their 
configuration rigidly. 
In reality, rigid adiabatic treatment of the spacecraft motion is no 
longer valid due to the intrinsic variation of arm-length caused by the 
Keplerian motion of three space crafts, as well as the tidal 
perturbation by the gravitational force of solar system planets 
\cite{Cornish:2003tz,Tinto:2003vj,Shaddock:2003bc}. 
In a rigorous sense, the time dependence of antenna pattern 
function cannot be described by the Euler rotation of antenna pattern 
function at rest frame (see footnote \ref{footnote:orbital motion}). 
Although the influence of arm-length variation is expected to be small 
in the low-frequency regime, alternative approach based on the other 
representation would be helpful to consider the GWB skymap beyond the 
low-frequency regime.

Here, we briefly discuss the reconstruction method based on the 
time-series representation (\ref{eq:def_of_C(f)}). 
The time-series representation is mathematically equivalent to the 
harmonic-Fourier representation under the rigid adiabatic treatment, 
but in general, no additional assumption for space craft configuration 
is invoked to derive the expression 
(\ref{eq:def_of_C(f)}), except for the premise that the time-dependence 
of the antenna pattern functions, i.e., the motion of each spacecraft, 
is well-known theoretically and/or observationally. 
Moreover, as we see below, the method based on the time-series
representation allows us to reconstruct directly the skymap $S_h$, 
without going through intermediate variables, e.g. $p_{\ell m}$. 
In this sense, the time-series representation would be superior to the 
harmonic-Fourier representation for practical purpose, although there 
still remains the same problem as discussed in Appendix 
\ref{appendix:on_the_degeneracy} concerning the {\it{degeneracy}} of the 
multipole coefficients.

In principle, the perturbative reconstruction scheme given in 
Sec.\ref{subsec:general_scheme} can be applicable to the map-making 
problem based on the time-series representation. 
To apply this, we discretize the integral expression 
(\ref{eq:def_of_C(f)}) so as to reduce it to a matrix form like 
equation (\ref{eq: c=Ap}).  For example, we discretize the celestial sphere 
$(\theta, \,\phi)$ into a regular $N$ meshes and also the continuous 
time-series into the regular $M$ grids. Then we have 
\begin{equation}
\widetilde{C}(t_i,\,f)= \sum_{j=1}^{N}\,\,
S_h(|f|,\,\,\mathbf{\Omega}_j)\,\,
\mathcal{F}^E(f,\,\mathbf{\Omega}_j;\,t_i)\, 
\frac{\Delta\mathbf{\Omega}_j}{4\pi}, \quad
(i=1,2,\cdots,\,M), 
\label{eq:discrete_eq}
\end{equation}
where $\Delta\mathbf{\Omega}_i=\sin\theta_i\,\Delta\theta_i\,\Delta\phi_i$. 
Here, we have ignored the noise contribution and dropped the subscript 
$_{IJ}$. For a reconstruction of low-frequency skymap, a large number 
of mesh and/or grid are not necessary. Typically, for the angular 
resolution up to $\ell=5$, it is sufficient to set 
$M=16$ and $N=16\time32$ (see Sec.\ref{subsec:reconstruction}).  
The above discretization procedure is repeated for the five output 
data of the correlation signals. 
Then, simply following the same procedure as in the case of the 
harmonic-Fourier representation, the least-squares method by SVD 
is applied to get the luminosity distribution of GWBs. 
Note that the meanings of the matrix $\mathbf{A}^{(i)}$, the vectors 
$\mathbf{c}^{(i)}$ and $\mathbf{p}^{(i)}$ in the perturbative expansion
are appropriately changed as follows. For the matrix $\mathbf{A}^{(i)}$, we have 
\begin{equation}
\mathbf{A}^{(i)} = \frac{1}{4\pi}
\left(
\begin{array}{cccc}
 \mathcal{F}^{(i)}(\mathbf{\Omega}_1, t_1)\,\Delta\mathbf{\Omega}_1 & 
 \mathcal{F}^{(i)}(\mathbf{\Omega}_2, t_1)\,\Delta\mathbf{\Omega}_2 & \cdots &
 \mathcal{F}^{(i)}(\mathbf{\Omega}_N, t_1)\,\Delta\mathbf{\Omega}_N \\
 \mathcal{F}^{(i)}(\mathbf{\Omega}_1, t_2)\,\Delta\mathbf{\Omega}_1 & 
 \mathcal{F}^{(i)}(\mathbf{\Omega}_2, t_2)\,\Delta\mathbf{\Omega}_2 & \cdots & 
 \mathcal{F}^{(i)}(\mathbf{\Omega}_N, t_2)\,\Delta\mathbf{\Omega}_N \\
 \vdots  & \vdots  & \ddots &\vdots \\
 \mathcal{F}^{(i)}(\mathbf{\Omega}_1, t_M)\,\Delta\mathbf{\Omega}_1 & 
 \mathcal{F}^{(i)}(\mathbf{\Omega}_2, t_M)\,\Delta\mathbf{\Omega}_2 & \cdots & 
 \mathcal{F}^{(i)}(\mathbf{\Omega}_N, t_M)\,\Delta\mathbf{\Omega}_N \\
\end{array}
\right), \quad\quad (i=2,\,3,\,\cdots) 
\end{equation}
The corresponding vectors $\mathbf{c}_{(i)}$ and $\mathbf{p}_{(i)}$ becomes
\begin{equation}
\mathbf{c}^{(i)} = 
\left(
\begin{array}{c}
\widetilde{C}^{(i)}(t_1) \\
\widetilde{C}^{(i)}(t_2) \\
\vdots  \\  
\widetilde{C}^{(i)}(t_M) 
\end{array}
\right), 
\qquad
\mathbf{p}^{(i)} = 
\left(
\begin{array}{c}
S_h^{(i)}(\Omega_1) \\
S_h^{(i)}(\Omega_2) \\
\vdots  \\  
S_h^{(i)}(\Omega_N) \\
\end{array}
\right), 
\end{equation}
where $\widetilde{C}^{(i)}(t)$ and $S_h^{(i)}(\Omega)$  the perturbative 
coefficients of $\widetilde{C}(t,f)$ and $S_h (|f|, \Omega)$ in power 
of $\hat{f}^i$, respectively.  Using these expressions, the
least-squares solution is constructed as 
$\mathbf{p}_{\rm approx}= 
\mathbf{p}_{\rm approx}^{(2)}+\mathbf{p}_{\rm approx}^{(3)}+\cdots$, 
with a help of equation (\ref{eq:perturbative_p=AC}). 
Then, the resultant expression $\mathbf{p}_{\rm approx}$ directly gives a 
GWB skymap in the ecliptic coordinate, i.e.,
$S_h(|f|,\,\mathbf{\Omega})$, not the multipole coefficients.

%
\section{Demonstration:  skymap of 
Galactic background}
\label{sec:demonstration}
%

Perturbative reconstruction scheme presented in the previous section is 
applicable to a general map-making problem for any kind of GWB sources. 
In this section, to see how our general scheme works in practice, we 
demonstrate the reconstruction of a GWB skymap focusing on a specific 
source of anisotropic 
GWB. An interesting example for LISA detector is a confusion-noise 
background produced by the Galactic population of unresolved binaries. 
After describing a model of Galactic GWB in Sec.\ref{subsec:GWB_model}, 
the expected signals for time-modulation data of self- and
cross-correlated TDIs are calculated in Sec.\ref{subsec:on_snr}.  
Based on these, the detectable Fourier components for time-modulation 
signals are discussed evaluating the signal-to-noise ratios. 
In Sec.\ref{subsec:reconstruction}, a reconstruction of the GWB skymap
is performed based on the harmonic-Fourier representation and the
time-series representation. The resultant values of the multipole
coefficients for Galactic GWBs are compared with those from the 
full-resolution skymap, taking account of the influence of the noises.

\subsection{A model of Galactic GWB}
\label{subsec:GWB_model}

For our interest of GWBs in the low-frequency regime with 
$f\lesssim f_* \simeq 9.54$ mHz, it is reasonable to assume that the 
anisotropic GWB spectrum $S_h(f,\mathbf{\Omega})$ is separately treated as 
$S_h(f,\mathbf{\Omega})=H(f)\,P(\mathbf{\Omega})$, as we mentioned. 
The power spectral density $H(f)$ is approximated by a power-law function like 
$H(f)= {\mathrm N } \,f^{\alpha}$, whose amplitude will be explicitly
given later. For illustrative purpose, we consider the simplest model of 
luminosity distribution $P(\mathbf{\Omega})$, in which the Galactic GWB
is described by an incoherent superposition of gravitational waves
produced by compact binaries whose spatial structure just traces the
Galactic stellar distribution observed via infrared photometry. 
We use the fitting model of Galactic stellar distribution given in 
\cite{Binney:1996sv}, which consists of triaxial bulge and disk components 
(see also \cite{Seto:2004ji}).  
The explicit functional form of the density distribution
$\rho(\mathbf{x})$ written in the Galactic coordinate system becomes 
\begin{eqnarray}
    \label{eq:rho_GWB}
&&    \rho(\mathbf{x}) =  \rho_{\rm disk}(\mathbf{x}) +  
    \rho_{\rm bulge}(\mathbf{x})\,\,;
    \label{eq:BGS_model} \\
&&\quad    \rho_{\rm bulge} = \frac{\rho_0}{(1+a/a_0)^{1.8}}\,\,
e^{-(a/a_{\rm m})^2},
    \nonumber \\
&&\quad    \rho_{\rm disk} = \left(\frac{e^{-|z|/z_0}}{z_0}
 + \alpha \,
\frac{e^{-|z|/z_1}}{z_1}\right) \,R_s\, e^{-R/R_{\rm s}}
\nonumber
\end{eqnarray}
with the quantities $a$ and $R$ defined by 
$a\equiv [x^2+(y/\eta)^2+(z/\zeta)^2 ]^{1/2}$ and 
$R\equiv (x^2 + y^2)^{1/2}$. Note that the $z$-axis is oriented to the north 
Galactic pole and the direction of the $x$-axis is $20^{\circ}$
different from the Sun-center line.
Here the parameters are given as follows: $\rho_0=624$, 
$a_{\rm m} = 1.9$ kpc, $a_0=100$ pc, $R_{\rm s}=2.5$ kpc, $z_0=210$ pc, 
$z_1=42$ pc, $\alpha = 0.27$, $\eta=0.5$ and $\zeta=0.6$. 
Provided the three-dimensional structure of stellar distribution 
$\rho(\mathbf{x})$, the angular distribution $P(\mathbf{\Omega})$ is 
obtained by projecting it onto the sphere in observed frame, i.e., 
ecliptic coordinate:  
\begin{equation}
    P(\mathbf{\Omega}) = C\,\int dr\,\,4\pi\,r^2\,\, 
    \frac{\rho(r,\mathbf{\Omega})}{r^2}, 
\end{equation}
where $C$ is a numerical constant normalized by  
$\int d\mathbf{\Omega}\,P(\mathbf{\Omega})=1$.   
The integration over $r$ is performed along a line-of-sight direction 
from a location of space interferometer to infinity.

\begin{figure}[t]
\begin{center}
 \includegraphics[width=10cm,angle=0,clip]{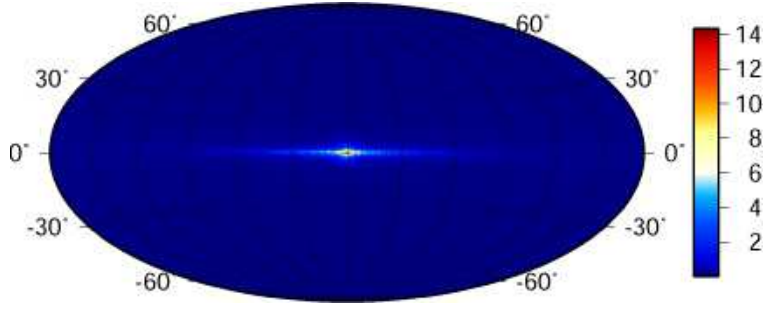}
\label{fig:fill_skymap}
\end{center}
\caption{Full-resolution 
      skymap of the Galactic gravitational-wave background shown 
      as Hammer-Aitoff map in Galactic coordinate. 
    }
\label{fig:full-resolution_skymap}
\end{figure}

\begin{figure}[t]
\begin{center}
 \includegraphics[width=8.3cm,clip]{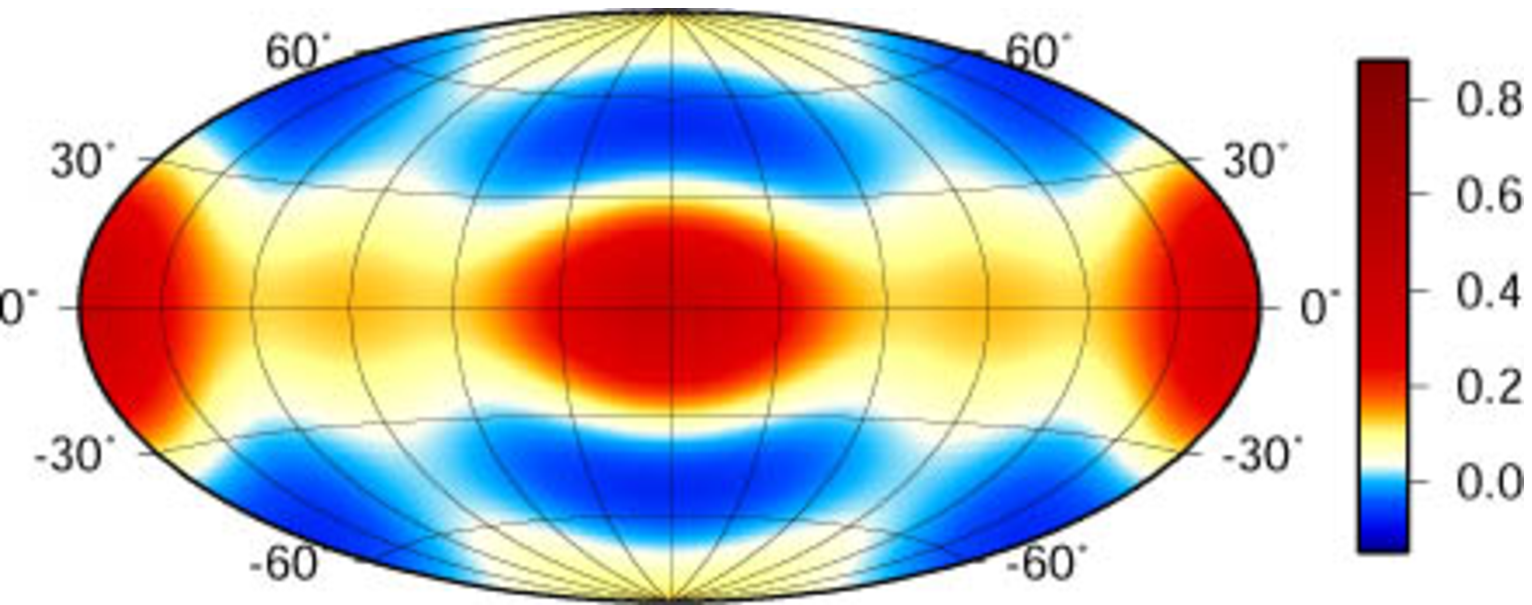}
\hspace*{0.3cm}
 \includegraphics[width=8.3cm,clip]{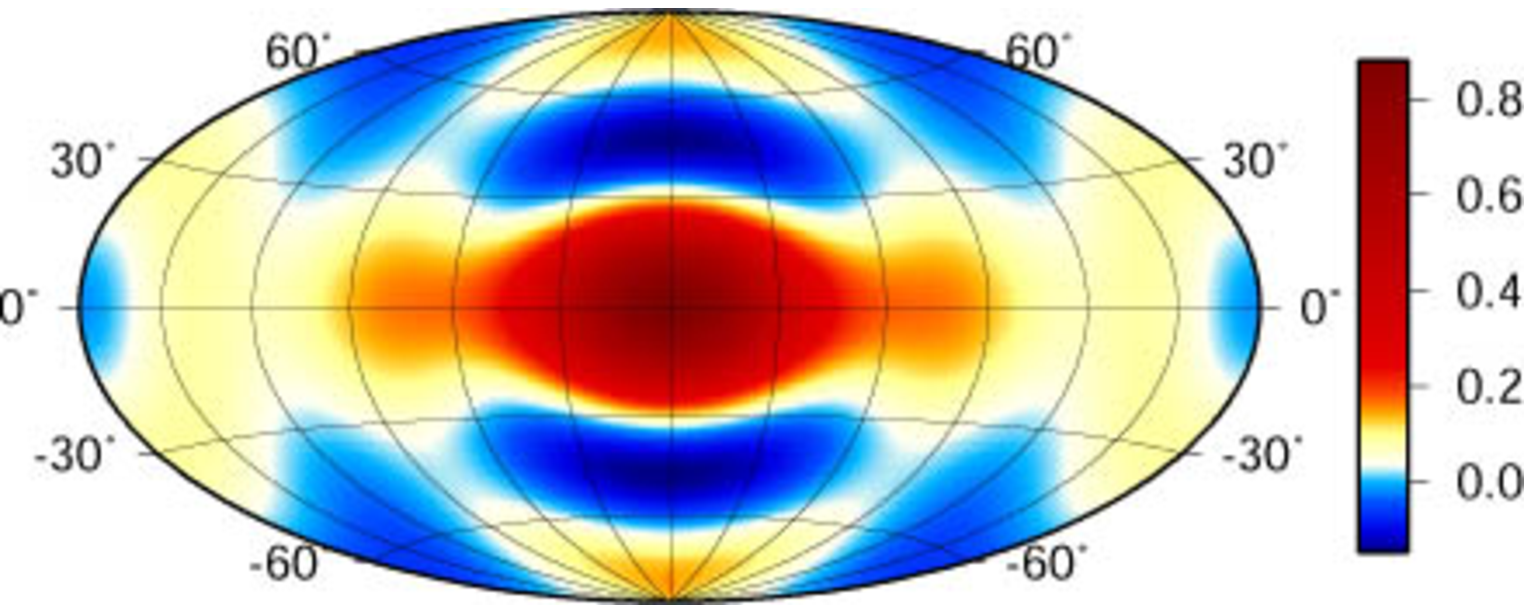}
\end{center}
    \caption{Expected images of GWB skymap by LISA based on 
      the low-frequency reconstruction scheme, which are both depicted in the 
      Galactic coordinate. {\it Left}: low-resolution skymap 
      created by truncating all the multipoles except for 
      $\ell=0$, $2$ and $4$ from the original full-resolution skymap. 
     {\it Right:} skymap created by truncating the higher multipoles 
      $\ell>6$.  
    }
     \label{fig:expected_skymap}
\end{figure}

In Fig.\ref{fig:full-resolution_skymap}, the projected intensity 
distribution of GWB is numerically obtained specifically in ecliptic 
coordinate and it is then transformed into Galactic coordinate, shown 
as the Hammer-Aitoff map.  A strong intensity peak is found around the 
Galactic center and the disk-like structure can be clearly seen. 
Notice that while the result depicted in
Fig.\ref{fig:full-resolution_skymap} represents a full-resolution
skymap, it cannot be attained from the perturbative reconstruction
scheme in low-frequency regime. For the antenna pattern functions of 
cross-correlated TDI signals up to the order $\mathcal{O}(\hat{f}^3)$, 
the detectable multipole moments of GWB are limited to $\ell\leq5$. 
Moreover, in the low-frequency limit $\mathcal{O}(\hat{f}^2)$, only 
the even modes $\ell=0$, $2$ and $4$ are measurable.
Thus, the reconstructed skymap would be rather miserable compared to 
the full skymap which contains the higher multipoles $\ell \gtrsim 30$ 
(see Fig.\ref{fig:sigma_gwb} in Appendix).  Taking account of these 
facts, in right panel of Fig.\ref{fig:expected_skymap}, we plot the 
low-resolution skymap, which was obtained from the full-resolution
skymap just dropping the higher multipole moments with $\ell>5$. 
In Appendix \ref{appendix:multipole of GWB}, with a help of the 
Fortran package of spherical harmonic analysis (\cite{Adams:2003}, see 
Appendix \ref{appendix:multipole of GWB}), the numerical values of 
the multipole coefficients $p_{\ell m}$ up to $\ell=5$ are computed 
and are summarized in Table \ref{tab:summay_multipole}. 
Also in the left panel, the odd modes are further subtracted and the 
remaining multipole moments are only $\ell=0$, $2$ and $4$. 
As a result, the fine structure around the bulge and the disk components 
is coarse-grained and the intensity of the GWB diminishes. 
It also shows some fake patterns with negative intensity. 
Nevertheless, one can clearly see the anisotropic structure of GWB
, which is mainly contributed from the gravitational-wave sources 
around the Galactic disk. With a perturbative reconstruction of 
low-frequency up to $\mathcal{O}(\hat{f}^3)$, one can roughly infer 
that the main sources of Galactic GWB comes from the Galactic center.

%
%
%
%
\subsection{Time-modulation signals and signal-to-noise ratio}
\label{subsec:on_snr}
%
%
%
%
%
%
Given the intensity distribution of GWB, one can calculate the 
cross-correlation signals observed via LISA, which are inherently 
time-dependent due to the orbital motion of the LISA detector.  
The effect of the annual modulation of the Galactic binary confusion 
noise on the LISA data analysis had been previously studied in  
\cite{Seto:2004ji} in the low-frequency limit $\hat{f}\ll1$. 
Recently, Monte Carlo simulations of Galactic GWB were carried out by 
several groups and the annual modulation of GWB intensity has been 
confirmed in a realistic setup with specific detector output 
\cite{Benacquista:2003th,Edlund:2005ye,Timpano:2005gm}.

\begin{figure}[t]
\begin{center}
 \includegraphics[width=5.7cm,clip]{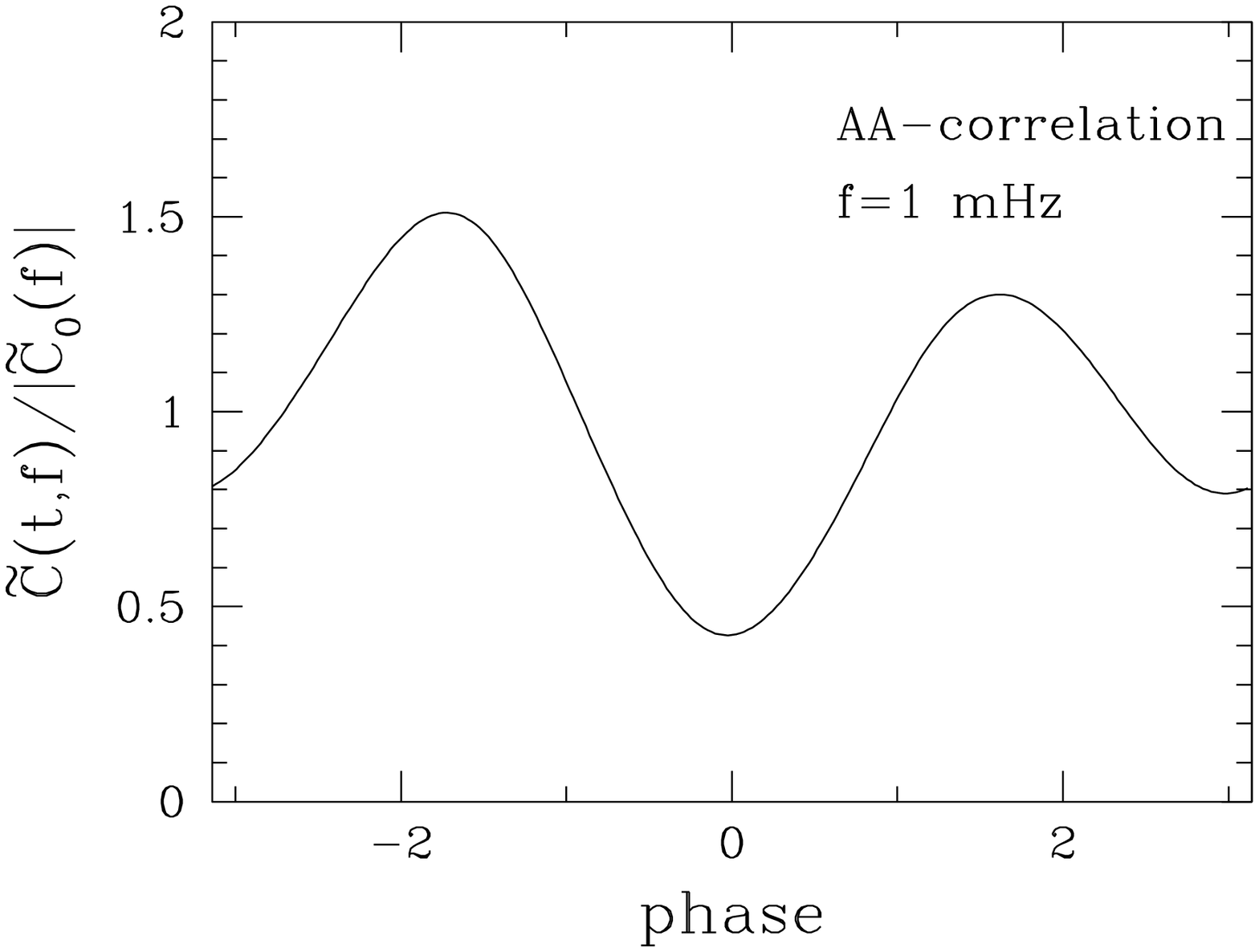}
\hspace*{0cm}
 \includegraphics[width=5.7cm,clip]{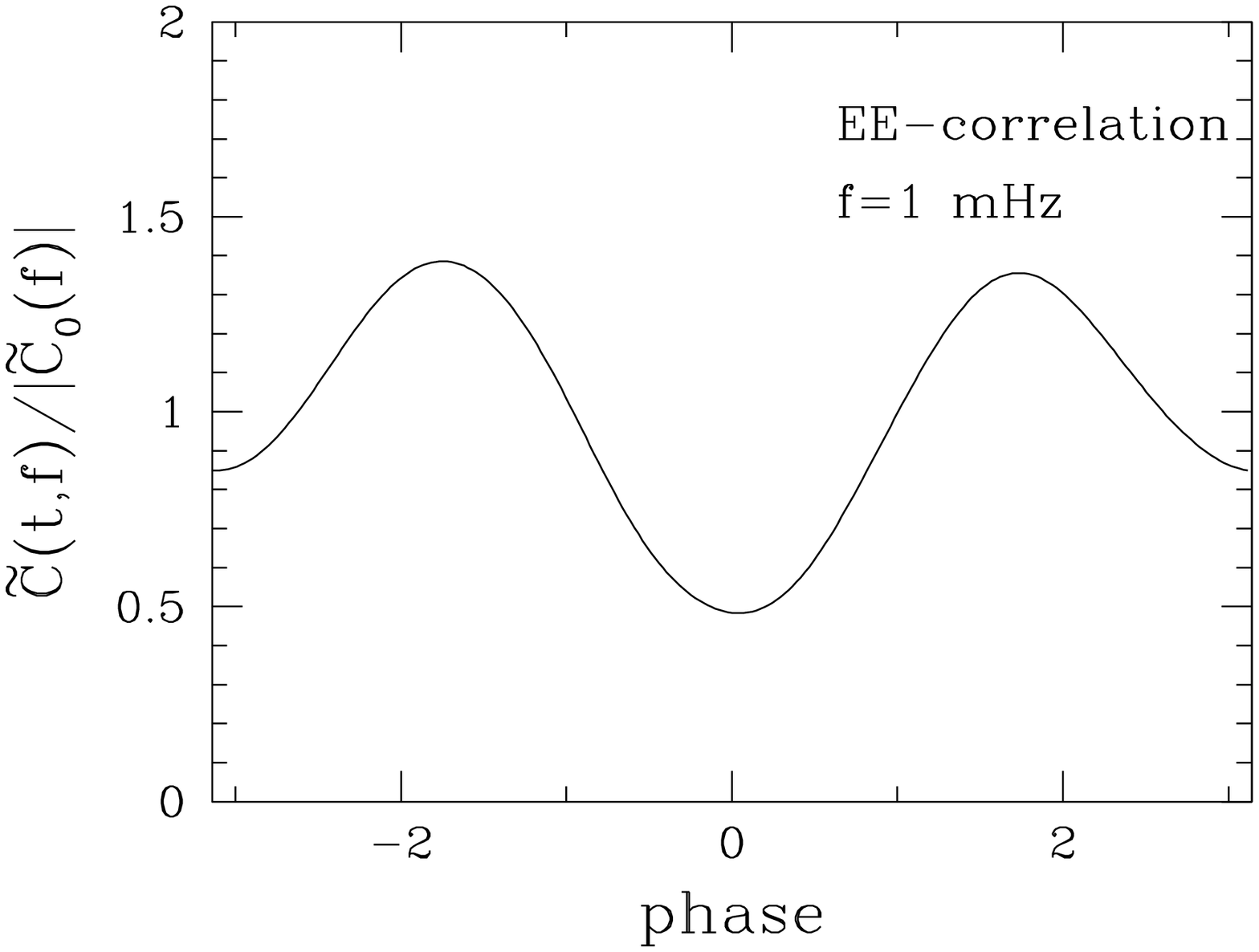}
\hspace*{0cm}
 \includegraphics[width=5.7cm,clip]{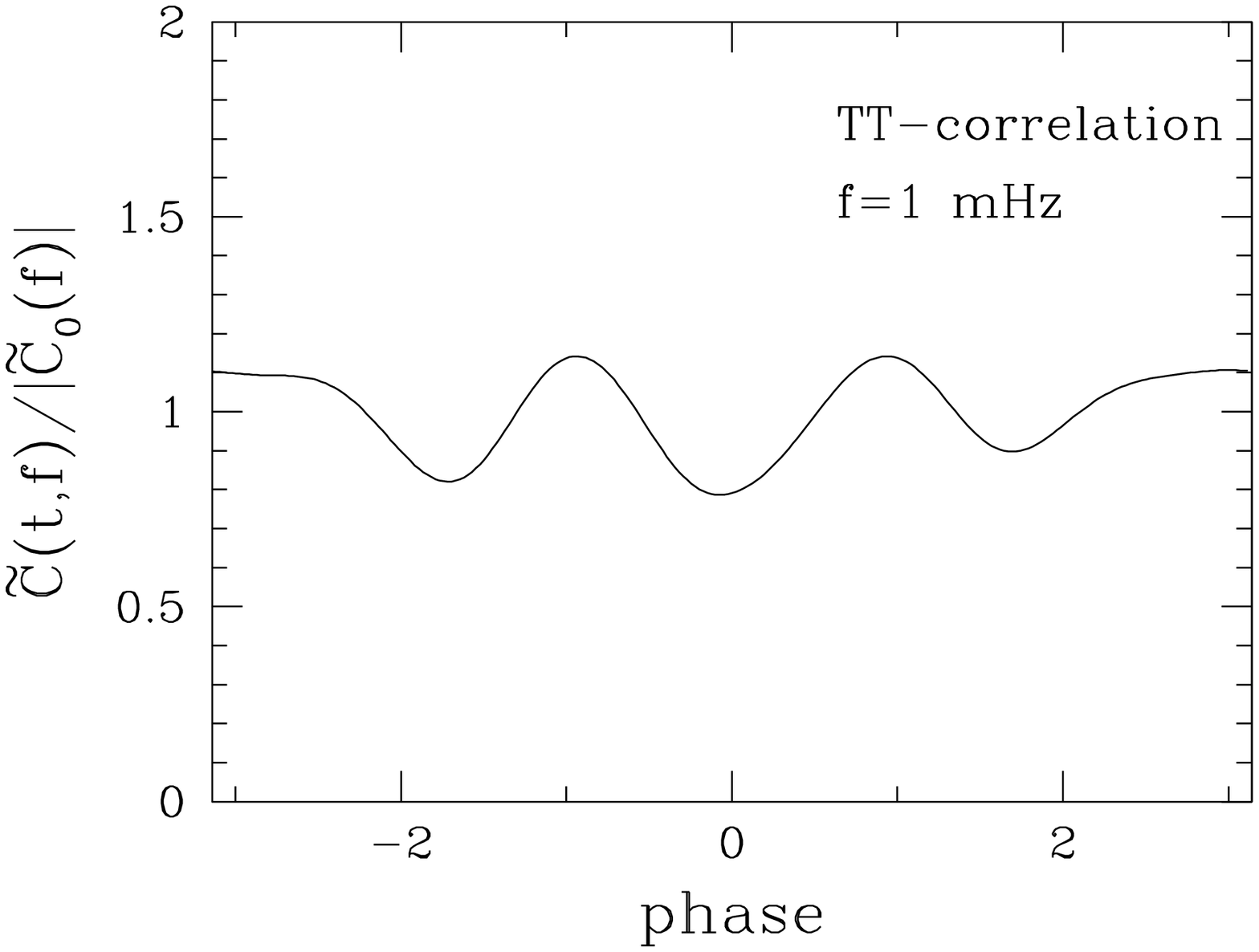}
\end{center}
\begin{center}
 \includegraphics[width=5.7cm,clip]{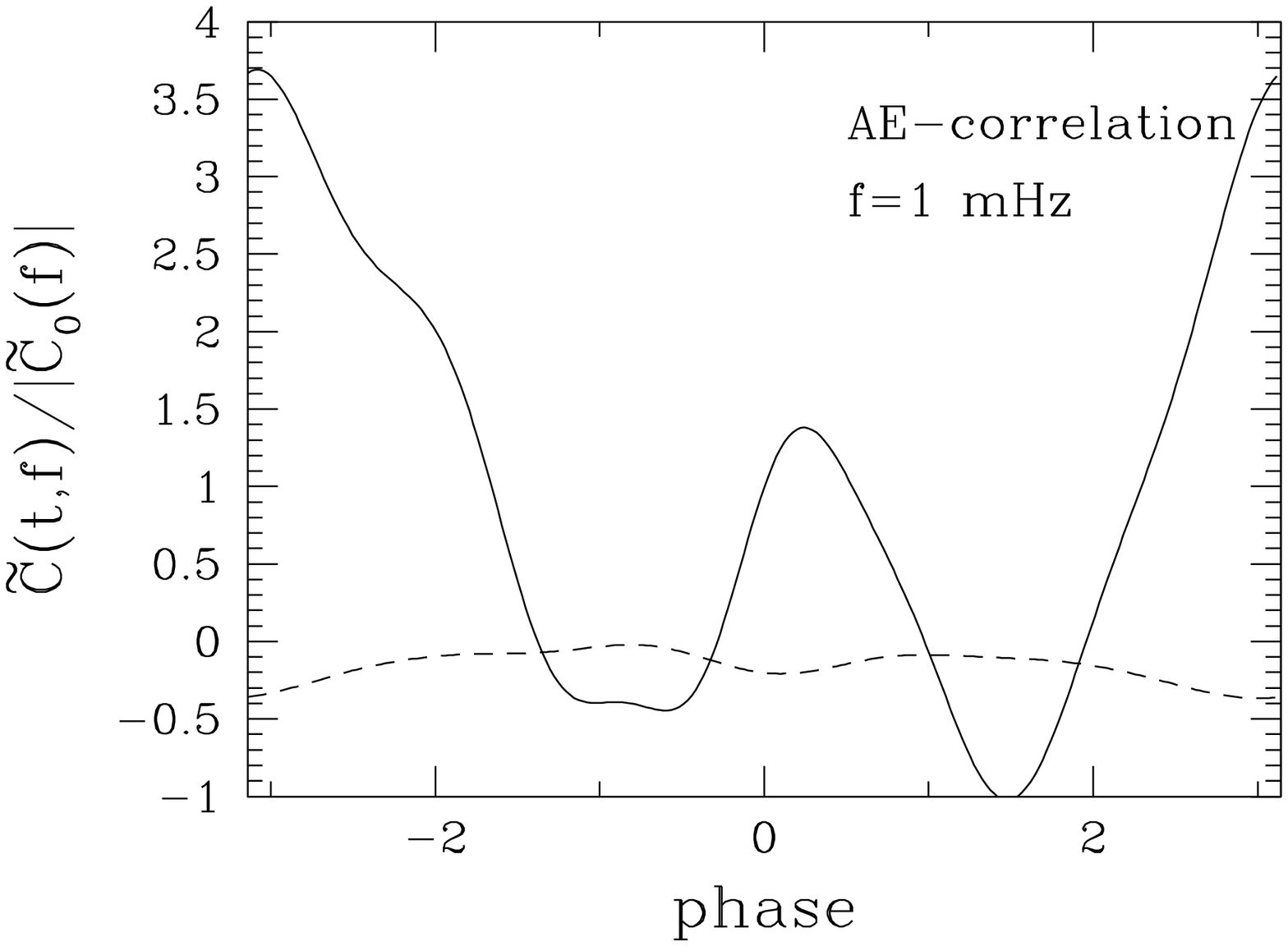}
\hspace*{0cm}
 \includegraphics[width=5.7cm,clip]{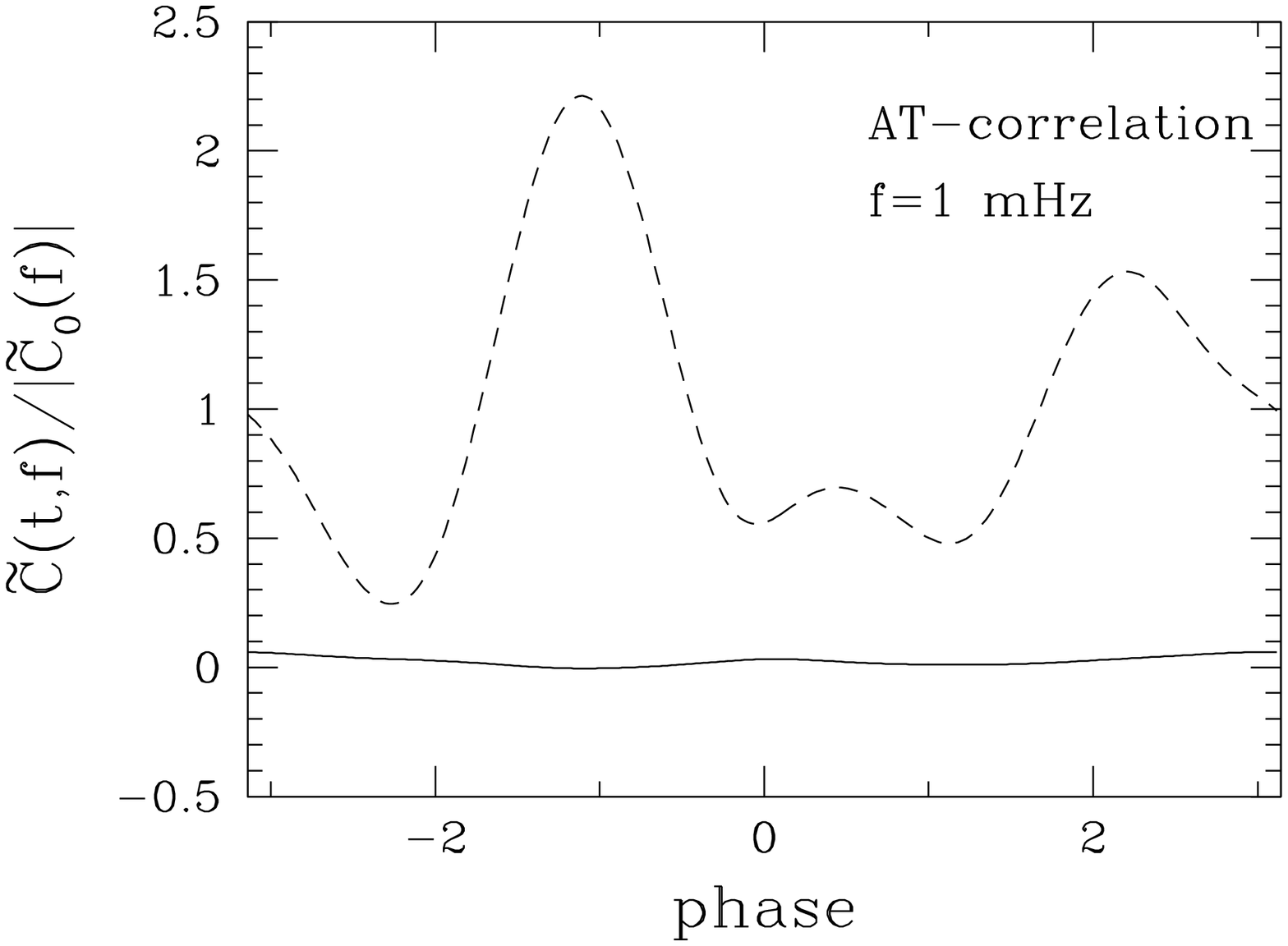}
\hspace*{0cm}
 \includegraphics[width=5.7cm,clip]{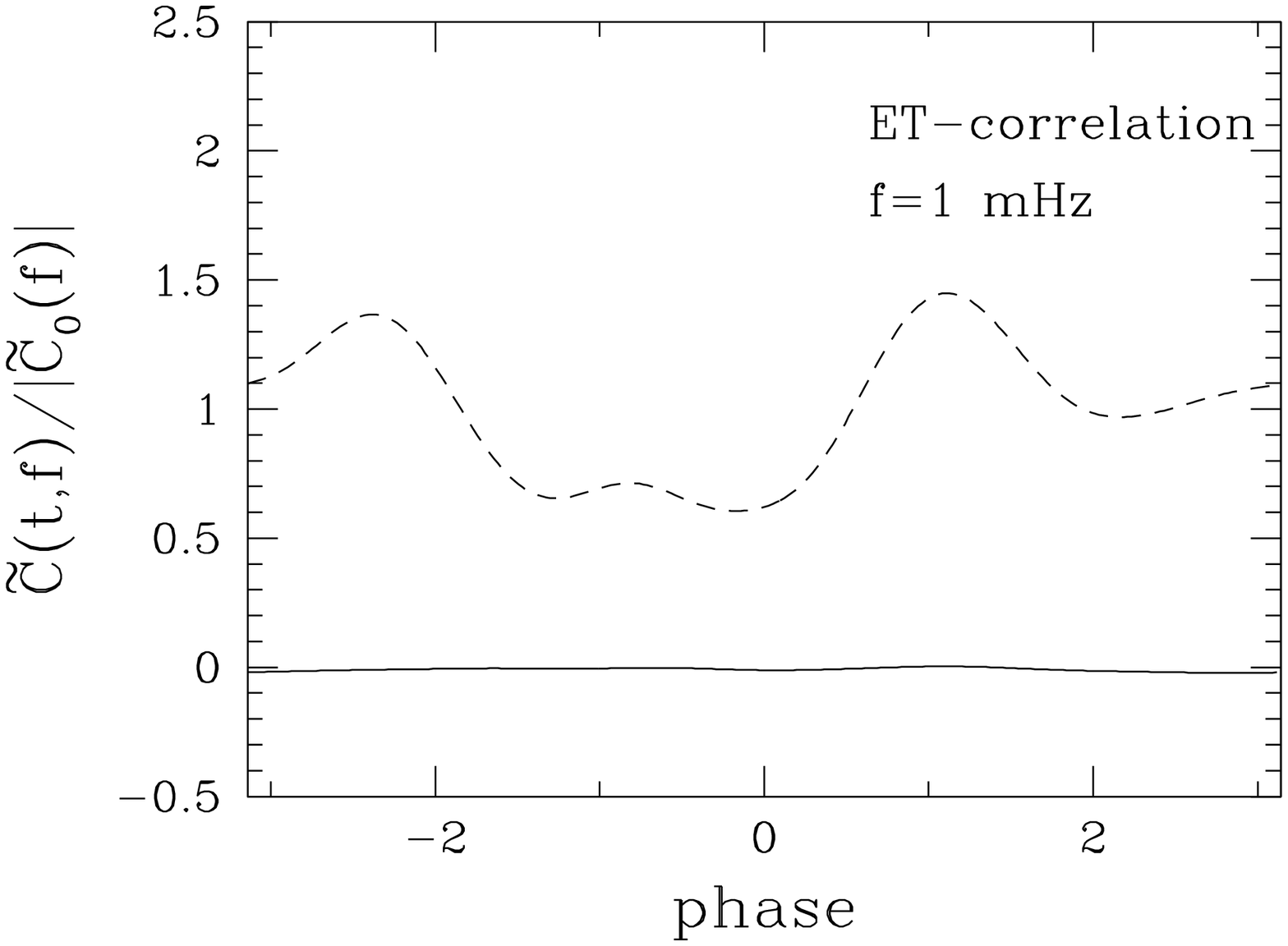}
\end{center}
    \caption{ 
    Annual time-modulation signals of self- and cross-correlation data 
    as function of orbital phase assuming the observed frequency 
    $\hat{f}=0.1$, i.e., $f\simeq1 \mathrm{mHz}$. 
    Here, the correlation signals 
    $\widetilde{C}_{IJ}(t,f)$ are plotted normalizing  
    with $k=0$ component of correlation signal $\widetilde{C}_{IJ,0}(f)$. 
    Solid and dashed lines represent the real and the imaginary 
    part of correlation signal, respectively. 
    }
    \label{fig:annual data}
\end{figure}

Owing to the expression (\ref{eq:def_of_C(f)}), the time-modulation 
signals $\widetilde{C}(t,f)$ neglecting the instrumental noises are
computed for optimal TDIs at the frequency $\hat{f}=0.1$, i.e., 
$f\simeq1 \mathrm{mHz}$ using the full expression of antenna pattern function 
(\ref{eq:def_of_antenna}). The results are then plotted as function of 
orbital phase. In Fig. \ref{fig:annual data}, six outputs of the 
self- and cross-correlation signals normalized to its time-averaged value, 
$\widetilde{C}(t,f)/|\widetilde{C}_0(f)|$ are shown. 
The solid and dashed lines denote the real and imaginary parts of 
the correlation signals, respectively.

The time modulation of these outputs basically reflects the spatial 
structure of GWB. As LISA orbits around the Sun, the direction normal 
to LISA's detector plane, which is the direction sensitive to the 
gravitational waves, sweeps across the Galactic plane. 
Since the response of the LISA detector to the gravitational 
waves along the $\mathbf{\hat{\Omega}}$ direction give the same 
response to the waves along the 
$-\mathbf{\hat{\Omega}}$ direction 
(see Eq.(\ref{eq:Y_lm expansion of S and F}) and the brief comment
there), the time-modulation signal is expected to have a bimodal 
structure, like $AA$- and $EE$-correlations. 
However, the actual modulation signals are more complicated, depending 
on the angular sensitivity of their antenna pattern functions, as well 
as the observed frequency. Further, the cross-correlation data can be 
generally complex variables, whose behaviors are different between 
real- and imaginary-parts.

Notice that the time-modulation signals presented in Fig. 
\ref{fig:annual data} are the results in an idealistic situation free 
from the noise contribution. In presence of the random noise, some of 
the correlation data which contain gravitational wave signals are 
overwhelmed by noise, which cannot be used for the reconstruction of 
GWB skymap. Hence, one must consider the signal-to-noise ratio (SNR) 
to discriminate available correlation data. 
To evaluate this, the output signals depicted in 
Fig.\ref{fig:annual data} are first transformed to its Fourier counterpart, 
$\widetilde{C}_k(f)$. The resultant Fourier components for each signal 
are then compared to the noise contributions. 
The SNR for each component is expressed in the following form 
(\cite{Seto:2004np}, see also 
\cite{Giampieri:1997,Ungarelli:2001xu}): 
\begin{equation}
\left(\frac{S}{N}\right)_k =\sqrt{2\,\,\Delta f\,\,T}\,\,
\frac{|\widetilde{C}_k|}{N_k}.  
\label{eq:def_SNR}
\end{equation}
Here we set the observational time to $T=10^8$sec and the bandwidth to 
$\Delta f=10^{-3}$Hz. The quantity $N_k$ represents the noise 
contribution. The important remark is that the noise contribution 
comes from not only the detector noise but also the randomness of 
the signal itself. The details of the analytic expression for $N_k$ 
will be given elsewhere \cite{Kudoh:2005inprep}. 
Here, as a crude estimate, we evaluate the noise contribution $N_k$ as 
\begin{equation}
N_k= \sqrt{2\,\,\Delta f\,\,T}\,\,S_n^{II}, \quad(I= A,~E,~T) 
\end{equation}
for $k=0$ component of self-correlation signals and 
\begin{equation}
N_k= \sqrt{\mbox{max}\left(\widetilde{C}_{II,0}\widetilde{C}_{JJ,0},
~\widetilde{C}_{II,0}S_n^{JJ},~
\widetilde{C}_{JJ,0}S_n^{II},~S_n^{II}S_n^{JJ}\right) }, 
\quad (I,J= A,~E,~T) 
\end{equation}
for cross-correlation signals and $k\neq0$ component of self-correlation signals. 
To estimate SNR, one further needs the spectral density for instrumental 
noise, i.e., $S_n^{\scriptscriptstyle\rm AA}(f)$, 
$S_n^{\scriptscriptstyle\rm EE}(f)$ and $S_n^{\scriptscriptstyle\rm
TT}(f)$. We use the expressions given in equation (58) of Paper I 
(see also \cite{Cornish:2001bb,Prince:2002hp})
\footnote{In equation (58) of Paper I, there are some typos in the
numerical values of $S_{\rm shot}(f)$ and $S_{\rm accel}(f)$. 
In the present paper, we adopt $S_{\rm shot}(f)=1.6\times10^{-41}$Hz$^{-1}$ 
and $S_{\rm accel}(f)=2.31\times10^{-41}(\mbox{mHz}/f)^4$Hz$^{-1}$ 
according to Refs.\cite{Bender:1998,Cornish:2001bb}.}.
Note that all the cross-correlated noise spectra such as 
$S_n^{\scriptscriptstyle\rm AE}(f)$ and $S_n^{\scriptscriptstyle\rm AT}(f)$ 
are exactly canceled.

\begin{figure}[t]
\begin{center}
 \includegraphics[width=5.7cm,clip]{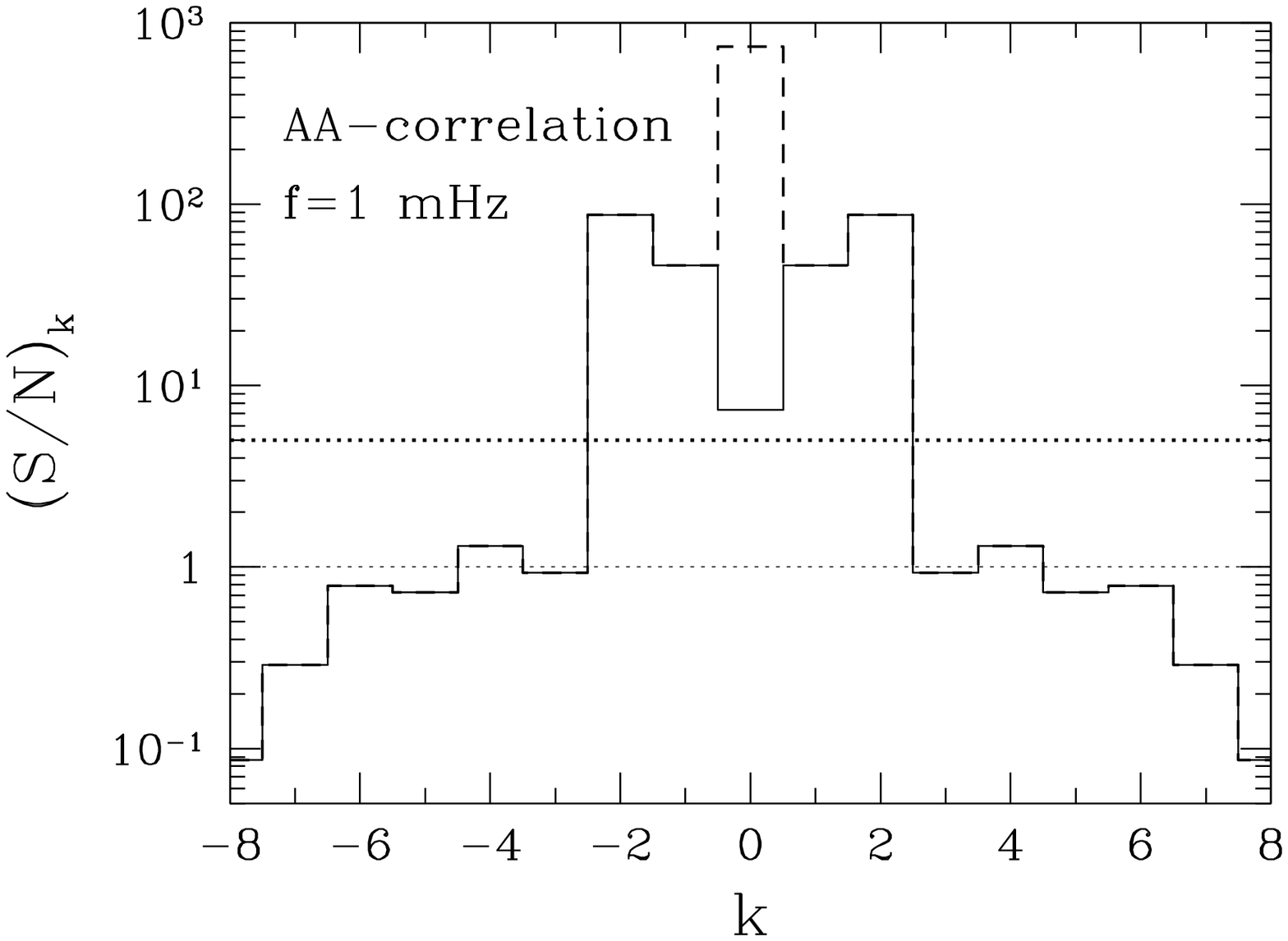}
\hspace*{0cm}
 \includegraphics[width=5.7cm,clip]{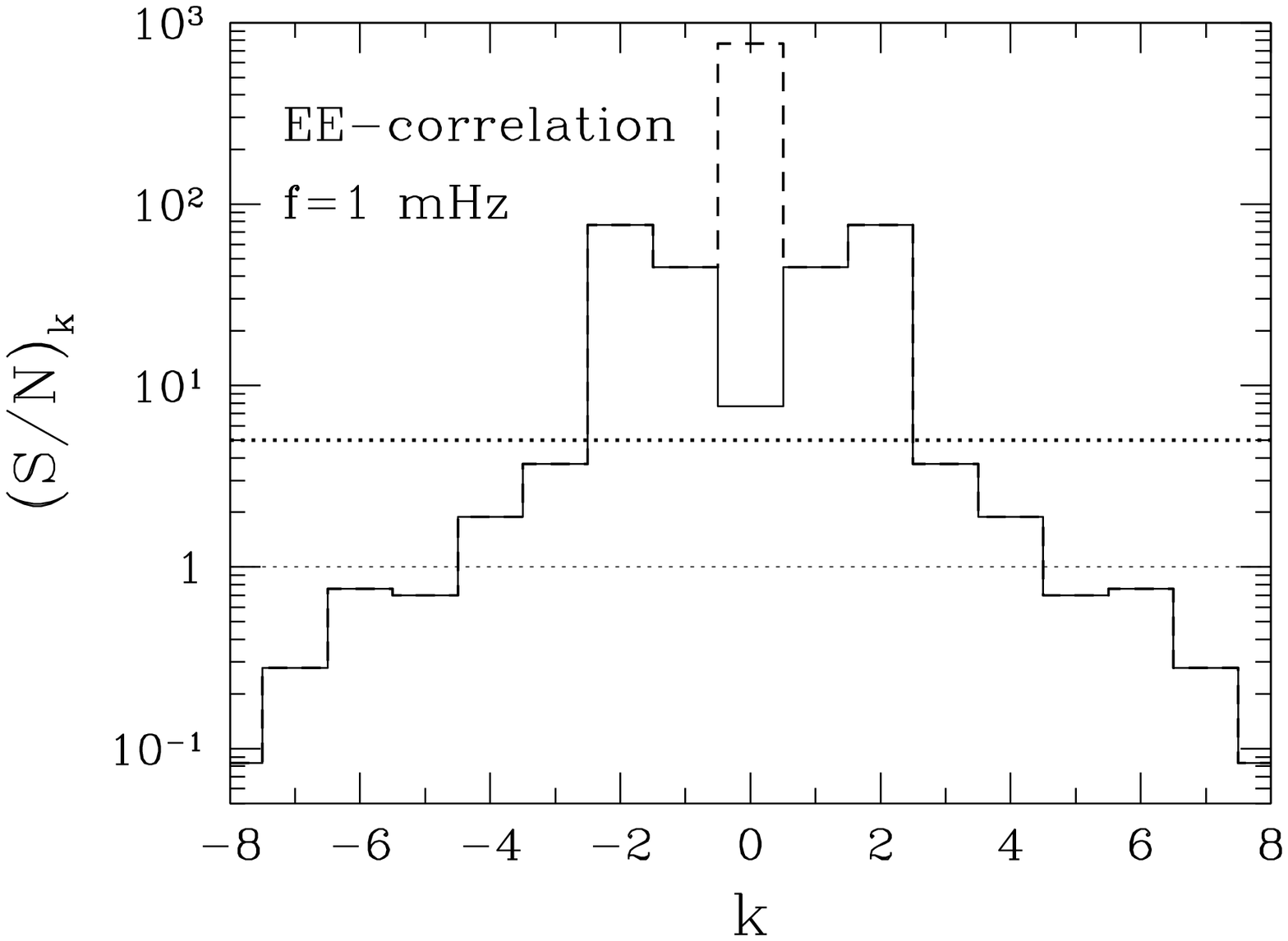}
\hspace*{0cm}
 \includegraphics[width=5.7cm,clip]{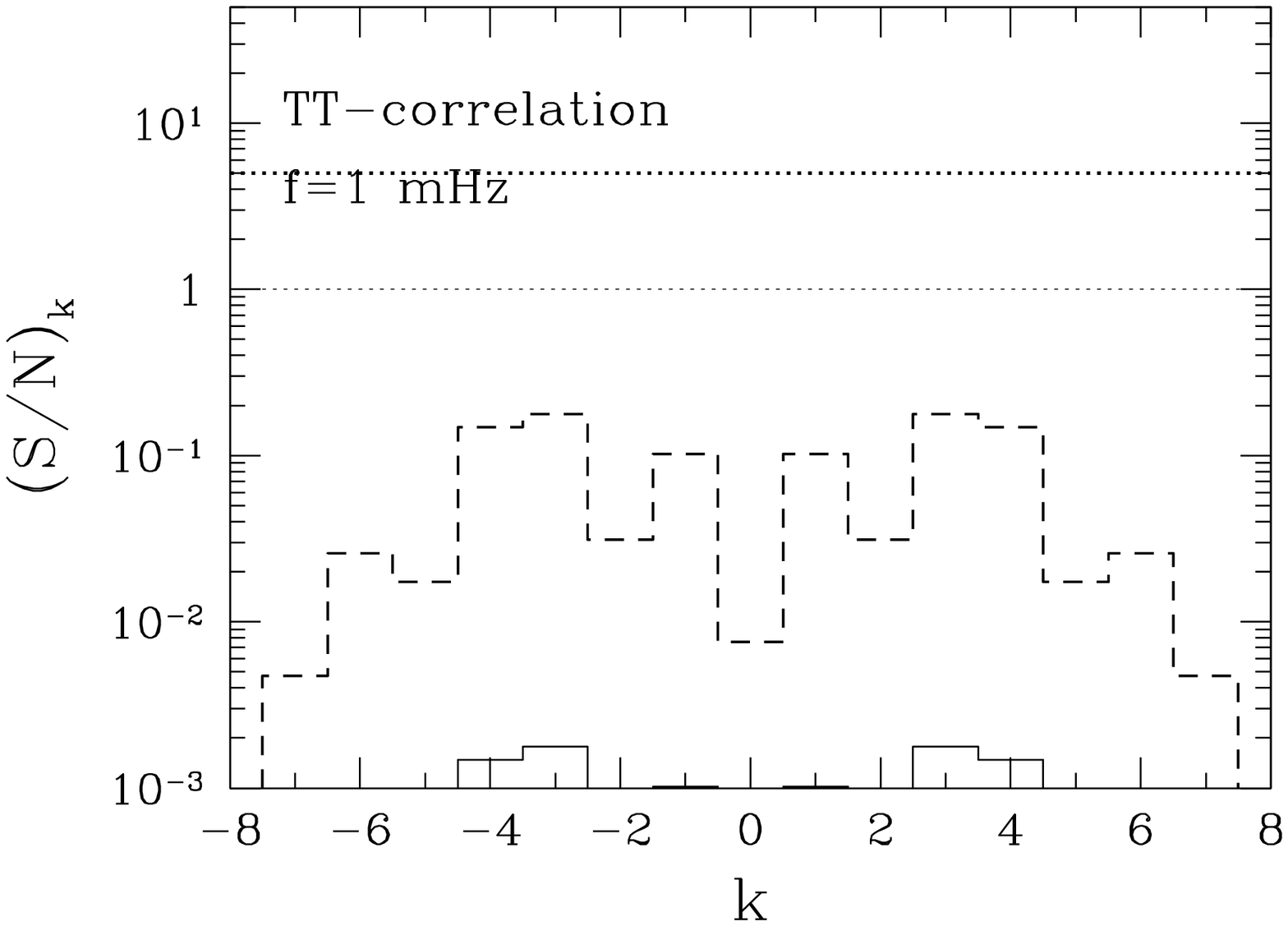}
\end{center}
\begin{center}
 \includegraphics[width=5.7cm,clip]{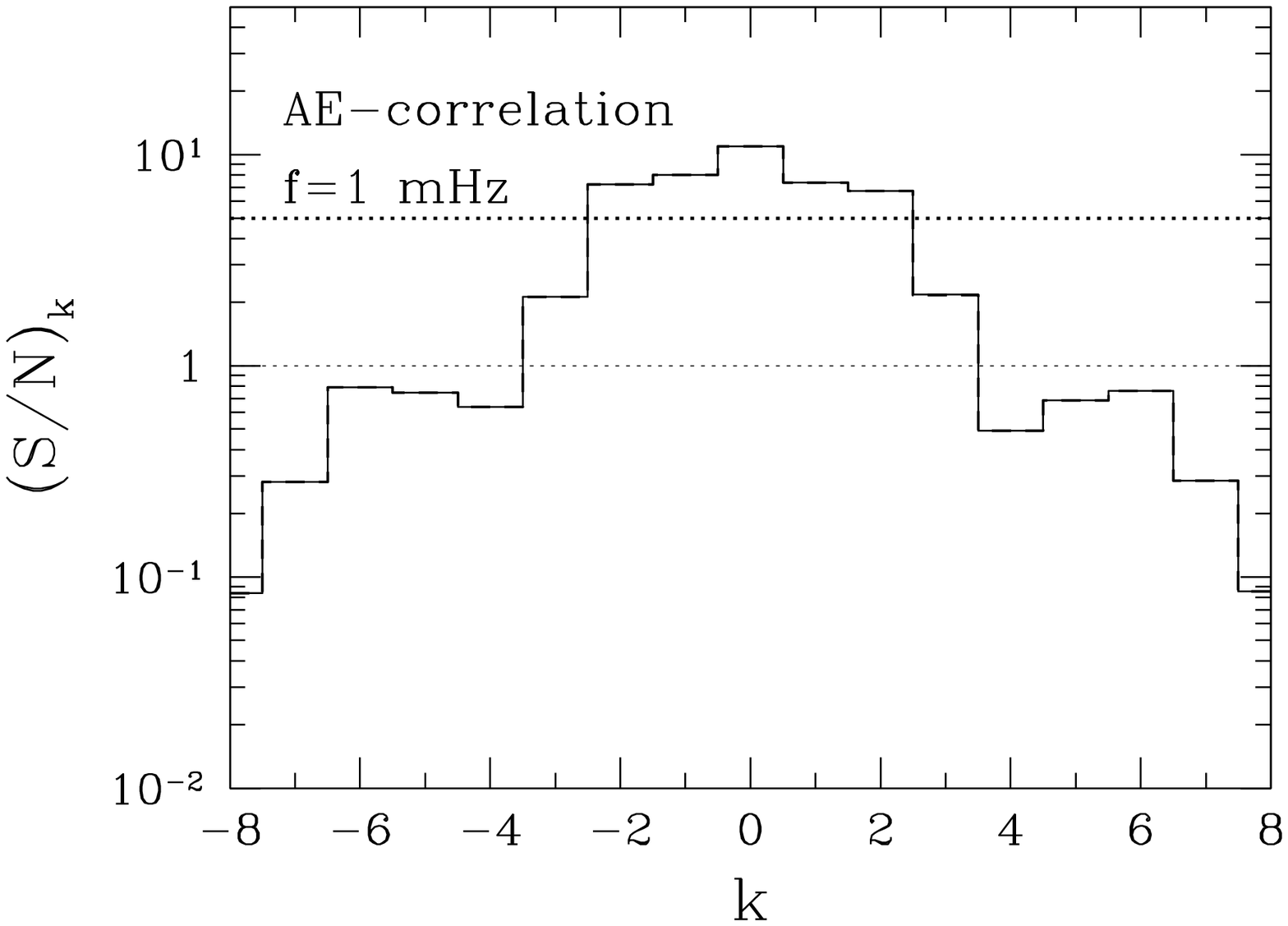}
\hspace*{0cm}
 \includegraphics[width=5.7cm,clip]{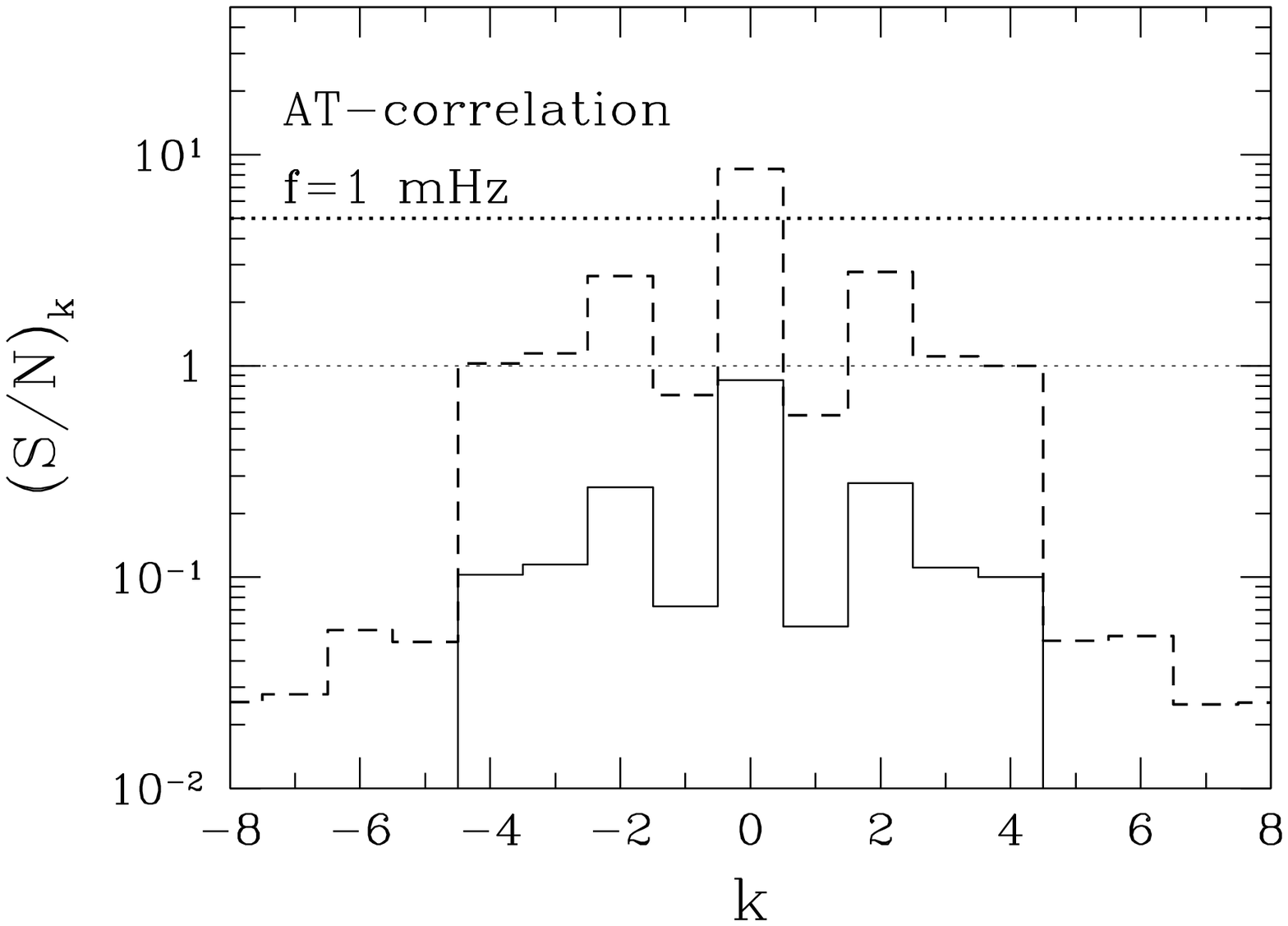}
\hspace*{0cm}
 \includegraphics[width=5.7cm,clip]{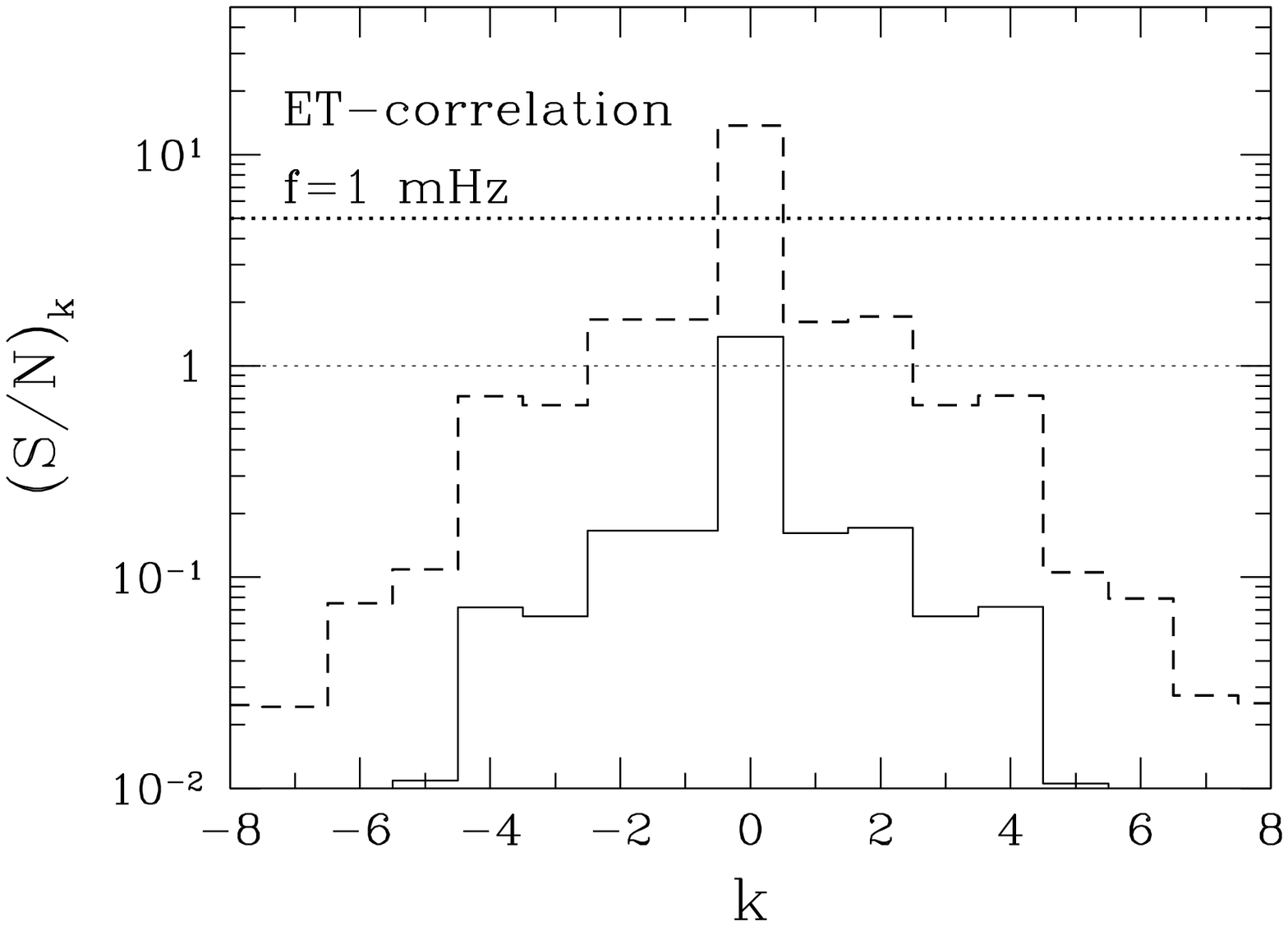}
\end{center}
    \caption{ 
      Signal-to-noise ratio for each Fourier component of  
      correlation signals, $(S/N)_k$ estimated at 
      $\hat{f}=0.1$ ($f\simeq1$mHz). The histogram depicted in solid line 
      indicates the case with the amplitude of Galactic GWB given by 
      $S_h^{1/2}=5\times10^{-19}$Hz$^{-1/2}$ at $f=1$mHz (case A),  
      while the histogram in dashed 
      line represents the signal-to-noise ratio for the case with 
      $S_h^{1/2}=5\times10^{-18}$Hz$^{-1/2}$ (case B). The thin-dashed,  
      and thick-dashed lines respectively denote 
      the lines of $(S/N)_k=1$ and $5$. 
       }
    \label{fig:SNR}
\end{figure}

In Fig.\ref{fig:SNR}, the SNRs for six output data are evaluated and 
are shown in the histogram as function of Fourier component, $k$. 
In these panels, thick-dotted lines show the detection limit of 
$(S/N)_k=5$, while the thin-dotted lines mean $(S/N)_k=1$. When 
evaluating the SNR, we specifically consider the two cases: 
\begin{description}
\item[Case A:]\quad realistic case in which 
the rms amplitude of GWB spectrum is given by 
$S_h^{1/2}=5\times10^{-19}$Hz$^{-1/2}$ at the frequency 
$f=1$mHz \footnote{The amplitude of GWB spectrum might be reduced 
by a factor of $2\sim5$ according to the recent numerical simulations 
\cite{Edlund:2005ye,Timpano:2005gm}. 
In case A, however, the noise contributions in SNR (\ref{eq:def_SNR}) 
used for reconstruction analysis are basically determined by the $k=0$ 
components of self-correlation signals, i.e., 
$N_k=\sqrt{\widetilde{C}_{II,0}\widetilde{C}_{JJ,0}}$. 
Hence, a slight change of the GWB amplitude would not alter the final results.}.
\item[Case B:]\quad optimistic case in which the rms amplitude of GWB 
spectrum is ten times larger than that in the realistic case, i.e., 
$S_h^{1/2}=5\times10^{-18}$Hz$^{-1/2}$ at $f=1$mHz.
\end{description}
The results of SNR are then shown in solid (case A) and dashed lines 
(case B), respectively.

As anticipated from the sensitivity of antenna pattern function in Table 
\ref{tab:summay_antenna}, the SNR for $TT$-correlation is much less than unity. 
Even in the optimistic case, the SNR is about ten times smaller than
unity and thus the $TT$-correlation data cannot be used for
reconstruction of the GWB skymap. Apart from this, the SNRs for the 
self-correlation signals $AA$ and $EE$ as well as for the
cross-correlation signal $AE$ are generally good compared to the 
cross-correlation data $AT$ and $ET$. With sufficient higher SNR of 
$(S/N)_k\geq5$, the available Fourier components of $AA$-, $EE$- and 
$AE$-correlations become 
$k=-2,\,-1,\,0,\,+1,\,+2$ in both realistic and optimistic cases. 
This is consistent with the previous estimates \cite{Seto:2004np}.
\footnote{The precise numerical values of the SNR for $AA$- and 
$EE$-correlations slightly differ from Ref.~\cite{Seto:2004np}. 
This is mainly because the Euler rotation angles for the LISA orbital 
motion given in Sec.\ref{subsec:correlation} are different from 
those in \cite{Seto:2004np}. } 
On the other hand, with a large amplitude of GWB spectrum (case B), 
only the $k=0$ component is accessible in the signal combinations of 
$AT$ and $ET$. This is mainly due to the fact that the sensitivity of 
the $T$-variable is poor at low-frequency and thereby the noise 
contribution $N_k$ becomes 
$\{\widetilde{C}_{AA,0}S_n^{\rm\scriptscriptstyle TT}\}^{1/2}$ or 
$\{\widetilde{C}_{EE,0}S_n^{\rm\scriptscriptstyle TT}\}^{1/2}$, 
which is much larger than   
$\{\widetilde{C}_{AA,0}\widetilde{C}_{TT,0}\}^{1/2}$ or 
$\{\widetilde{C}_{EE,0}\widetilde{C}_{TT,0}\}^{1/2}$.

Thus, the available Fourier components of correlation data used for 
the reconstruction of the skymap would be severely restricted in practice.  
Under such restricted situation, the deconvolution problem of the linear system 
(\ref{eq:deconvolution}) tends to be \textit{under-determined}. 
Nevertheless, as it will be shown below, one can determine the $\ell=0$, 
$2$ and $4$ modes of multipole coefficients of the Galactic GWB with 
sufficiently small errors. 
In addition, with the $k=0$ components of $AT$- and $ET$-correlations, 
the odd moments $\ell=1$ and $3$ can be recovered.

%
%
%
%
%
%
%
   \subsection{Reconstruction of a skymap}
\label{subsec:reconstruction}

Keeping the remarks on the SNR estimation in previous subsection in
mind, we now proceed to a reconstruction analysis and test the validity
of perturbative reconstruction method presented in Sec. \ref{sec:method}. 
For this purpose, in addition to the analysis in the under-constrained cases 
(case A and B) mentioned above, we also consider the over-constrained 
case as an illustrative example.

\begin{table}[b]
\caption{
Components of $\widetilde{C}_k$ used for reconstruction analysis based 
on the harmonic-Fourier representation}
\begin{ruledtabular}
\begin{tabular}{c|cccccc}
 & AA & EE & TT & AE & AT & ET 
\\
\hline
Over-determined case & $-2\leq k\leq +2$ & $-2\leq k \leq +2$ & none 
& $-8 \leq k \leq +8$ & $-4 \leq k \leq +4$ & $-4 \leq k \leq +4$ 
\\
Under-determined case & & & & & & \\
case A & $-2\leq k\leq +2$ & $-2\leq k \leq +2$ & none 
& $-2 \leq k \leq +2$ & none & none 
\\
case B & $-2\leq k\leq +2$ & $-2\leq k \leq +2$ & none 
& $-2 \leq k \leq +2$ & $k=0$ & $k=0$
\end{tabular}
\label{tab:components_used}
\end{ruledtabular}
\end{table}

\subsubsection{Over-determined case}
\label{subsubsec:over-determined}

\textit{(I) Harmonic-Fourier representation} \quad
Let us first focus on a very idealistic situation that the noise 
contributions are entirely neglected. In such a case, all the components 
of self- and cross-correlation data $\widetilde{C}_k$ are available to 
the reconstruction analysis. 
In practice, however, it is sufficient to consider some restricted 
components among all available data. 
Here, to make a skymap with multipoles of $\ell\leq5$, we use the 
$k=-2\sim+2$ components of self-correlation signals  
$AA$ and $EE$, the $k=-4\sim+4$ components of cross-correlation signals 
$AT$ and $ET$, and $k=-8\sim+8$ components of $AE$-signal 
(see Table \ref{tab:components_used}). Even in this case, the linear 
system (\ref{eq:deconvolution}) is still over-determined and the 
least-squares approximation by SVD is potentially powerful to obtain 
the multipole coefficients of anisotropic GWB. 
The procedure of reconstruction analysis is the same one as presented in 
Fig.\ref{fig:flowchart}. 
For the output signals of harmonic-Fourier representation, we use the 
data $\widetilde{C}_k$ observed at the frequencies 
$\hat{f}=0.05$ and $0.15$, in addition to the data for our interest at 
$\hat{f}=0.1$. 
Collecting these multi-frequency data, the perturbative expansion form 
of the vector $\mathbf{c}(f)$ is specified up to the third order in 
$\hat{f}$ and the coefficients $\mathbf{c}^{(i)}$ are determined 
(Appendix \ref{appendix:SVD}).

\begin{figure}[t]
\begin{center}
 \includegraphics[width=8.3cm,clip]{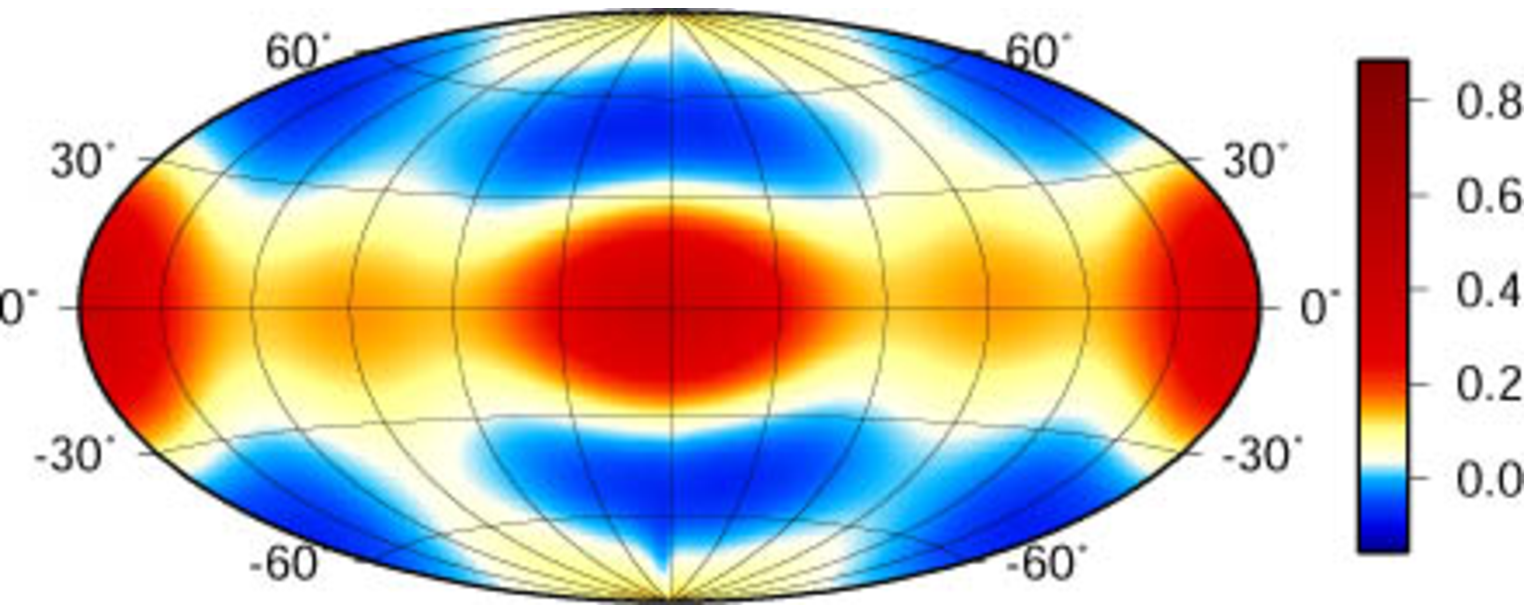}
\hspace{0.3cm}
 \includegraphics[width=8.3cm,clip]{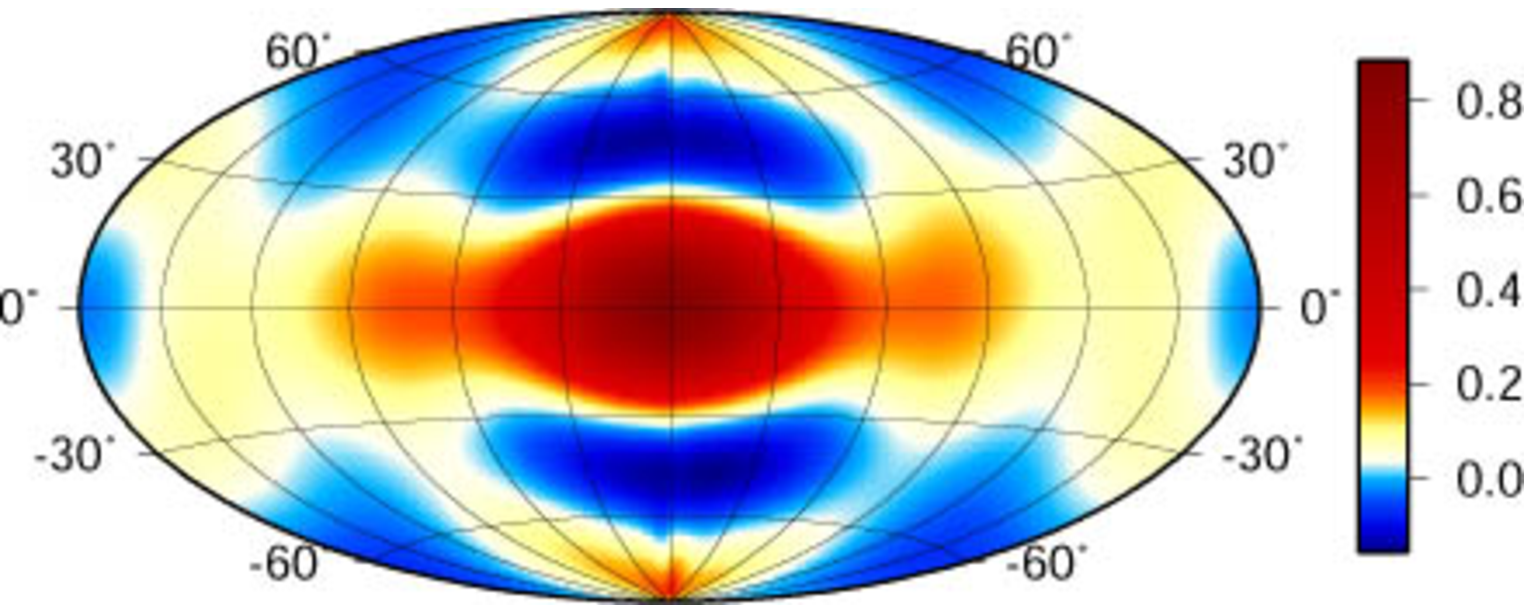}
\end{center}
   \caption{Reconstructed skymap from the time-modulation signals in 
      the {\it{over-determined}} case. The results were obtained 
      by the method based on harmonic-Fourier representation and are 
      plotted in the Galactic coordinate. 
      The left panel shows the leading-order result,  
      where $\ell=0$, $2$ and $4$ modes are only reconstructed, while  
      the right panel represents the result including all reconstructed 
      multipoles ($\ell\leq5$). 
     }
    \label{fig:reconstructed_skymap1}
\end{figure}

\begin{figure}[tp]
\begin{center}
 \includegraphics[width=8.3cm,clip]{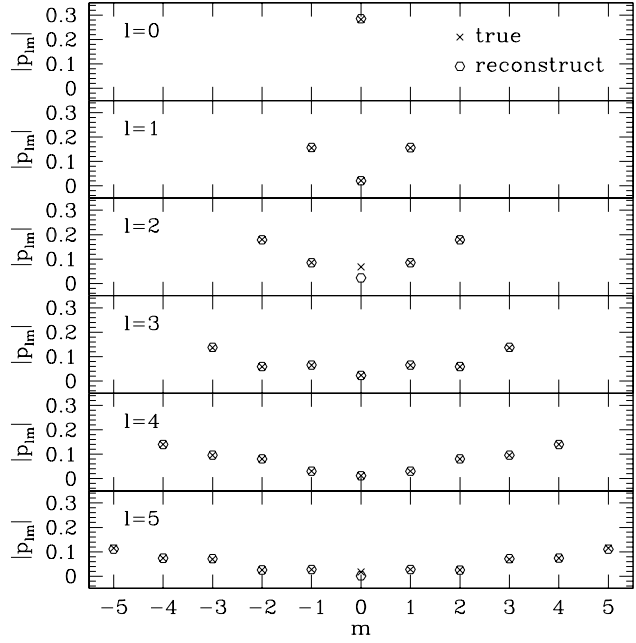}
\end{center}
 \caption{
 Reconstructed values of multipole coefficients $|p_{\ell m}|$  
 in over-determined case.  These numerical values are 
 evaluated in the ecliptic frame. 
 The open circles represent the reconstruction results, while the 
 crosses mean the true values, which are obtained from the spherical 
 harmonic expansion of full-resolution skymap 
   (see Table \ref{tab:summay_multipole}).} 
   \label{fig:reconstructed_plm1}
\end{figure}

In Fig.\ref{fig:reconstructed_skymap1}, the reconstructed results of 
multipole coefficients $p_{\ell m}$ are converted to the projected 
intensity distribution 
(\ref{eq:Y_lm expansion of S and F}) and are shown as Hammer-Aitoff map 
in Galactic coordinate. 
Left panel shows the skymap reconstructed from the lowest-order signals 
of self- and cross-correlation data 
$\mathbf{c}^{(2)}$, which only includes the $\ell=0$, $2$ and $4$ modes, 
while right panel is the result taking account of the leading-order correction 
$\mathbf{c}^{(3)}$. Comparing those with the expected skymap shown in 
Fig.\ref{fig:expected_skymap}, the reconstruction seems almost perfect. 
In Fig. \ref{fig:reconstructed_plm1}, numerical values of the
reconstructed multipole coefficients are compared with the true values 
listed in Table \ref{tab:summay_multipole}. The agreement between the 
reconstruction results ({\it open circles}) and the true values 
({\it crosses}) is quite good and the fractional errors are well within 
a few percent except for $p_{20}$ and $p_{50}$. 
A remarkable fact is that the monopole and the quadrupole values can be 
reproduced reasonably well despite the presence of the degeneracy 
mentioned in Appendix.\ref{appendix:on_the_degeneracy}. 
This is just an accidental result. The (small) discrepancy in the ``recovered'' 
$p_{20}$ can be ascribed to the expression (\ref{eq:fake_solution}). 
After all, the least-squares method by SVD provides a robust
reconstruction method when the linear system (\ref{eq:deconvolution}) 
becomes over-determined.


\begin{figure}[t]
\begin{center}
 \includegraphics[width=8cm,clip]{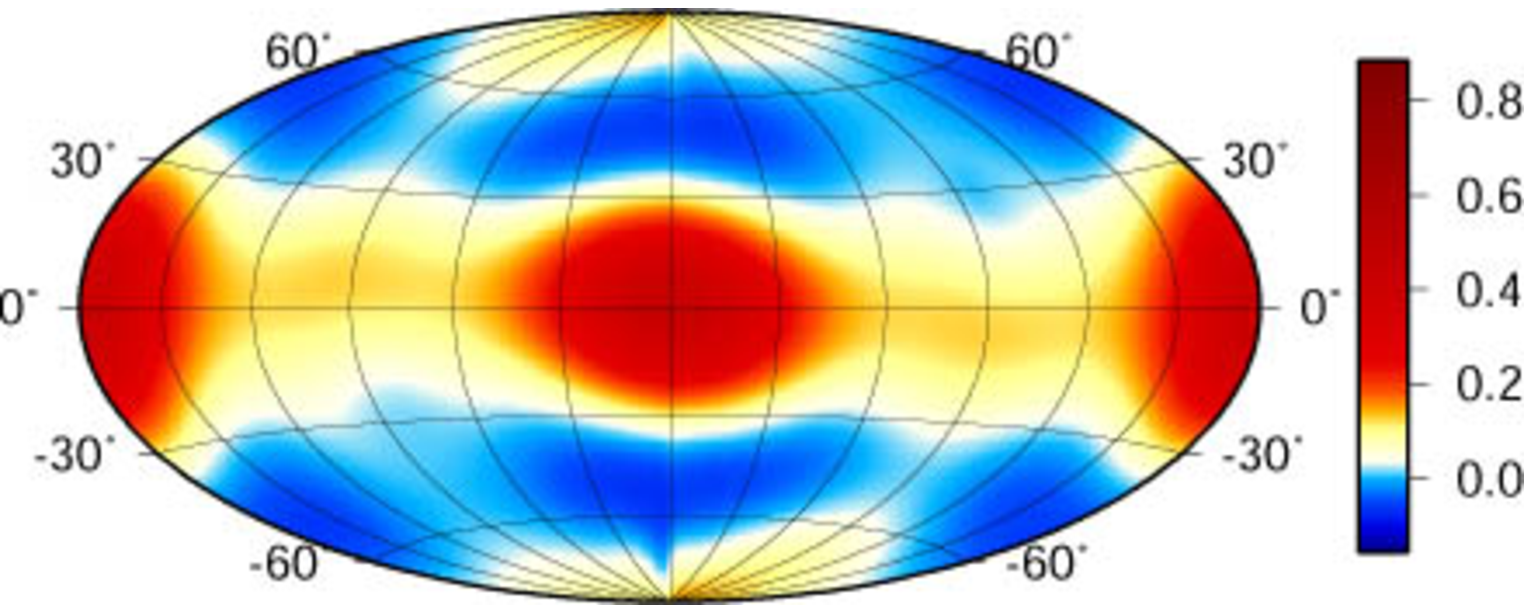}
 \includegraphics[width=8cm,clip]{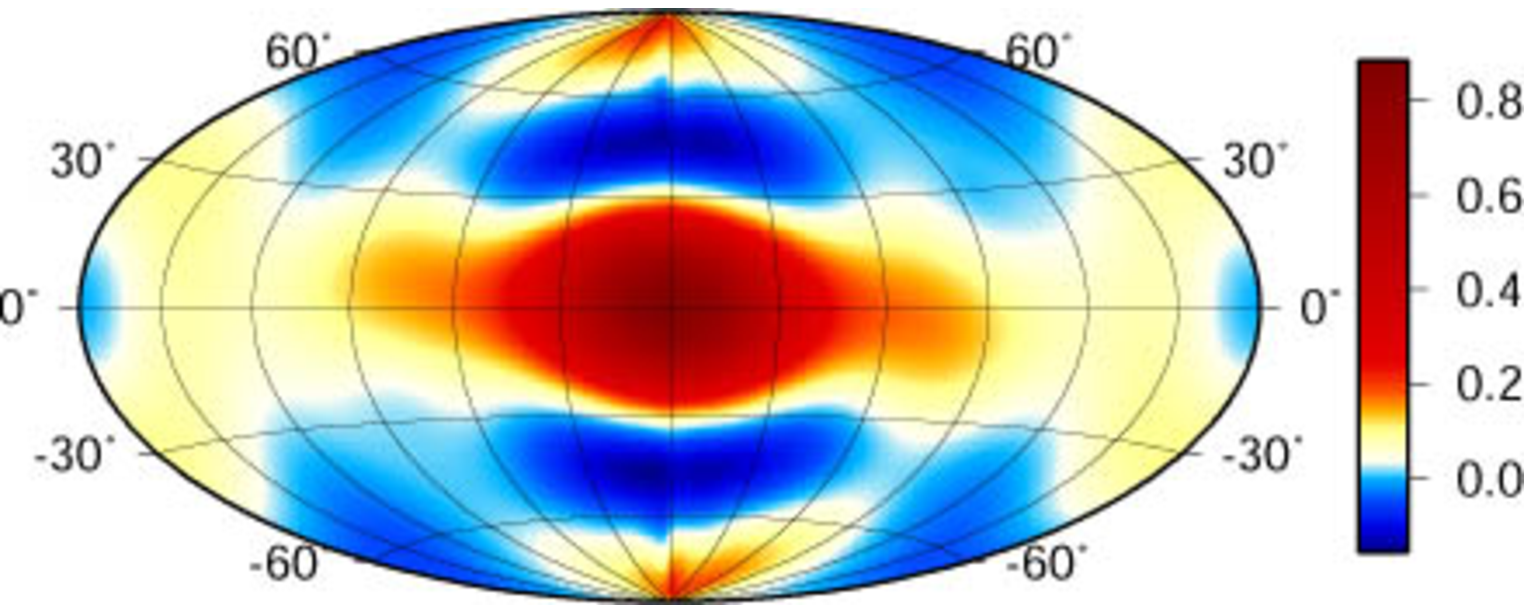}
\end{center}
 \caption{Reconstructed skymap from the time-modulation signals 
 based on the time-series representation. Here, 
 to perform the 
 reconstruction analysis, the number of grid and/or mesh 
 for the discretized data set (\ref{eq:discrete_eq}) 
 was specified as $M=16$ in time and $N=17\times32$ in 
 spherical coordinate.  
 The left panel shows the lowest-order result 
 in which only the $\ell=0$, $2$ and $4$ modes are included, 
 while  the right panel represents the result taking account of 
 the leading-order correction to the antenna pattern functions,  
 which includes the multipoles, $\ell \leq5$. 
 }
 \label{fig:reconstructed skymap2}
\end{figure}

\textit{(II) Time-series representation} \quad 
The successful reconstruction of the GWB skymap can also be achieved 
by the alternative approach based on the time-series representation 
(\ref{eq:def_of_C(f)}). Following the procedure presented in 
Sec.\ref{subsec:time-domain}, the intensity skymap of the GWB is 
directly obtained and the results taking account of the lowest-order 
and the leading-order contributions to the antenna pattern functions 
are shown in left and right panels in 
Fig.\ref{fig:reconstructed skymap2}, respectively. 
Here, to create the discretized data set (\ref{eq:discrete_eq}), the
number of grid and/or mesh was specified as $M=16$ in time and 
$N=17\times32$ in spherical coordinate. The time-series data of antenna 
pattern functions were numerically generated based on the full analytic 
expressions given in Sec.\ref{subsec:antenna_pattern} under assuming
that the arm-length of the three space crafts are rigidly kept fixed. 
The reconstructed skymap reasonably agrees with 
Fig.\ref{fig:reconstructed_skymap1} as well as Fig.\ref{fig:expected_skymap}. 
Although the situation considered here is very idealistic and thus the 
results in Fig.\ref{fig:reconstructed skymap2} should be regarded as 
just a preliminary one, one expects that the methodology based on the 
time-series representation is potentially powerful even when the rigid 
adiabatic treatment of space craft motion becomes inadequate. 
To discuss its effectiveness, a further investigation is needed. 
The details of the analysis including the effects of arm-length 
variation will be presented elsewhere.

\subsubsection{Under-determined case}
\label{subsubsec:under-determined}

\begin{figure}[t]
\begin{center}
 \includegraphics[width=8.3cm,clip]{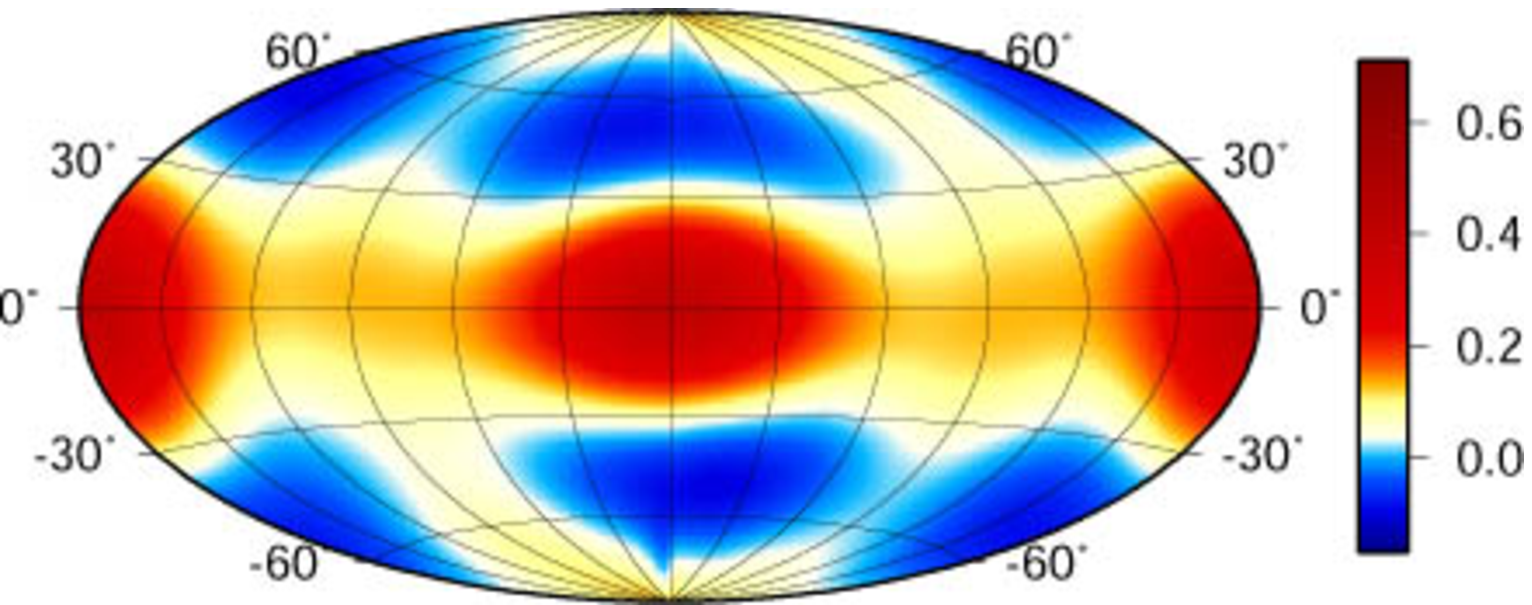}
\end{center}
\begin{center}
 \includegraphics[width=8.3cm,clip]{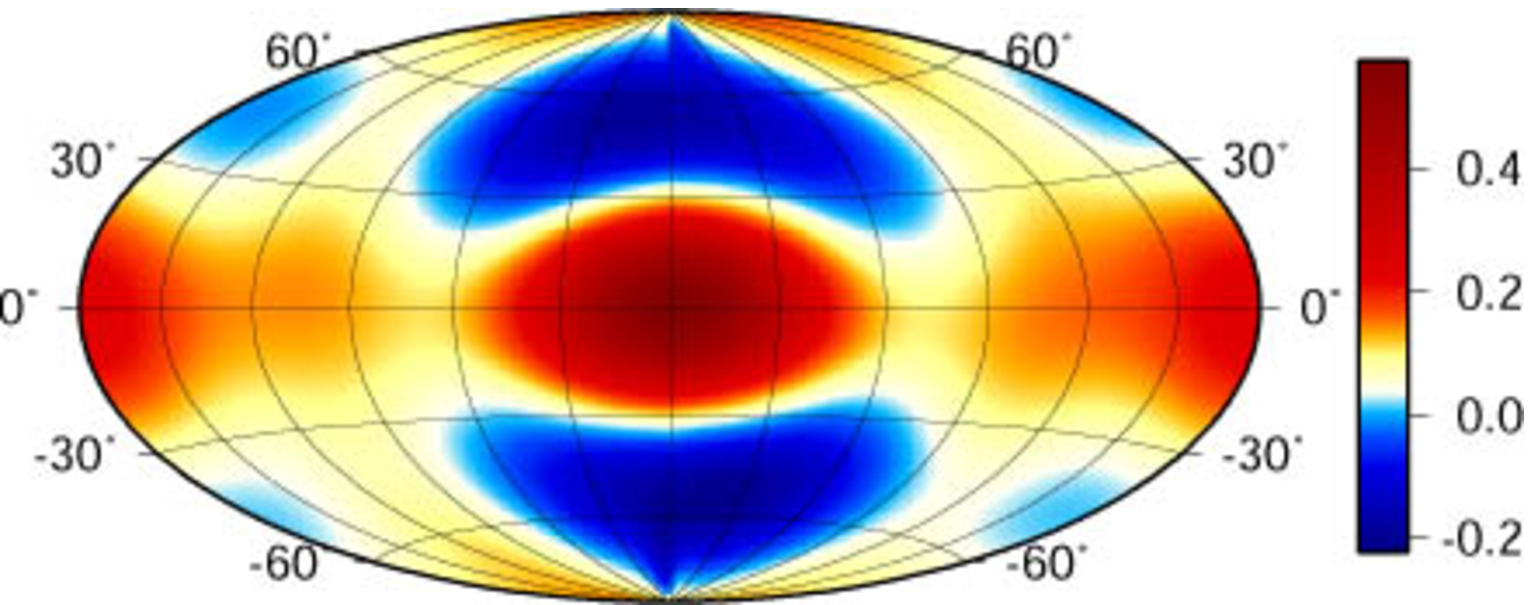}
\hspace*{0.3cm}
 \includegraphics[width=8.3cm,clip]{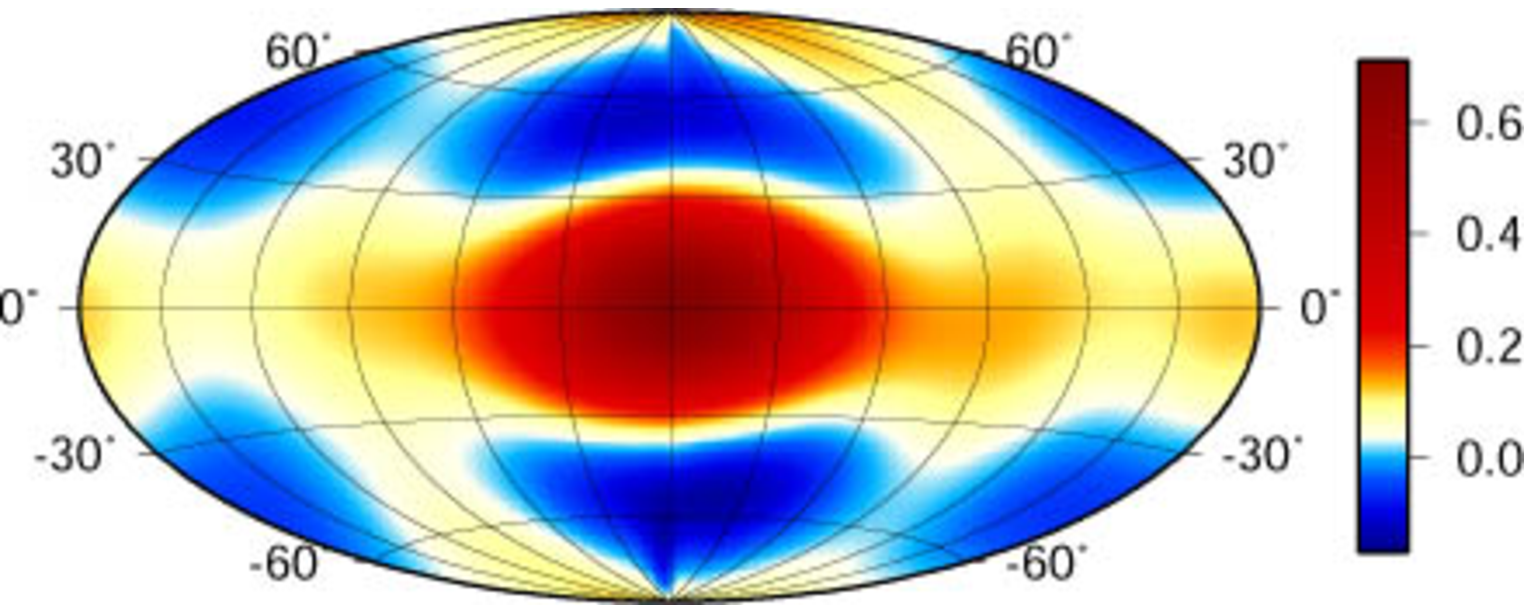}
\end{center}
 \caption{Reconstructed skymap from the time-modulation signals 
 using the restricted Fourier components in {\it{under-determined}} cases 
 (case A, B). These are all plotted in Galactic coordinate. 
 The upper panel shows the result from lowest-order 
 analysis in which only the multipole coefficients of 
 $\ell=0,\,2$ and $4$ modes are considered. 
 The bottom panels are the intensity map taking account of the 
 leading-order correction including the multipoles, $\ell<5$. Left and 
 right panels respectively show the results obtained from the case A and B.  
 Note that the available number of Fourier components was restricted in 
 the reconstruction analysis according to Table \ref{tab:components_used}. }
    \label{fig:reconstructed skymap3}
\end{figure}

\begin{figure}[ht]
\begin{center}
 \includegraphics[width=8.3cm,clip]{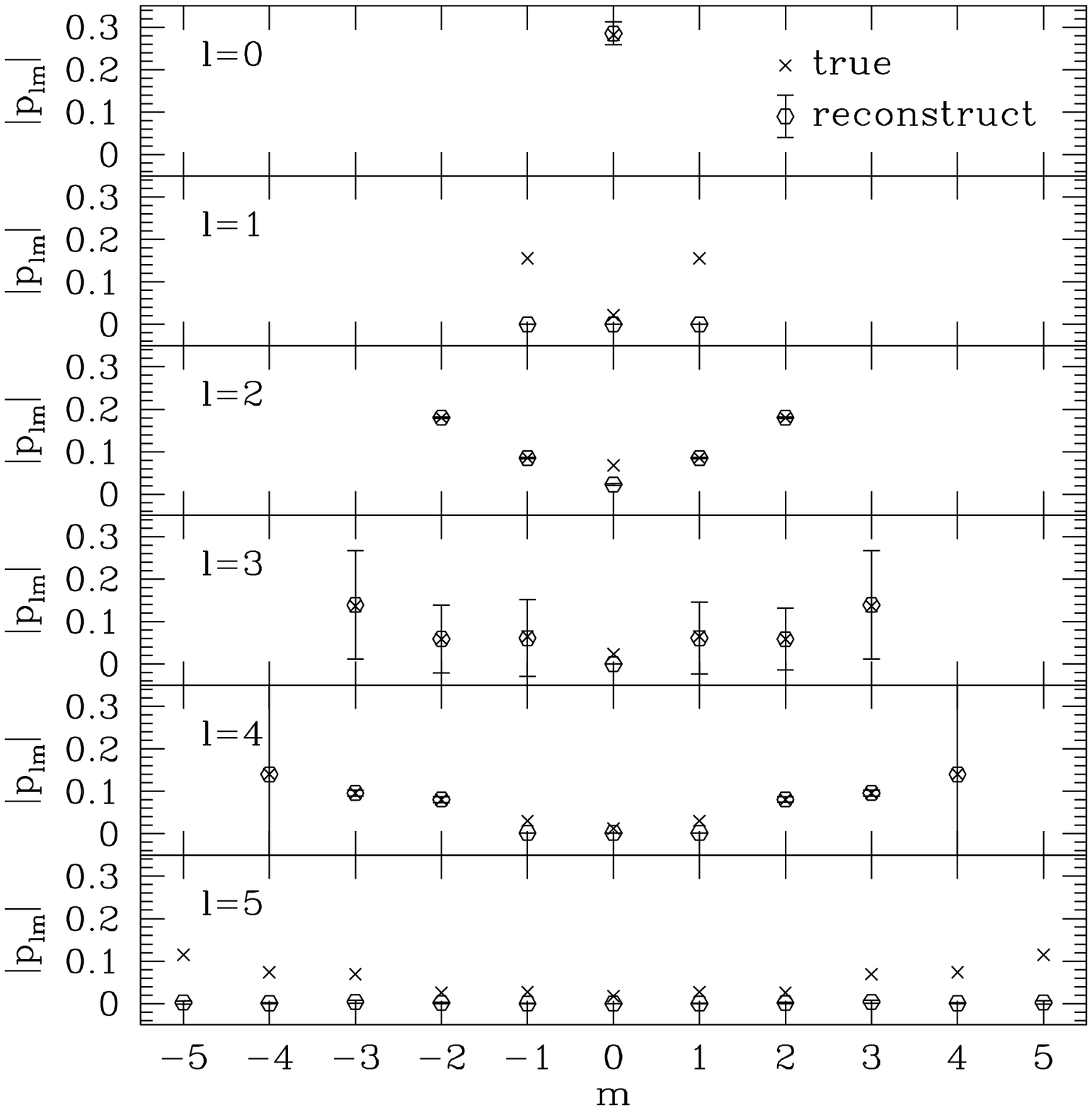}
\hspace*{0.3cm}
 \includegraphics[width=8.3cm,clip]{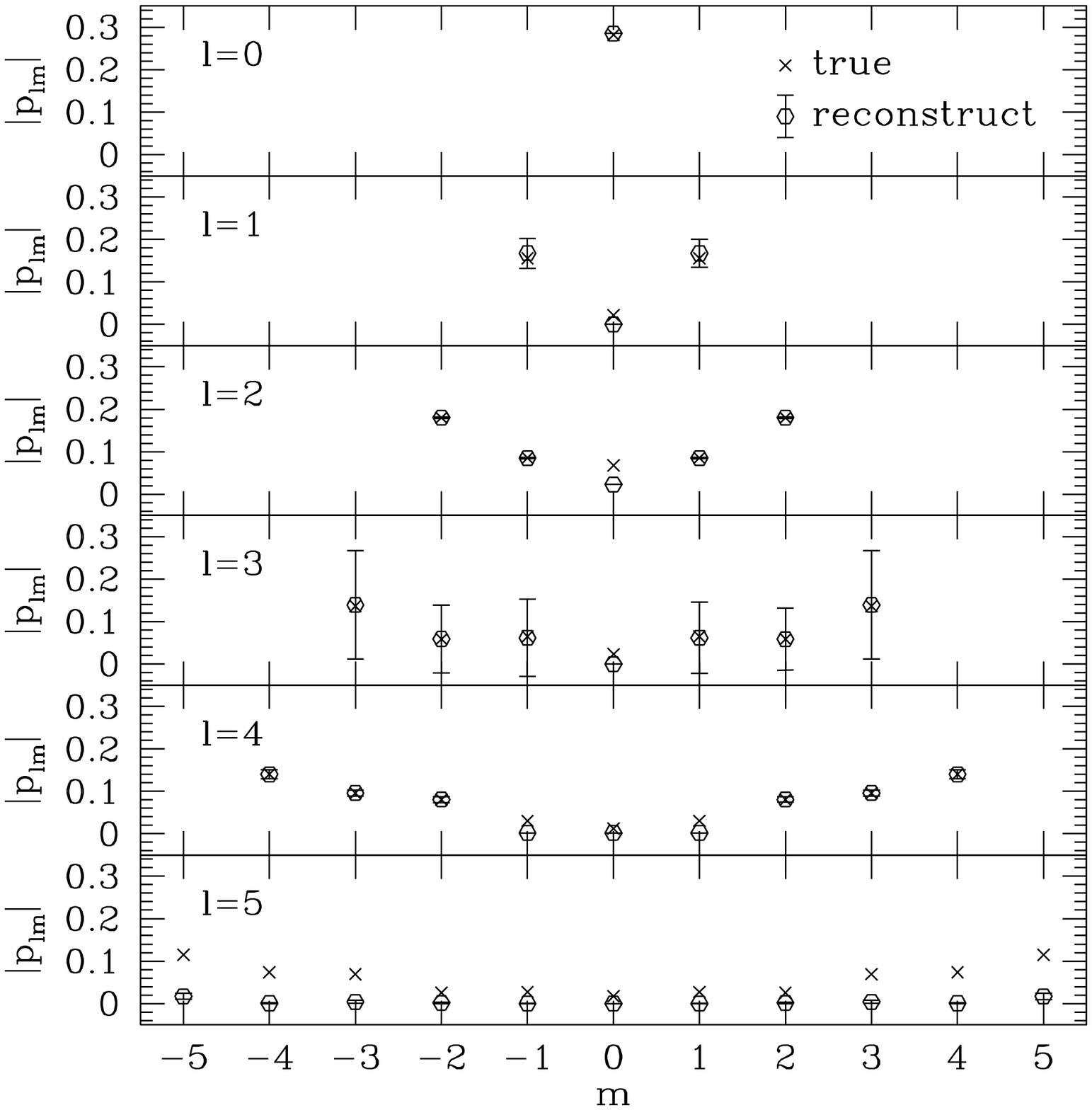}
\end{center}
 \caption{Reconstructed values of the multipole coefficients 
 $|p_{\ell m}|$ in the under-constrained cases. 
 The numerical values are evaluated in the ecliptic coordinates. 
 Left (right) panel shows the result in case A (case B). The open
 circles with error bar represent the reconstruction results, while 
 the crosses mean the true values. The error bars represent the
 statistical error in the signal processing. The evaluation of the error 
 is discussed in Appendix \ref{appendix:statistical_error}. 
 }
 \label{fig:reconstructed_plm3}
\end{figure}

Turning to focus on the analysis based on the harmonic-Fourier 
representation, we next consider the under-constrained case in which 
the number of available Fourier components is restricted due to the 
noises (case A and B discussed in Sec.\ref{subsec:on_snr}). 
Fig.\ref{fig:reconstructed skymap3} shows the reconstructed images of 
GWB intensity map in Galactic coordinate, free from the noises but ]
restricting the number of Fourier components according to Table 
\ref{tab:components_used}. 
The top panel is the intensity map obtained from the lowest-order signals 
$\mathbf{c}^{(2)}$.  
Since the accessible Fourier components are basically the same in 
the lowest-order analysis, the same results are obtained between case A and B. 
On the other hand, the bottom panels of Fig.\ref{fig:reconstructed
skymap3} represent the skymap reconstructed from both 
$\mathbf{c}^{(2)}$ and $\mathbf{c}^{(3)}$, which indicate that the
different images of GWB skymap are obtained depending on the available 
number of Fourier components (or the amplitude of GWB spectrum); 
case A ({\it left}) and case B ({\it right}). 
In Fig.\ref{fig:reconstructed_plm3}, numerical values of the 
reconstructed multipoles $p_{\ell m}$ in ecliptic coordinate are 
summarized together with the statistical errors. 
The statistical errors were roughly estimated according to the
discussion in Appendix \ref{appendix:statistical_error} based on the SNR 
[Eq. (\ref{eq:def_SNR})].

It turns out that the case A fails to reconstruct all the dipole
moments, while they can be somehow reproduced in the case B. 
This is because the cross-correlation data $AT$ and $ET$ which include 
the information about $\ell=1$ modes were not used in the reconstruction 
analysis in case A. Although the $\ell=5$ modes of multipole
coefficients 
were not reproduced well in both cases, their contributions to the
intensity distribution are not large. Hence, the reconstructed GWB 
skymap in case B roughly matches the expected intensity map in 
Fig.\ref{fig:expected_skymap} and the visual impression becomes better 
than case A. This readily implies that large amplitude of the GWB 
spectrum is required for a correct reconstruction of a skymap, 
in practice. However, it should be emphasized that the present 
reconstruction technique can work well in the under-determined cases. 
Even in the realistic situation with smaller amplitude of GWB 
spectrum (case A), the reconstructed skymap including the multipoles 
$\ell<5$ still shows a disk-like structure, which may provide an
important clue to discriminate between the Galactic and the
extragalactic GWBs.

%
%
%
%
%
%
%
%
%
%
%
%
    \section{Conclusion and Discussion}
    \label{sec:conclusion}
%
%
%
%
In this paper, we have presented a perturbative reconstruction method to 
make a skymap of GWB observed via space interferometer.  
The orbital motion of the detector makes the output signals of GWB 
time-dependent due to the presence of anisotropies of GWB. 
Since the output signals of GWB are obtained through an all-sky integral 
of primary signals convolving with an antenna pattern function of 
gravitational-wave detectors, the time dependence of output data can 
be used to reconstruct the luminosity distribution of GWBs under full 
knowledge of detector's antenna pattern functions. 
Focusing on the low-frequency regime, we have explicitly given a 
non-parametric reconstruction method based on both the 
\textit{harmonic-Fourier} and the \textit{time-series} representation. 
With a help of low-frequency expansion of the antenna pattern functions, 
the least-squares approximation by SVD enables us to obtain the
multipole coefficients of GWB or direct intensity map even when the
system becomes over-determined or under-determined. 
For illustrative purpose, the reconstruction analysis of the GWB skymap 
has been demonstrated for the confusion-noise background of Galactic 
binaries around the low-frequency $f=1$mHz. It then turned out that the 
space interferometer LISA free from the noises is capable of making a 
skymap of Galactic GWB with angular resolution up to the multipoles, 
$\ell=5$. For more realistic case based on the estimation of 
signal-to-noise ratios, the system tends to become under-determined 
and the number of Fourier components used in reconstruction analysis 
would be severely restricted. 
Nevertheless, the resultant skymap still contains the information 
of the multipoles up to $\ell<5$, from which one can infer that 
the main sources of GWB come from the Galactic center.

Although the present paper focuses on the reconstruction of the GWB 
skymap from the LISA, the methodology discussed in Sec.\ref{sec:method} 
is quite general and is also applicable to the reconstruction of 
low-frequency skymap obtained from the ground detectors as well as 
the other space interferometers, provided their antenna pattern functions. 
Despite a wide applicability of the present method, however, accessible 
multipole moments of low-frequency GWB are generally restricted to lower 
multipoles due to the properties of antenna pattern functions.  
This would be generally true in any gravitational-wave detectors 
at the frequencies $f<f_*=c/(2\pi L)$, where $L$ is arm-length of 
single detector or separation between the two detectors. 
Hence, with the limited angular resolution, only the present methodology 
may not give a powerful constraint on the luminosity distribution of
GWBs. In this respect, we need to combine the other techniques such as 
the parametric reconstruction method, in which we specifically assume an 
explicit functional form of the luminosity distribution characterized by 
the finite number of model parameters and determine them through the 
likelihood analysis.  The data analysis strategy to give a tight
constraint on the luminosity distribution is definitely a very important 
issue and it must be considered urgently. 
Nevertheless, we note that detectable multipole moments depend
practically on the signal-to-noise ratios.  Since the signal-to-noise 
ratios are determined both from the amplitude of GWB and detector's 
intrinsic noises, a further feasibility study is necessary in order 
to clarify the detectable multipole moments correctly. 
With improved data analysis strategy, it might be even possible that 
the angular resolution of GWB skymap becomes better than that of LISA 
considered in the present paper.

Another important issue on the map-making problem is to consider the 
reconstruction of skymap beyond the low-frequency regime, where the 
antenna pattern functions give a complicated response to the anisotropic 
GWBs and thereby the angular resolution of GWB map can be improved 
\cite{Kudoh:2004he}. 
Since the low-frequency expansion cannot be used there, a new 
reconstruction technique should be devised to extract the information 
of anisotropic GWBs. 
Further, in the case of LISA, the effect of arm-length variation becomes 
important and the rigid adiabatic treatment of the detector response 
cannot be validated. Hence, as emphasized in
Sec.\ref{subsec:time-domain}, it would become essential to consider the 
reconstruction method based on the time-series representation, in which 
no additional assumptions for spacecraft configuration and/or motion are 
needed to compute the antenna pattern functions. The analysis concerning 
this issue will be presented elsewhere. 
%
%
%
%
%
%
%
\begin{acknowledgments}
We would like to thank N. Seto for valuable comments on the estimation of
signal-to-noise ratios. 
We also thank Y. Himemoto and T. Hiramatsu for discussions and comments, 
T. Takiwaki and K. Yahata for the technical support to plot the skymap. 
The work of H.K. is supported by the 
Grant-in-Aid for Scientific Research of Japan Society for Promotion of Science. 
\end{acknowledgments} 
%
%
%
%
%
%
\appendix
\section{Multipole coefficients of antenna 
pattern functions in low-frequency regime}
\label{appendix:Mulipole coefficients}
%
%
%
%
%
In this appendix, based on the analytic expression (21) of paper I, 
the multipole moments for antenna pattern functions of optimal TDIs 
are calculated at detector's rest frame. 
Using the low-frequency approximation $\hat{f}\ll1$, we present the 
perturbative expressions up to the forth order in $\hat{f}$.

To evaluate the antenna pattern functions, we must first specify the 
directional unit vectors $\mathbf{a}$, $\mathbf{b}$ and $\mathbf{c}$, 
which connect respective spacecrafts.
Here, we specifically choose (Fig. 1 and Eq.(28) of Paper I): 
\begin{eqnarray} 
\mathbf{a} = 
-\frac{\sqrt{3}}{2}\,\mathbf{x}  + \frac{1}{2}\,\mathbf{y} ,
\quad\quad
\mathbf{b} = -\mathbf{y} ,
\quad\quad
\mathbf{c} = 
\frac{\sqrt{3}}{2}\,\mathbf{x}  + \frac{1}{2}\,\mathbf{y} ,
\nonumber
\end{eqnarray} 
where the vectors $\mathbf{x}$ and $\mathbf{y}$ respectively denote 
the unit vectors parallel to the $x$- and $y$-axes in detector's rest 
frame (see Fig.1 of Paper I). 
Then, based on the expression (\ref{eq:def AET mode}), the antenna 
pattern functions of the optimal TDIs, $\mathcal{F}_{IJ}$ 
$(I,J=A,E,T)$ are analytically computed at detector's rest frame. 
Their multipole moments become 
\begin{equation}
a_{\ell m}(\hat{f}) = \int_0^{\pi}d\theta\,\int_0^{2\pi}\,d\phi\,
\sin\theta\,\,Y_{\ell m}^*(\theta,\phi)\,\,\mathcal{F}(\hat{f},\theta,\phi),
\end{equation}
which are expressed as a function of normalized frequency 
$\hat{f}=f/f_*$. 
Here, for definiteness, we write down the explicit form of the harmonic 
functions $Y_{\ell m}$:  
\begin{eqnarray}
Y_{\ell}^m (\theta,\phi) 
= \sqrt{\frac{2\ell +1}{4\pi}\,\frac{(\ell-m)!}{(\ell+m)!}} \,\,
P_{\ell}^m(\cos\theta)\,\, e^{i m \phi}.
\label{eq:harmonic_Y_lm}
\end{eqnarray}

The analytic expressions for the multipole moments $a_{\ell m}$ are 
generally intractable due to the complicated form of the antenna 
pattern functions. In the low-frequency regime, however, the 
perturbative treatment regarding $\hat{f}$ as a small expansion 
parameter is applied to derive an analytic expression of multipole 
moments. Below, we summarize the perturbation results up to the fourth 
order in $\hat{f}$:  
\begin{eqnarray}
{\mathcal F}_{AA} :
&&
a_{00}= \frac{2\sqrt{\pi}}{5}\,\hat{f}^2 - 
\frac{211\sqrt{\pi}}{1260}\,\hat{f}^4,
\quad
a_{20}=\frac{4}{7}\sqrt{\frac{\pi}{5}}\,\hat{f}^2- 
\frac{16}{63}\sqrt{\frac{\pi}{5}}\,\hat{f}^4
\quad
a_{22}= - \frac{1-i \,\sqrt{3}}{756}
      {\sqrt{\frac{15\pi }{2}}}\,{\hat f}^4,
\cr
&&
a_{40}= \frac{\sqrt{\pi}}{105}\,{\hat f}^2 +
\frac{13\sqrt{\pi}}{9240}\,{\hat f}^4, 
\quad
a_{42}= \frac{13 (1-i{\sqrt{3}})}{5544} \sqrt{\frac{\pi}{10}}\,{\hat f}^4, 
\quad
a_{44}=-\frac{1+i \sqrt{3}}{6}\sqrt{\frac{\pi}{70}}\,{\hat f}^2 
+\frac{13(1+i \sqrt{3})}{792}\sqrt{\frac{5\pi}{14}}\,{\hat f}^4 
\cr
&&
a_{60}= \frac{1}{1848}\sqrt{\frac{\pi}{13}} \,{\hat f}^4 ,
\qquad
a_{62}=  -\frac{(1 -i\,\sqrt{3} )}{4752} 
\sqrt{\frac{3\pi }{455}}\,{\hat f}^4 ,
\qquad
a_{64}= -\frac{1+i\sqrt{3}}{396} \sqrt{\frac{\pi }{182}}\,{\hat f}^4,
\end{eqnarray}
for self-correlated antenna pattern, $\mathcal{F}_{AA}$.   Note that 
the multipole moments of ${\mathcal F}_{EE}$ are related to those of 
$\mathcal{F}_{AA}$ through the relations,  
$a_{\ell m}^{\scriptscriptstyle\rm EE}=
a_{\ell m}^{\scriptscriptstyle\rm AA}$ for $m=0,\pm6,\pm12,\pm18,\cdots$ and 
$a_{\ell m}^{\scriptscriptstyle\rm EE}=
-a_{\ell m}^{\scriptscriptstyle\rm AA}$ for $m=\pm2,\pm4,\pm 8,\cdots$ 
(see Paper I). The multipole moments of cross-correlation signals are 
\begin{eqnarray}
{\mathcal F}_{AE} :
&&
a_{22}= \frac{3+ i\sqrt{3}}{756} \sqrt{\frac{5\pi}{2}}\,{\hat f}^4,
\quad
a_{33}=\frac{1}{18}  \sqrt{\frac{7\pi}{15}}\,{\hat f}^3 ,
\quad
a_{42}= - \frac{ 13(\sqrt{3}+i)}{5544} \sqrt{\frac{\pi }{10}}\,{\hat f}^4,
\quad
\cr
&& 
a_{44}= \frac{\sqrt{3} -i}{6}\sqrt{\frac{\pi}{70}}\,\hat{f}^2  
-\frac{13(\sqrt{3} -i)}{792}\sqrt{\frac{5\pi}{14}}\,\hat{f}^4,   
\quad
a_{53}=  \frac{1}{18}  \sqrt{\frac{\pi }{1155}}\,{\hat f}^3,
\quad
a_{62}= \frac{\sqrt{3} + i}{4752}\sqrt{ \frac{3\pi}{455} }\,{\hat f}^4,
\cr
&&
a_{64}=  \frac{ \sqrt{3}-i }{396}  \sqrt{\frac{\pi }{182}}\,{\hat f}^4,
\end{eqnarray}

\begin{eqnarray}
{\mathcal F}_{AT } :
&&
a_{11}=  \frac{ ( 1 +i  \sqrt{3} ) \sqrt{\pi}}{168}\,{\hat f}^3,
\quad
a_{22}= -\frac{\sqrt{3}+i}{864}\sqrt{\frac{3\pi }{5}}\,{\hat f}^4,
\quad
a_{31}= \frac{1+i\sqrt{3}}{72}{\sqrt{\frac{\pi }{14}}}\, {\hat f}^3   ,
\quad
\cr
&& 
a_{42}=  \frac{ \sqrt{3}+i}{3168}{\sqrt{\frac{\pi }{5}}}\,{\hat f}^4  ,
\quad
a_{44}=  \frac{17(\sqrt{3}-i)}{3168}\sqrt{\frac{\pi}{35}}\,{\hat f}^4,
\quad
a_{51}= \frac{ 1 +i \sqrt{3}}{1008}{\sqrt{\frac{\pi }{55}}}\,{\hat f}^3 ,
\cr
&&
a_{55}= \frac{1 -i\sqrt{3}}{72}{\sqrt{\frac{3\pi }{154}}}\,{\hat f}^3,
\quad
a_{62}= \frac{\sqrt{3}+i}{4752}{\sqrt{\frac{3\pi }{910}}}\,{\hat f}^4,
\quad
a_{64}= \frac{\sqrt{3} - i}{3168}{\sqrt{\frac{\pi }{91}}}\,{\hat f}^4.
\end{eqnarray}
The multipole moments of antenna pattern function $\mathcal{F}_{ET}$ are 
also obtained from those of $\mathcal{F}_{AT}$ using the relation, 
$ a_{\ell m}^{ET} = - (i/\sqrt{3})\, \tan \left(m\pi/3\right) a_{\ell m
}^{AT}$, as shown in Paper I.

Finally, we also list the multipole moments of self-correlated antenna 
pattern $\mathcal{F}_{TT}$, which are all higher-order contribution with 
$\mathcal{O}(\hat{f}^4)$:  
\begin{eqnarray}
{\mathcal F}_{TT} :
&&
a_{00}=  \frac{\sqrt{\pi} }{504}\,{\hat f}^4 ,
\qquad
a_{40}= - \frac{\sqrt{\pi} }{1584}\,{\hat f}^4 ,
\qquad
a_{60}= - \frac{1}{11088} \sqrt{\frac{\pi}{13}}\,{\hat f}^4,
\qquad
a_{66}= - \frac{1}{48} \sqrt{\frac{\pi }{3003}}\,{\hat f}^4 .
\end{eqnarray}
%
%
%
%
%
%
\section{Remarks on the degeneracy between 
monopole and quadrupole moments for LISA measurement of 
GWB anisotropy}
\label{appendix:on_the_degeneracy}
%
%
%
%
%
%
%
%
%
The reconstruction method based on the harmonic-Fourier representation 
(\ref{eq:deconvolution}) presented in Sec.\ref{subsec:general_scheme} 
has been first discussed by Cornish 
\cite{Cornish:2001hg,Cornish:2002bh}. Later, Seto \& Cooray 
\cite{Seto:2004np} considered the reconstruction of a skymap in 
the low-frequency \textit{limit} using the optimal set of TDI 
signals $A$ and $E$. In this case, the accessible multipole moments 
$p_{\ell m}$ are $\ell=0,\,2$ and $4$ (see Table \ref{tab:summay_antenna}).  
Seto \& Cooray explicitly wrote down the expressions for linear equation 
(\ref{eq:deconvolution}) in the case of the self-correlation signals 
(i.e., AA-, EE-correlations) and showed that the output data with 
sufficient statistical significance are only $\widetilde{C}_0$,  
$\widetilde{C}_{\pm 1}$ and $\widetilde{C}_{\pm 2}$. 
Further, they found that the multipole coefficients 
$p_{00}$ and $p_{20}$ cannot be separately determined due to the 
degeneracy associated with a specific combination between the 
Wigner $D$ matrices and the antenna pattern functions.

Here, we explicitly point out the origin of this degeneracy and 
discuss its influences on the reconstruction of GWB skymap. 
Using the multipole coefficients of antenna patterns in 
Appendix \ref{appendix:Mulipole coefficients}, the lowest-order contribution 
to the linear system 
(\ref{eq:deconvolution}) for AA- and EE-correlations becomes 
\begin{eqnarray}
\widetilde{C}^{(2)}_{AA,0} 
&=&\frac{1}{4\sqrt{\pi}}\left\{
 \frac{2}{5}\,p_{00} - \frac{1}{14\sqrt{5}}\,p_{20}
 -\frac{37}{13440}\,p_{40} 
- \frac{27}{512\sqrt{70}}\sum_{m=-4,4} 
b_{4m}^{\scriptscriptstyle ({\mathrm{AA}})}
\,p_{4m} \right\},
\label{eq:degenerate_CAA0}
\\
\widetilde{C}^{(2)}_{EE,0} 
&=&\frac{1}{4\sqrt{\pi}}\left\{
 \frac{2}{5}\,p_{00} - \frac{1}{14\sqrt{5}}\,p_{20}
-\frac{37}{13440}\,p_{40}
+ \frac{27}{512\sqrt{70}}\sum_{m=-4,4} 
b_{4m}^{\scriptscriptstyle ({\mathrm{AA}})}
\,p_{4m} \right\}.  
\label{eq:degenerate_CEE0}
\end{eqnarray}
with $b^{\scriptscriptstyle ({\mathrm{AA}})}_{44}= 1-i\sqrt{3}= 
[b^{\scriptscriptstyle ( {\mathrm{AA}} )}_{4,-4}]^*$.  
Here, we only show the relevant components of $\widetilde{C}_k$ which 
contains the multipole coefficients $p_{00}$ and $p_{20}$. 
The above expressions include the multipole coefficients of 
$\ell=4$, which are all determined separately from the lowest-order 
contribution of $AE$-correlation, $\widetilde{C}_{AE,k}^{(2)}$.  
Thus, apart from the octupole moments, equations
(\ref{eq:degenerate_CAA0}) and (\ref{eq:degenerate_CEE0}) clearly show 
the presence of degeneracy between the remaining multipole coefficients, 
$p_{00}$ and $p_{20}$.

In a language of the least-squares method by SVD, this degeneracy 
implies that the sub-system in the matrix equation
(\ref{eq:perturbative_c=Ap}) 
containing the coefficients $p_{00}$ and $p_{20}$ becomes under-determined. 
In this case, the least-squares method by SVD cannot correctly produce 
the approximate solutions for $p_{00}$ and $p_{20}$, although it still 
provides some ``approximate'' solution. 
From equations (\ref{eq:degenerate_CAA0}) and
(\ref{eq:degenerate_CEE0}), 
the only meaningful equation for $p_{00}$ and $p_{20}$ is now reduced to 
\begin{equation}
\widetilde{C}=\frac{1}{4\sqrt{\pi}}\left(
 \frac{2}{5}\,p_{00} - \frac{1}{14\sqrt{5}}\,p_{20}\right),  
\end{equation}
where the numerical constant $\widetilde{C}$ represents a collection of
the 
irrelevant terms of $\ell=4$ modes, which are separately determined from
the linear system in $AE$-correlation.  If one naively applies the 
least-squares method to the above system, a very tight relation is obtained:  
\begin{eqnarray}
p_{00} = \frac{7840\sqrt{\pi}}{789}\,\, \widetilde{C},  
\quad \quad
p_{20} = \frac{280\sqrt{5\pi}}{789}\,\, \widetilde{C}.      
\label{eq:fake_solution}
\end{eqnarray}

The presence of degenerate coefficients may be a big obstacle in 
constructing the skymap as well as in determining the normalization 
factor of GWB spectrum. In principle, this degeneracy can be broken 
when we consider the higher-order terms of $\mathcal{O}(\hat{f}^4)$.
However, these terms are generally small and irrelevant for the 
reconstruction analysis in the low-frequency regime. 
In this sense, the reconstruction of low-frequency skymap is, in nature, 
incomplete and the other additional information for monopole or
quadrupole moment is required to make a full skymap. Nevertheless, it 
should be emphasized that with a high signal-to-noise ratio, the other 
remaining multipole coefficients can be all determined by the
least-squares solution by SVD, independently of the above degeneracy. 
Moreover, in cases with $p_{00} \gg p_{20}$, which is usually 
satisfied, the least-squares solution (\ref{eq:fake_solution}) provides 
a modest estimate of the degenerate coefficients $p_{00}$ and $p_{20}$ 
because of the hierarchy of the coefficients in Eq. (\ref{eq:fake_solution}). 
This point has been explicitly demonstrated in Sec.\ref{sec:demonstration}. 
%
%
%

%
%
%
\section{Data sets and least-squares method}
\label{appendix:SVD}
 
In this appendix, we describe in more details how to obtain the 
multipole moments of GWBs from many data streams by applying the 
least-squares method. 
Here we specifically focus on the situation considered in 
Sec. \ref{subsubsec:over-determined}. That is, the data sets that we use are 
(i) the $k=-2 \sim+2$ components of self-correlation signals $AA$ and $EE$, 
(ii) $k=-8\sim+8$ components of $AE$-signal, and (iii) the $k=-4\sim+4$ 
components of cross-correlation signals $AT$ and $ET$.  According to the 
general strategy given in Sec. \ref{subsec:general_scheme} (see Eq. 
(\ref{eq:perturbative_c=Ap})), we first combine the leading data streams 
of (i) and (ii) which correspond to $i=2$ in Eq. (\ref{eq:expand_vec_c}).  
The combined data sets consist of the following matrices. 
\begin{equation}
  \left(
  \begin{array}{c}
   \tilde{C}_{+2 ,AA}^{(2)}  \\
   \tilde{C}_{+1 ,AA}^{(2)}  \\
   \tilde{C}_{-1 ,AA}^{(2)}  \\
   \tilde{C}_{-2 ,AA}^{(2)}  \\
   \tilde{C}_{+2 ,EE}^{(2)}  \\
    \vdots    \\
   \tilde{C}_{-2 ,EE}^{(2)}  \\   
  \end{array}
\right)
=
\left(
\begin{array}{ccccccccccccc}
 0  &0     &0     & * &  \\
 0  &0     & *    & 0 &  \\
 0  &*     &0     & 0 &  \\ 
 *  &0     &0     & 0 &  \\
 0  &0     &0     & * &  \\
 0  &0     &*     & 0 &  \\
 0  &*     &0     & 0 &  \\ 
 *  &0     &0     & 0 &  \\ 
\end{array} 
\right)
    \left(
  \begin{array}{c}
    p_{2 2}  \\
    p_{2 1}  \\
    p_{2,-1}   \\
    p_{2,-2}   \\
  \end{array}
\right), 
\quad
  \left(
  \begin{array}{c}
   \tilde{C}_{+8 ,AE}^{(2)}  \\
   \tilde{C}_{+7 ,AE}^{(2)}  \\
    \vdots    \\
   \tilde{C}_{0 ,AE}^{(2)}  \\
    \vdots    \\
   \tilde{C}_{-7 ,AE}^{(2)}  \\
   \tilde{C}_{-8 ,AE}^{(2)}  \\
  \end{array}
\right)
=
\left(
\begin{array}{ccccccccccccc}
 0  &      &&      & * &  \\
    &      &&\Rdots&   &  \\
    &\Rdots&&      &   &  \\ 
 *  &      &&      & * &  \\
    &      &&\Rdots&   &  \\
    &\Rdots&&      &   &  \\ 
 *  &      &&      & 0 &  \\ 
\end{array} 
\right)
    \left(
  \begin{array}{c}
    p_{4 4}  \\
    p_{4 3}  \\
    \vdots    \\
    p_{4 0}    \\
    \vdots    \\
    p_{4,-3}  \\
    p_{4,-4}  \\
  \end{array}
\right), 
\end{equation}
where $*$ represents a non-vanishing complex number. Note that for 
simplicity we have not included the data $\tilde{C}_{0 ,AA}^{(2)}$ 
and  $\tilde{C}_{0 ,EE}^{(2)}$, which contain degeneracy between 
the multipole moments (see Sec. \ref{appendix:on_the_degeneracy}). 
The matrices in the right hand side correspond to $\mathbf{A}^{(i)}$ 
in Eq. (\ref{eq:expand_vec_c}). These sparse forms of matrix are typical 
for the present problem. 
We perform the singular value decomposition with respect to these sparse 
matrices, and construct pseudo-inverse matrices which give the 
least-squares solutions of $p_{\ell m}$. 
In a similar way the least-squares solutions of $\ell=\mathrm{odd}$ 
modes are obtained from the following matrix equation,  
\begin{equation}
  \left( 
  \begin{array}{c}  
   \tilde{C}_{+8 ,AE}^{(3)}  \\
    \vdots    \\
   \tilde{C}_{-8 ,AE}^{(3)}  \\
   \tilde{C}_{4 ,AT}^{(3)}  \\
    \vdots    \\
   \tilde{C}_{-4 ,AT}^{(3)}  \\
   \tilde{C}_{4  ,ET}^{(3)}  \\
    \vdots    \\
   \tilde{C}_{-4 ,ET}^{(3)}  \\
  \end{array}  
\right)  
=
\left(
\begin{array}{ccccccccccccc}
    &      &                           &  &  \\
    &      &  \mathbf{A}_{\mathrm{AE}}^{(3)}    &  &  \\
    &      &      &   &  \\ 
    &      &  \mathbf{A}_{\mathrm{AT}}^{(3)}    &  &  \\
    &      &      &   &  \\
    &      &  \mathbf{A}_{\mathrm{ET}}^{(3)}    &  &  \\
    &      &      &   &  \\ 
\end{array} 
\right)
    \left(
  \begin{array}{c}
    p_{11}    \\
    p_{10}    \\
    p_{1,-1}   \\
    p_{33}    \\
    \vdots    \\
    p_{3,-3}  \\
    p_{5 5}   \\
    \vdots    \\
    p_{5, -5}  \\
  \end{array}
\right),  
\end{equation}
where we have symbolically written the matrix ${\mathbf{A}}^{(3)}$ due 
to space limitation. 
%
%
%
%
%
\section{Multipole coefficients for gravitational-wave 
  backgrounds}
\label{appendix:multipole of GWB}
%
%
%
%
Here, we give the multipole coefficients for normalized intensity 
distribution $P(\mathbf{\Omega})$ of galactic GWB presented in 
Sec.\ref{subsec:GWB_model}. 
To evaluate this, we first create the intensity map $P(\mathbf{\Omega})$ 
with $129\times256$ regular grids of celestial sphere $(\theta,\,\phi)$. 
Then, the spherical harmonic expansion of the intensity map is
numerically carried out using the SPHEREPACK 3.1 package 
\cite{Adams:2003}.  
Table \ref{tab:summay_multipole} summarizes the multipole coefficients 
$p_{\ell m}$ up to $\ell=5$, which are relevant to the analysis in 
Sec.\ref{sec:demonstration}. Note also that 
$p_{\ell,-m}=(-1)^m p_{\ell m}^*$.

\begin{figure}[ht]
    \begin{center}
    \includegraphics[width=8.3cm,clip]{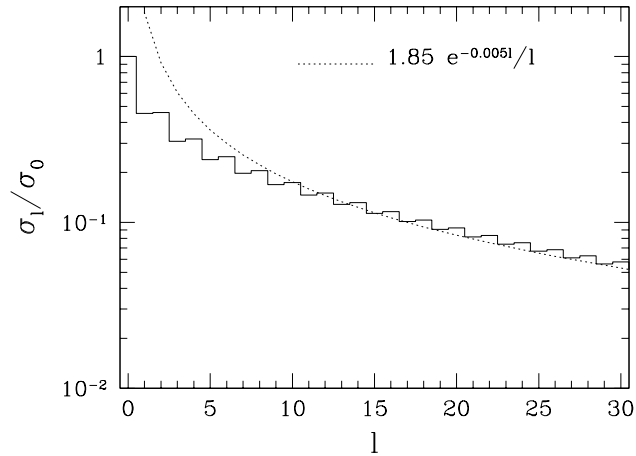}
    \end{center}
    \caption{Normalized angular power of galactic GWB anisotropy, 
$\sigma_{\ell}/\sigma_0$ as a function of multipole $\ell$. 
      \label{fig:sigma_gwb}}
\end{figure}

To characterize the contribution of the $\ell$-th moment to the galactic 
GWB, we introduce the angular power defined by 
\begin{eqnarray}
\sigma_{\ell}=\left\{ \frac{1}{2\ell+1}\,\, 
\sum_{m=-\ell}^{\ell} |p_{\ell m}|^2 \right\}^{1/2},  
\end{eqnarray}
which is rotationally invariant \cite{Kudoh:2004he}. 
Figure \ref{fig:sigma_gwb} shows the normalized angular power, 
$\sigma_{\ell}/\sigma_0$ up to $\ell=30$. 
The dominant contribution to the intensity of galactic GWB comes from
the multipoles with $\ell\lesssim 4$, however, the asymptotic behavior 
at higher multipoles is very slow and can be fitted by 
$\sigma_{\ell}/\sigma_{0} \sim 1.85\, e^{-0.005\ell}/\ell$ 
(dotted line in Fig.\ref{fig:sigma_gwb}), which turns out to be a good 
approximation even to much higher multipoles, $\ell\sim100$. 

\begin{table}[t]
\caption{
Multipole coefficients for galactic GWB. 
$p_{\ell m}$ with $m<0$ is given by 
$p_{\ell m}= (-1)^m p^*_{\ell,-m}$.
}
\begin{tabular}{cc|rrr}
$\ell$ & $m$ & Re[$p_{\ell m}$] & Im[$p_{\ell m}$] \\
\hline\hline
0 & 0 & 0.282  & 0 \\
\hline
1 & 0 & -0.0215 & 0 \\
1 & 1 & 0.00876& -0.156 \\
\hline
2 & 0 & -0.0681 & 0 \\
2 & 1 & -0.0828 & 0.0216 \\
2 & 2 & -0.180 & -0.0124 \\
\hline
3 & 0 & 0.0231 & 0 \\
3 & 1 & 0.00419 & 0.0648 \\
3 & 2 & 0.0212 & 0.0544 \\
3 & 3 & -0.0170 & 0.135 \\
\hline
4 & 0 & 0.0120 & 0 \\
4 & 1 & -0.0205 & -0.0224 \\
4 & 2 & 0.0807 & -0.00274 \\
4 & 3 & 0.0936 & -0.0198 \\
4 & 4 & 0.139 & 0.0194 \\
\hline
5 & 0 & -0.0172 & 0 \\
5 & 1 & 0.000230 & -0.0272 \\
5 & 2 & -0.0250 & -0.00640 \\
5 & 3 & -0.00336 & -0.0693 \\
5 & 4 & -0.0177 & -0.0719 \\
5 & 5 & 0.0222 & -0.1.13 \\
\hline
\hline
\end{tabular}
\label{tab:summay_multipole}
\end{table}
%
%
%
%
%
%
%
%
\section{Estimation of statistical error 
  in reconstruction analysis}
\label{appendix:statistical_error}

In the reconstruction analysis based on the harmonic-Fourier
representation, the statistical errors plotted in 
Fig.\ref{fig:reconstructed_plm3} are roughly estimated as follows. 
In the presence of the noises, the least-squares solution given 
in equation (\ref{eq:perturbative_p=AC}) becomes 
\begin{equation}
\mathbf{p}^{(i)}_{\rm approx} =\left[\mathbf{A}^{(i)}\right]^+ \cdot 
\left\{\mathbf{c}^{(i)}+\mathbf{s}_{\rm n}^{(i)}\right\}, 
\end{equation}
where the additional term $\mathbf{s}_{\rm n}^{(i)}$ represents the 
noise contributions to the $i$-th order coefficient of perturbative 
expansion for $\mathbf{c}(f)$. 
The root-mean-square amplitude of the error $\Delta\mathbf{p}^{(i)}$ 
is then defined by taking the ensemble average of the random noises as 
$\Delta\mathbf{p}^{(i)}
\equiv \left\langle|\mathbf{p}^{(i)} 
-\langle\mathbf{p}^{(i)}\rangle|^2\right\rangle$, which gives the 
errors in multipole coefficient $\Delta p_{\ell m}$. 
The $j$-th components of the vector $\Delta\mathbf{p}^{(i)}$ becomes 
\begin{equation}
|\Delta\mathbf{p}^{(i)}_{{\rm approx},j}|^2 =
\left[\mathbf{A}^{(i)}\right]^+_{jk} \left[\mathbf{A}^{(i)}\right]^{+*}_{jk} \,
\langle |\mathbf{s}_{{\rm n},k}^{(i)}|^2\rangle.
\end{equation}
Here, the variance $\langle |\mathbf{s}_{\rm n,k}^{(i)}|^2\rangle$ 
roughly corresponds to the quadrature of the vector $\mathbf{c}^{(i)}$ 
divided by the signal-to-noise ratio:  
\begin{equation}
\mathbf{S}_{{\rm n},k}^{(i)}= \left\{\alpha\,\,
\frac{\displaystyle \mathbf{c}_k^{(i)}}
{\displaystyle \mathbf{(s/r)}_k}\right\}^2.  
\label{eq:error}
\end{equation}
In the above expression, the vector $\mathbf{(s/r)}$ represents the SNR, 
each component of which is the quantity $(S/N)_k$ defined in 
(\ref{eq:def_SNR}) just for the same component of the vector 
$\mathbf{c}^{(i)}$. 
Notice that the factor $\alpha$ is multiplied in equation
(\ref{eq:error}) by hand. 
The reason why we have introduced the factor $\alpha$ is as follows. 
First note that the quantity $(S/N)_k$ basically reflects the 
signal-to-noise ratio for the most dominant term in the perturbative 
expansion for the signal 
$\tilde{C}_k(f)$ (see Eq.(\ref{eq:expand_vec_c})). 
For the $AE$-correlation,  $(S/N)_k$ gives the signal-to-noise 
ratio for second-order term, i.e., $\mathbf{c}^{(2)}$. 
For the $AT$-correlation,  $(S/N)_k$ reflects the signal-to-noise ratio for 
$\mathbf{c}^{(3)}$. 
In our reconstruction analysis, the higher-order contributions to the 
$AE$-correlation, $\mathbf{c}^{(3)}$ is used to make the skymap 
with multipoles $\ell\leq5$, which can be estimated by analyzing the 
multi-frequency data. In general, the signal from higher-order 
contribution is weaker than the lowest-order term, leading to the 
calibration error. The significance of this error would be enhanced 
by the factor roughly proportional to $\hat{f}^{-(i-2)}$ for the $i$-th 
order terms of $AE$-correlation. Hence, in order to mimic this, the 
factor $\alpha$ is multiplied and set to $\alpha=\hat{f}^{-(i-2)}$. 
In the case examined in Fig.\ref{fig:reconstructed_plm3}, we adopt 
$\alpha=\hat{f}^{-1}=10$ for third-order cross-correlation data 
$\tilde{C}_{{\rm AE},k}^{(3)}$.   
Otherwise we set $\alpha=1$. As a result, statistical errors of $\ell=3$ 
modes become larger than those in $\ell=$even modes 
(see Fig.\ref{fig:reconstructed_plm3}). Note that the multipole 
coefficients with $\ell=1$ in case A and with $\ell=5$ in both case 
are completely degenerate and cannot be recovered by reconstruction 
analysis. Hence, the statistical errors were not evaluated.

\section{Computational method of multipole coefficients}

In this Appendix, we give a brief description of the numerics of 
calculating the multipole coefficients using the SPHEREPACK 3.1 
package \cite{Adams:2003}.  
The traditional spherical harmonic transform of a scalar function is 
\begin{eqnarray}
 \Psi (\theta,\phi) = \sum_{\ell,m} p_{\ell m} Y_{\ell m} (\theta,\phi).
\label{eq:Psi}
\end{eqnarray}
The spherical harmonics are given by equation 
(\ref{eq:harmonic_Y_lm}) and the associated Legendre polynomials are 
\begin{eqnarray}
 P^m_{\ell} (x) = \frac{(-1)^m}{2^\ell \ell!} (1-x^2)^{m/2} \frac{d^{m+\ell}}
{dx^{m+\ell}}(x^2-1)^\ell.
\label{eq:legendre}
\end{eqnarray}
On the other hand, numerical computation of spherical harmonic 
expansion is performed with the subroutine \verb|shaec| in 
SPHEREPACK 3.1 package, which directly gives the following expansion form: 
\begin{eqnarray}
\Psi(\theta,\phi) &=& \sqrt{ \frac{\pi}{2} }
        \sum_{\ell=0}^\infty \sum_{m=0}^{\infty}{}'
        P^m_\ell (\cos\theta) \gamma_{\ell m} 
\left[ \alpha_{\ell m} \cos m \phi - \beta_{\ell m} \sin m \phi \right], 
\label{eq:spherepack expansion}
\end{eqnarray}
where the prime notation on the sum indicates that the first term 
corresponding to $m=0$ is multiplied by the factor $1/2$. 
To relate the coefficients $\alpha_{\ell m}$ and $\beta_{\ell m}$ with 
$p_{\ell m}$, the expansion (\ref{eq:Psi}) is compared with the 
expression (\ref{eq:spherepack expansion}), leading to  
\begin{eqnarray}
p_{\ell m } &=&   \sqrt{ \frac{\pi}{2} } 
(\alpha_{\ell m} + i \beta_{\ell m} ),
\cr
(-1)^m  p_{\ell,- m} &=&  \sqrt{ \frac{\pi}{2} } (\alpha_{\ell m} - i 
\beta_{\ell m}), 
\label{eq:alpha_ellm to p_ellm}
\end{eqnarray}
for $m >0$, and 
\begin{equation}
p_{\ell 0} =  \sqrt{ \frac{\pi}{2} } \alpha_{\ell 0}, 
\label{eq:alpha_ellm to p_ellm2}
\end{equation}
for $m=0$. Hence, with a help of these expressions, one can read 
off the multipole coefficients $p_{\ell m}$ from the expansion 
formula (\ref{eq:spherepack expansion}). 
Notice that the associated Legendre polynomials used in the 
SPHEREPACK 3.1 package differ from (\ref{eq:legendre}) by a 
factor $(-1)^m$ so that one must multiply this factor to the 
transformation law (\ref{eq:alpha_ellm to p_ellm}) and 
(\ref{eq:alpha_ellm to p_ellm2}).

%
%
%
%
%
%



\begin{thebibliography}{37}
\expandafter\ifx\csname natexlab\endcsname\relax\def\natexlab#1{#1}\fi
\expandafter\ifx\csname bibnamefont\endcsname\relax
  \def\bibnamefont#1{#1}\fi
\expandafter\ifx\csname bibfnamefont\endcsname\relax
  \def\bibfnamefont#1{#1}\fi
\expandafter\ifx\csname citenamefont\endcsname\relax
  \def\citenamefont#1{#1}\fi
\expandafter\ifx\csname url\endcsname\relax
  \def\url#1{\texttt{#1}}\fi
\expandafter\ifx\csname urlprefix\endcsname\relax\def\urlprefix{URL }\fi
\providecommand{\bibinfo}[2]{#2}
\providecommand{\eprint}[2][]{\url{#2}}

\bibitem[{\citenamefont{Seto et~al.}(2001)\citenamefont{Seto, Kawamura, and
  Nakamura}}]{Seto:2001qf}
\bibinfo{author}{\bibfnamefont{N.}~\bibnamefont{Seto}},
  \bibinfo{author}{\bibfnamefont{S.}~\bibnamefont{Kawamura}}, \bibnamefont{and}
  \bibinfo{author}{\bibfnamefont{T.}~\bibnamefont{Nakamura}},
  \bibinfo{journal}{Phys. Rev. Lett.} \textbf{\bibinfo{volume}{87}},
  \bibinfo{pages}{221103} (\bibinfo{year}{2001}), \eprint{astro-ph/0108011}.

\bibitem[{BBO(2003)}]{BBO:2003}
\bibinfo{journal}{URL, http://universe.gsfc.nasa.gov/program/bbo.html}
  (\bibinfo{year}{2003}).

\bibitem[{\citenamefont{Ungarelli and
  Vecchio}(2001{\natexlab{a}})}]{Ungarelli:2000jp}
\bibinfo{author}{\bibfnamefont{C.}~\bibnamefont{Ungarelli}} \bibnamefont{and}
  \bibinfo{author}{\bibfnamefont{A.}~\bibnamefont{Vecchio}},
  \bibinfo{journal}{Phys. Rev.} \textbf{\bibinfo{volume}{D63}},
  \bibinfo{pages}{064030} (\bibinfo{year}{2001}{\natexlab{a}}),
  \eprint{gr-qc/0003021}.

\bibitem[{\citenamefont{Smith et~al.}(2005)\citenamefont{Smith, Kamionkowski,
  and Cooray}}]{Smith:2005mm}
\bibinfo{author}{\bibfnamefont{T.~L.} \bibnamefont{Smith}},
  \bibinfo{author}{\bibfnamefont{M.}~\bibnamefont{Kamionkowski}},
  \bibnamefont{and} \bibinfo{author}{\bibfnamefont{A.}~\bibnamefont{Cooray}}
  (\bibinfo{year}{2005}), \eprint{astro-ph/0506422}.

\bibitem[{\citenamefont{Hils et~al.}(1990)\citenamefont{Hils, Bender, and
  Webbink}}]{Hils:1990hg}
\bibinfo{author}{\bibfnamefont{D.}~\bibnamefont{Hils}},
  \bibinfo{author}{\bibfnamefont{P.~L.} \bibnamefont{Bender}},
  \bibnamefont{and} \bibinfo{author}{\bibfnamefont{R.~F.}
  \bibnamefont{Webbink}}, \bibinfo{journal}{Astrophys. J.}
  \textbf{\bibinfo{volume}{360}}, \bibinfo{pages}{75} (\bibinfo{year}{1990}).

\bibitem[{\citenamefont{Bender and Hils}(1997)}]{Bender:1997bc}
\bibinfo{author}{\bibfnamefont{P.~L.} \bibnamefont{Bender}} \bibnamefont{and}
  \bibinfo{author}{\bibfnamefont{D.}~\bibnamefont{Hils}},
  \bibinfo{journal}{Class. Quant. Grav.} \textbf{\bibinfo{volume}{14}},
  \bibinfo{pages}{1439} (\bibinfo{year}{1997}).

\bibitem[{\citenamefont{Nelemans et~al.}(2001)\citenamefont{Nelemans,
  Yungelson, and Portegies~Zwart}}]{Nelemans:2001hp}
\bibinfo{author}{\bibfnamefont{G.}~\bibnamefont{Nelemans}},
  \bibinfo{author}{\bibfnamefont{L.~R.} \bibnamefont{Yungelson}},
  \bibnamefont{and} \bibinfo{author}{\bibfnamefont{S.~F.}
  \bibnamefont{Portegies~Zwart}}, \bibinfo{journal}{Astron. Astrophys.}
  \textbf{\bibinfo{volume}{375}}, \bibinfo{pages}{890} (\bibinfo{year}{2001}),
  \eprint{astro-ph/0105221}.

\bibitem[{\citenamefont{Benacquista et~al.}(2004)\citenamefont{Benacquista,
  DeGoes, and Lunder}}]{Benacquista:2003th}
\bibinfo{author}{\bibfnamefont{M.~J.} \bibnamefont{Benacquista}},
  \bibinfo{author}{\bibfnamefont{J.}~\bibnamefont{DeGoes}}, \bibnamefont{and}
  \bibinfo{author}{\bibfnamefont{D.}~\bibnamefont{Lunder}},
  \bibinfo{journal}{Class. Quant. Grav.} \textbf{\bibinfo{volume}{21}},
  \bibinfo{pages}{S509} (\bibinfo{year}{2004}), \eprint{gr-qc/0308041}.

\bibitem[{\citenamefont{Edlund et~al.}(2005)\citenamefont{Edlund, Tinto,
  Krolak, and Nelemans}}]{Edlund:2005ye}
\bibinfo{author}{\bibfnamefont{J.~A.} \bibnamefont{Edlund}},
  \bibinfo{author}{\bibfnamefont{M.}~\bibnamefont{Tinto}},
  \bibinfo{author}{\bibfnamefont{A.}~\bibnamefont{Krolak}}, \bibnamefont{and}
  \bibinfo{author}{\bibfnamefont{G.}~\bibnamefont{Nelemans}},
  \bibinfo{journal}{Phys. Rev.} \textbf{\bibinfo{volume}{D71}},
  \bibinfo{pages}{122003} (\bibinfo{year}{2005}), \eprint{gr-qc/0504112}.

\bibitem[{\citenamefont{Timpano et~al.}(2005)\citenamefont{Timpano, Rubbo, and
  Cornish}}]{Timpano:2005gm}
\bibinfo{author}{\bibfnamefont{S.~E.} \bibnamefont{Timpano}},
  \bibinfo{author}{\bibfnamefont{L.~J.} \bibnamefont{Rubbo}}, \bibnamefont{and}
  \bibinfo{author}{\bibfnamefont{N.~J.} \bibnamefont{Cornish}}
  (\bibinfo{year}{2005}), \eprint{gr-qc/0504071}.

\bibitem[{\citenamefont{Giampieri and
  Polnarev}(1997{\natexlab{a}})}]{Giampieri:1997}
\bibinfo{author}{\bibfnamefont{G.}~\bibnamefont{Giampieri}} \bibnamefont{and}
  \bibinfo{author}{\bibfnamefont{A.~G.} \bibnamefont{Polnarev}},
  \bibinfo{journal}{Mon. Not. R. Astron. Soc.} \textbf{\bibinfo{volume}{291}},
  \bibinfo{pages}{149} (\bibinfo{year}{1997}{\natexlab{a}}).

\bibitem[{\citenamefont{Giampieri and
  Polnarev}(1997{\natexlab{b}})}]{Giampieri:1997ie}
\bibinfo{author}{\bibfnamefont{G.}~\bibnamefont{Giampieri}} \bibnamefont{and}
  \bibinfo{author}{\bibfnamefont{A.~G.} \bibnamefont{Polnarev}},
  \bibinfo{journal}{Class. Quant. Grav.} \textbf{\bibinfo{volume}{14}},
  \bibinfo{pages}{1521} (\bibinfo{year}{1997}{\natexlab{b}}).

\bibitem[{\citenamefont{Allen and Ottewill}(1997)}]{Allen:1997gp}
\bibinfo{author}{\bibfnamefont{B.}~\bibnamefont{Allen}} \bibnamefont{and}
  \bibinfo{author}{\bibfnamefont{A.~C.} \bibnamefont{Ottewill}},
  \bibinfo{journal}{Phys. Rev.} \textbf{\bibinfo{volume}{D56}},
  \bibinfo{pages}{545} (\bibinfo{year}{1997}), \eprint{gr-qc/9607068}.

\bibitem[{\citenamefont{Ungarelli and
  Vecchio}(2001{\natexlab{b}})}]{Ungarelli:2001xu}
\bibinfo{author}{\bibfnamefont{C.}~\bibnamefont{Ungarelli}} \bibnamefont{and}
  \bibinfo{author}{\bibfnamefont{A.}~\bibnamefont{Vecchio}},
  \bibinfo{journal}{Phys. Rev.} \textbf{\bibinfo{volume}{D64}},
  \bibinfo{pages}{121501} (\bibinfo{year}{2001}{\natexlab{b}}),
  \eprint{astro-ph/0106538}.

\bibitem[{\citenamefont{Cornish}(2001)}]{Cornish:2001hg}
\bibinfo{author}{\bibfnamefont{N.~J.} \bibnamefont{Cornish}},
  \bibinfo{journal}{Class. Quant. Grav.} \textbf{\bibinfo{volume}{18}},
  \bibinfo{pages}{4277} (\bibinfo{year}{2001}), \eprint{astro-ph/0105374}.

\bibitem[{\citenamefont{Cornish}(2002{\natexlab{a}})}]{Cornish:2002bh}
\bibinfo{author}{\bibfnamefont{N.~J.} \bibnamefont{Cornish}},
  \bibinfo{journal}{Class. Quant. Grav.} \textbf{\bibinfo{volume}{19}},
  \bibinfo{pages}{1279} (\bibinfo{year}{2002}{\natexlab{a}}).

\bibitem[{\citenamefont{Seto}(2004)}]{Seto:2004ji}
\bibinfo{author}{\bibfnamefont{N.}~\bibnamefont{Seto}}, \bibinfo{journal}{Phys.
  Rev.} \textbf{\bibinfo{volume}{D69}}, \bibinfo{pages}{123005}
  (\bibinfo{year}{2004}), \eprint{gr-qc/0403014}.

\bibitem[{\citenamefont{Seto and Cooray}(2004)}]{Seto:2004np}
\bibinfo{author}{\bibfnamefont{N.}~\bibnamefont{Seto}} \bibnamefont{and}
  \bibinfo{author}{\bibfnamefont{A.}~\bibnamefont{Cooray}}
  (\bibinfo{year}{2004}), \eprint{astro-ph/0403259}.

\bibitem[{\citenamefont{Kudoh and Taruya}(2005{\natexlab{a}})}]{Kudoh:2004he}
\bibinfo{author}{\bibfnamefont{H.}~\bibnamefont{Kudoh}} \bibnamefont{and}
  \bibinfo{author}{\bibfnamefont{A.}~\bibnamefont{Taruya}},
  \bibinfo{journal}{Phys. Rev.} \textbf{\bibinfo{volume}{D71}},
  \bibinfo{pages}{024025} (\bibinfo{year}{2005}{\natexlab{a}}),
  \eprint{gr-qc/0411017}.

\bibitem[{\citenamefont{Cutler}(1998)}]{Cutler:1997ta}
\bibinfo{author}{\bibfnamefont{C.}~\bibnamefont{Cutler}},
  \bibinfo{journal}{Phys. Rev.} \textbf{\bibinfo{volume}{D57}},
  \bibinfo{pages}{7089} (\bibinfo{year}{1998}), \eprint{gr-qc/9703068}.

\bibitem[{\citenamefont{Moore and Hellings}(2002)}]{Moore:1999zw}
\bibinfo{author}{\bibfnamefont{T.~A.} \bibnamefont{Moore}} \bibnamefont{and}
  \bibinfo{author}{\bibfnamefont{R.~W.} \bibnamefont{Hellings}},
  \bibinfo{journal}{Phys. Rev.} \textbf{\bibinfo{volume}{D65}},
  \bibinfo{pages}{062001} (\bibinfo{year}{2002}), \eprint{gr-qc/9910116}.

\bibitem[{\citenamefont{Peterseim et~al.}(1997)\citenamefont{Peterseim,
  Jennrich, Danzmann, and Schutz}}]{Peterseim:1997ic}
\bibinfo{author}{\bibfnamefont{M.}~\bibnamefont{Peterseim}},
  \bibinfo{author}{\bibfnamefont{O.}~\bibnamefont{Jennrich}},
  \bibinfo{author}{\bibfnamefont{K.}~\bibnamefont{Danzmann}}, \bibnamefont{and}
  \bibinfo{author}{\bibfnamefont{B.~F.} \bibnamefont{Schutz}},
  \bibinfo{journal}{Class. Quant. Grav.} \textbf{\bibinfo{volume}{14}},
  \bibinfo{pages}{1507} (\bibinfo{year}{1997}).

\bibitem[{\citenamefont{Takahashi and Nakamura}(2003)}]{Takahashi:2003wm}
\bibinfo{author}{\bibfnamefont{R.}~\bibnamefont{Takahashi}} \bibnamefont{and}
  \bibinfo{author}{\bibfnamefont{T.}~\bibnamefont{Nakamura}},
  \bibinfo{journal}{Astrophys. J.} \textbf{\bibinfo{volume}{596}},
  \bibinfo{pages}{L231} (\bibinfo{year}{2003}), \eprint{astro-ph/0307390}.

\bibitem[{\citenamefont{Edmonds}(1957)}]{Edmonds:1957}
\bibinfo{author}{\bibfnamefont{A.~R.} \bibnamefont{Edmonds}},
  \bibinfo{journal}{{\it{Angular Momentum in Quantum Mechanics}} Princeton
  University Press} p.~\bibinfo{pages}{59} (\bibinfo{year}{1957}).

\bibitem[{\citenamefont{Prince et~al.}(2002)\citenamefont{Prince, Tinto,
  Larson, and Armstrong}}]{Prince:2002hp}
\bibinfo{author}{\bibfnamefont{T.~A.} \bibnamefont{Prince}},
  \bibinfo{author}{\bibfnamefont{M.}~\bibnamefont{Tinto}},
  \bibinfo{author}{\bibfnamefont{S.~L.} \bibnamefont{Larson}},
  \bibnamefont{and} \bibinfo{author}{\bibfnamefont{J.~W.}
  \bibnamefont{Armstrong}}, \bibinfo{journal}{Phys. Rev.}
  \textbf{\bibinfo{volume}{D66}}, \bibinfo{pages}{122002}
  (\bibinfo{year}{2002}), \eprint{gr-qc/0209039}.

\bibitem[{\citenamefont{Nayak et~al.}(2003)\citenamefont{Nayak, Pai,
  Dhurandhar, and Vinet}}]{Nayak:2002ir}
\bibinfo{author}{\bibfnamefont{K.~R.} \bibnamefont{Nayak}},
  \bibinfo{author}{\bibfnamefont{A.}~\bibnamefont{Pai}},
  \bibinfo{author}{\bibfnamefont{S.~V.} \bibnamefont{Dhurandhar}},
  \bibnamefont{and} \bibinfo{author}{\bibfnamefont{J.-Y.} \bibnamefont{Vinet}},
  \bibinfo{journal}{Class. Quant. Grav.} \textbf{\bibinfo{volume}{20}},
  \bibinfo{pages}{1217} (\bibinfo{year}{2003}), \eprint{gr-qc/0210014}.

\bibitem[{\citenamefont{Cornish}(2002{\natexlab{b}})}]{Cornish:2001bb}
\bibinfo{author}{\bibfnamefont{N.~J.} \bibnamefont{Cornish}},
  \bibinfo{journal}{Phys. Rev.} \textbf{\bibinfo{volume}{D65}},
  \bibinfo{pages}{022004} (\bibinfo{year}{2002}{\natexlab{b}}),
  \eprint{gr-qc/0106058}.

\bibitem[{\citenamefont{Armstrong et~al.}(1999)\citenamefont{Armstrong,
  Estabrook, and Tinto}}]{Armstrong:1999}
\bibinfo{author}{\bibfnamefont{J.~W.} \bibnamefont{Armstrong}},
  \bibinfo{author}{\bibfnamefont{F.~B.} \bibnamefont{Estabrook}},
  \bibnamefont{and} \bibinfo{author}{\bibfnamefont{M.}~\bibnamefont{Tinto}},
  \bibinfo{journal}{Astrophys. J.} \textbf{\bibinfo{volume}{527}},
  \bibinfo{pages}{814} (\bibinfo{year}{1999}).

\bibitem[{\citenamefont{Krolak et~al.}(2004)\citenamefont{Krolak, Tinto, and
  Vallisneri}}]{Krolak:2004xp}
\bibinfo{author}{\bibfnamefont{A.}~\bibnamefont{Krolak}},
  \bibinfo{author}{\bibfnamefont{M.}~\bibnamefont{Tinto}}, \bibnamefont{and}
  \bibinfo{author}{\bibfnamefont{M.}~\bibnamefont{Vallisneri}}
  (\bibinfo{year}{2004}), \eprint{gr-qc/0401108}.

\bibitem[{\citenamefont{Press et~al.}(2002)\citenamefont{Press, Teukolsky,
  Vetterling, and Flannery}}]{Press:NRC++}
\bibinfo{author}{\bibfnamefont{W.~H.} \bibnamefont{Press}},
  \bibinfo{author}{\bibfnamefont{S.~A.} \bibnamefont{Teukolsky}},
  \bibinfo{author}{\bibfnamefont{W.~T.} \bibnamefont{Vetterling}},
  \bibnamefont{and} \bibinfo{author}{\bibfnamefont{B.~P.}
  \bibnamefont{Flannery}}, \bibinfo{journal}{{\it{Numerical Recipes in C++}}
  (Cambridge University Press)}  (\bibinfo{year}{2002}).

\bibitem[{\citenamefont{Cornish and Hellings}(2003)}]{Cornish:2003tz}
\bibinfo{author}{\bibfnamefont{N.~J.} \bibnamefont{Cornish}} \bibnamefont{and}
  \bibinfo{author}{\bibfnamefont{R.~W.} \bibnamefont{Hellings}},
  \bibinfo{journal}{Class. Quant. Grav.} \textbf{\bibinfo{volume}{20}},
  \bibinfo{pages}{4851} (\bibinfo{year}{2003}), \eprint{gr-qc/0306096}.

\bibitem[{\citenamefont{Tinto et~al.}(2004)\citenamefont{Tinto, Estabrook, and
  Armstrong}}]{Tinto:2003vj}
\bibinfo{author}{\bibfnamefont{M.}~\bibnamefont{Tinto}},
  \bibinfo{author}{\bibfnamefont{F.~B.} \bibnamefont{Estabrook}},
  \bibnamefont{and} \bibinfo{author}{\bibfnamefont{J.~W.}
  \bibnamefont{Armstrong}}, \bibinfo{journal}{Phys. Rev.}
  \textbf{\bibinfo{volume}{D69}}, \bibinfo{pages}{082001}
  (\bibinfo{year}{2004}), \eprint{gr-qc/0310017}.

\bibitem[{\citenamefont{Shaddock}(2004)}]{Shaddock:2003bc}
\bibinfo{author}{\bibfnamefont{D.~A.} \bibnamefont{Shaddock}},
  \bibinfo{journal}{Phys. Rev.} \textbf{\bibinfo{volume}{D69}},
  \bibinfo{pages}{022001} (\bibinfo{year}{2004}), \eprint{gr-qc/0306125}.

\bibitem[{\citenamefont{Binney et~al.}(1997)\citenamefont{Binney, Gerhard, and
  Spergel}}]{Binney:1996sv}
\bibinfo{author}{\bibfnamefont{J.}~\bibnamefont{Binney}},
  \bibinfo{author}{\bibfnamefont{O.}~\bibnamefont{Gerhard}}, \bibnamefont{and}
  \bibinfo{author}{\bibfnamefont{D.}~\bibnamefont{Spergel}},
  \bibinfo{journal}{Mon. Not. R. Astron. Soc.} \textbf{\bibinfo{volume}{288}},
  \bibinfo{pages}{365} (\bibinfo{year}{1997}), \eprint{astro-ph/9609066}.

\bibitem[{\citenamefont{Adams and Swarztrauber}(2003)}]{Adams:2003}
\bibinfo{author}{\bibfnamefont{J.~C.} \bibnamefont{Adams}} \bibnamefont{and}
  \bibinfo{author}{\bibfnamefont{P.~N.} \bibnamefont{Swarztrauber}},
  \bibinfo{journal}{http://www.scd.ucar.edu/css/software/spherepack/}
  (\bibinfo{year}{2003}).

\bibitem[{\citenamefont{Kudoh and
  Taruya}(2005{\natexlab{b}})}]{Kudoh:2005inprep}
\bibinfo{author}{\bibfnamefont{H.}~\bibnamefont{Kudoh}} \bibnamefont{and}
  \bibinfo{author}{\bibfnamefont{A.}~\bibnamefont{Taruya}}
  (\bibinfo{year}{2005}{\natexlab{b}}), \eprint{in preparation}.

\bibitem[{\citenamefont{Bender and {\it{et. al.},}}(1998)}]{Bender:1998}
\bibinfo{author}{\bibfnamefont{P.~L.} \bibnamefont{Bender}} \bibnamefont{and}
  \bibinfo{author}{\bibnamefont{{\it{et. al.},}}}, \bibinfo{journal}{LISA
  Pre-Phase A Report}  (\bibinfo{year}{1998}).

\end{thebibliography}

\end{document}